\pdfoutput=1
\documentclass[11pt]{article}

\font\grande=cmr9.5 scaled \magstep4
\font\medio=cmr9.5 scaled \magstep2
\outer\def\beginsection#1\par{\medbreak\bigskip
      \message{#1}\leftline{\bf#1}\nobreak\medskip
\vskip-\parskip
      \noindent}
\usepackage{graphicx} 
\usepackage{mathrsfs}
\usepackage[overload]{textcase}
\begin{document}

\bibliographystyle{unsrt}

\titlepage

\vspace{1cm}
\begin{center}
{\grande Baryogenesis,  magnetogenesis}\\
\vspace{0.5 cm}
{\grande and the strength of anomalous interactions}\\
\vspace{1cm}
Massimo Giovannini \footnote{e-mail address: massimo.giovannini@cern.ch}\\
\vspace{1cm}
{{\sl Department of Physics, CERN, 1211 Geneva 23, Switzerland }}\\
\vspace{0.5cm}
{{\sl INFN, Section of Milan-Bicocca, 20126 Milan, Italy}}
\vspace*{1cm}
\end{center}
\vskip 0.3cm
\centerline{\medio  Abstract}
\vskip 0.1cm
The production of the hypermagnetic gyrotropy is investigated under the assumption that the gauge coupling smoothly evolves during a quasi-de Sitter phase and then flattens out in the radiation epoch by always remaining perturbative. In the plane defined by the strength of the anomalous interactions and by the rate of evolution of the gauge coupling the actual weight of the pseudoscalar interactions turns out to be always rather modest if major deviations from the homogeneity are to be avoided during the inflationary phase.  Even if the gauge power spectra are related by duality only in the absence of anomalous contributions, an approximate duality symmetry  constrains the late-time form of the hypermagnetic power spectra. Since the hypermagnetic gyrotropy associated with the modes reentering prior to the phase transition must be released into fermions later on,  the portions of  the parameter space where the obtained baryon asymmetry is close to the observed value are the most relevant for the present ends.  For the same range of parameters the magnetic power spectra associated with the modes reentering after symmetry breaking may even be of the order of a few hundredths of a nG over typical length scales comparable with the Mpc prior to the collapse of the protogalaxy.  

\noindent
\vspace{5mm}
\vfill
\newpage
\renewcommand{\theequation}{1.\arabic{equation}}
\setcounter{equation}{0}
\section{Introduction}
\label{sec1}
If parity is globally broken the average of the scalar product of the bulk velocity of the plasma with the corresponding vorticity does not vanish (i.e. $\langle \vec{v} \cdot (\vec{\nabla} \times \vec{v}) \rangle \neq 0$) and the the kinetic energy of 
the charged liquid can be transferred  to the magnetic field: in this case the resulting Ohmic currents will be directed in average 
along the magnetic field itself \cite{moffat,parker,zeldovich}. The quantity $\vec{v} \cdot (\vec{\nabla} \times \vec{v})$  (sometimes referred to as kinetic gyrotropy) measures the amount of parity breaking of the plasma and in the context of mean-field dynamos Vainshtein and Zeldovich \cite{zeld1,zeld2} (see also \cite{kaza,kraich}) 
introduced also the notion of {\em magnetic gyrotropy} by noting that the only possible pseudoscalar quadratic in $\vec{B}$ must be of the form $\vec{B}\cdot \vec{\nabla} \times \vec{B}$.  While the turbulent dynamo theory aims at producing the magnetic gyrotropy from its kinetic counterpart, in this paper we shall scrutinize the opposite process and discuss the direct production of magnetic gyrotropy during an inflationary stage of expansion with the purpose of building concrete models where the magnetogenesis constraints and the baryogenesis requirements are simultaneously satisfied in a single dynamical framework at weak coupling. 
 
The hydromagnetic dynamos are one of the key ingredients for the generation of large-scale magnetic fields in turbulent environments \cite{moffat,parker,zeldovich} 
but since the typical scale of the gravitational collapse of the protogalaxy is of the order of the Mpc, 
the  initial conditions of the large-scale magnetism must be somehow included in a wider cosmological picture, as lucidly suggested by Hoyle  \cite{ONEa} in the past century.  While the first realizations of this idea demanded drastic departures from the isotropy of the background at early times  \cite{ONEb,ONEc,ONEd}, according to the tenets of the inflationary paradigm the  primeval anisotropies and inhomogeneities of the expansion are washed out as soon as the inflationary event horizon is formed \cite{TWOa,TWOb,TWOc,TWOd} (see also \cite{TWOe} for a general introduction to the motivations of the inflationary paradigm).  
Even if there exist particularly contrived inflationary scenarios somehow compatible with an early anisotropy,  the  gradient expansion pioneered in Refs. \cite{TWOf,TWOg} suggests that  in conventional inflationary models any finite portion of the universe gradually loses the 
memory of an initially imposed anisotropy or inhomogeneity so that the universe attains the observed regularity regardless of the initial boundary conditions. Following the first formulations of the inflationary paradigm the attention has then been turned to the possibility that the gauge fields are  parametrically amplified in the early Universe and eventually behave as vector random fields that do not break the spatial isotropy. In this context the problem is however associated with the invariance under Weyl rescaling that forbids any efficient amplification of gauge fields and chiral fermions in conformally flat background geometries \cite{THREEb,THREEc}.  

One of the first attempts to break Weyl invariance in a cosmological setting by relying on the
pseudoscalar coupling of an Abelian gauge field has been described in Refs. \cite{FOURa,FOURb} where  
it was argued that in the case of an oscillating axion-like field the 
modes that are substantially amplifed during a quasi-de Sitter stage are the ones for which $k  \geq {\mathcal O}(a H)$ 
where $H$ denotes the Hubble rate, $a$ is the scale factor and $k$ is the comoving wavenumber (see also \cite{FOURc}  for a more complete 
scenario).  Even if the pseudoscalar coupling only leads to weak breaking of Weyl invariance, it efficiently modifies the topological properties of the hypermagnetic flux lines in the electroweak plasma \cite{FIVEa0}, as as originally discussed in \cite{FIVEa1} by taking into account the chemical potentials associated with the finite density effects. If 
the gyrotropy  sufficiently large the produced Chern-Simons condensates may decay and eventually produce the baryon asymmetry \cite{FIVEa1,FIVEb1,FIVEb2,FIVEb3,FIVEb4,FIVEc}. These  gyrotropic and helical fields play also a key role in various aspects of anomalous 
magnetohydrodynamics \cite{SIXa}. In the collisions of heavy ions this phenomenon is often dubbed chiral magnetic effect \cite{SIXa,SIXb} (see also \cite{SIXe,SIXf}) even if there are some differences between the formulation of anomalous magnetohydrodynamics \cite{SIXc,SIXd} (where Ohmic and chiral currents are concurrently present)
 and the standard chiral magnetic effects (where only chiral currents are customarily discussed).
 
The purpose of this paper is to address the origin of large-scale magnetism and of the baryon 
asymmetry of the Universe (BAU) in a unified dynamical framework by considering a more general variant of the scenario 
suggested in Refs. \cite{FOURa,FOURb,FOURc}. The idea is to produce the hypermagnetic gyrotropy during the quasi-de Sitter stage of expansion while the gauge coupling remains always perturbative.  This suggestion goes back to the two-step model of Ref. \cite{FIVEa0} where in a first step the  amplification of the gauge fields took place for typical scales larger than the effective horizon and, in the second step,  the combined evolution of the chemical potentials (and of other pseusoscalar fields)  twisted the hypercharge flux lines inside the 
electroweak horizon by ultimately producing hypermagnetic knots. Since the actual amplification of the gauge fields is controlled by the evolution of the gauge coupling,
to address simultaneously the baryogenesis and the magnetogenesis\footnote{Some time ago the problems related to the generation 
of large-scale magnetic fields has been dubbed {\em magnetogenesis} \cite{SEVENa}. Since then this terminology 
has been widely employed even if the problem itself is much older and, as already mentioned, can be traced back to the suggestions of Hoyle \cite{ONEa} (see also \cite{moffat,parker,zeldovich}.} problems the general form of the curved-space action 
will be taken in the following form:
\begin{equation}
S_{gauge} = - \frac{1}{16 \pi} \int d^{4} x \, \sqrt{- G} \, \biggl[ \lambda \,Y_{\alpha\beta} \, Y^{\alpha\beta} + \overline{\lambda} \, Y_{\alpha\beta} \, \widetilde{\,\,Y\,\,}^{\alpha\beta} \biggr].
\label{action1}
\end{equation}
The evolution of some scalar degree of freedom like the inflaton or some other spectator field may enter 
directly the definition of $\lambda$ and $\overline{\lambda}$.  The  logic conventionally applied to this 
kind of problems would suggest to fix the couplings and then proceed further: this is not the strategy 
 pursued here. On the contrary,  instead of fixing the couplings and then deducing the 
potential phenomenological implications, the  strength of the pseudoscalar 
interactions will be treated a free parameter that must be simultaneously 
compatible with the critical density constraints and with the perturbative evolution of the gauge coupling. 
The considerations reported here are therefore quite general since what ultimately matters is the overall 
evolution of the gauge coupling and their interplay with the anomalous terms; in this sense it could even 
be possible that $\lambda$ and $\overline{\lambda}$ depend on a general function of the scalar curvature 
or of some other curvature invariant (see e.g. \cite{EIGHTd}).

While both $\lambda$ and $\overline{\lambda}$ break Weyl invariance when $\overline{\lambda} \to 0$  the duality symmetry \cite{EIGHTa,EIGHTb,EIGHTc}  must be recovered. In the absence of sources duality rotates field strengths into their duals (i.e. tensors into pseudotensors) and the dual limit of the action (\ref{action1}) (i.e. $\overline{\lambda}\to 0$) is a useful guide for the correct determination asymptotes of the mode functions. Instead of committing ourselves to a specific model 
we shall instead assume a given evolution 
for the gauge coupling and then deduce the limits on the strength of the anomalous terms.
During the quasi-de Sitter stage of expansion the strength of the anomalous interactions 
(i.e. $\overline{\lambda}$) will turn out to be strongly constrained by the critical density 
bound but the gyrotropic configurations of the hypermagnetic fields will still be 
comparable with the values required to seed the BAU.

Since the scenario examined here is not conventional it is useful to remind the more standard perspectives of the problem. 
Through the years different realisations of the original Sakharov idea \cite{sak} have been proposed and the standard lore of baryogenesis stipulates that during a strongly first-order electroweak phase transition the expanding bubbles are nucleated while the baryon number is violated by sphaleron processes \cite{CKN1}. In the light of the current value of the Higgs mass to produce a sufficiently strong (first-order) phase transition and to get enough $CP$ violation at the 
bubble wall, the standard electroweak theory must be appropriately extended. Various suggestions exist along this 
direction the most common being the addition of an extra (singlet) scalar field, the presence of a supplementary Higgs doublet and the inclusion of higher-dimensional operators associated with the Higgs sector. Not to mention the possibility of light stops in the supersymmetric extensions of the minimal standard model.
A  complementary lore for the generation of the BAU is leptogenesis (see e.g. \cite{CKN2}) which can be conventionally realized thanks to heavy Majorana neutrinos decaying out of equilibrium and producing an excess of lepton 
number ($L$ in what follows). The excess in $L$ can lead to the observed baryon number thanks to sphaleron interactions violating $(B+ L)$.  We suggest here that the BAU could be the result of the decay of the hypermagnetic gyrotropy. While the $SU_{L}(2)$ anomaly is typically responsible for $B$ and $L$  nonconservation via instantons and sphalerons, the $U_{Y}(1)$ anomaly might lead to the transformation of the infrared modes of the hypercharge field into fermions. Fo this reason the production of the BAU demands, in this context, the dynamical generation of the gyrotropic 
configurations of the hypermagnetic field as argued,  in Refs. \cite{FIVEa0,FIVEa1} (see also \cite{FIVEb1,FIVEb2,FIVEb3,FIVEb4}). Even if the idea explored here is admittedly less conventional it certainly provides a useful playground for the potential unification of the baryogenesis conditions and of the magnetogenesis requirements.

The layout of this paper is therefore the following. The classical and the quantum 
descriptions of the problem will be introduced in section \ref{sec2} while the general form of the power spectra and of their dual limits will be presented in section \ref{sec3}.  In section \ref{sec4} we shall analyze the general framework where the gauge coupling increases during inflation and then flattens out at a tuneable rate. After computing all the relevant power spectra in the different dynamical regimes the  bounds on the anomalous terms will be  derived. The dual description (implying that the gauge coupling decreases during a quasi-de Sitter stage) will be studied in section \ref{sec5}. The results of sections \ref{sec4} and \ref{sec5} provide a direct test of the duality symmetry and of its partial breaking induced by the anomalous terms. In section \ref{sec6} the obtained results will be 
considered along a more phenomenological perspective by analyzing the relevant scales that reenter 
the effective horizon. At different times during the radiation phase 
the smaller scales that are inside the electroweak horizon will reenter first: these scales  affect
 the generation of the BAU and will be discussed in the first part of section \ref{sec6}. Just 
before matter-radiation equality the reentry of the scales ${\mathcal O}(\mathrm{Mpc})$ takes place: these scales will be 
crucial for the magnetogenesis requirements and they will be discussed in the second part of section \ref{sec6}.
Section \ref{sec7} contains our concluding considerations. To avoid extensive digressions some of the technical results that are 
relevant for the derivations have been relegated to the appendices \ref{APPA} and \ref{APPB}.

\renewcommand{\theequation}{2.\arabic{equation}}
\setcounter{equation}{0}
\section{Classical and quantum descriptions}
\label{sec2}
A useful parametrization of the action (\ref{action1}) follows when the inverse 
of $\lambda$ is identified with the gauge coupling by setting 
$e^2 = 4 \pi/\lambda$. In this way the action gets modified as: 
 \begin{equation}
S_{gauge} = \int d^{4} x \, \sqrt{- G} \, \biggl[ - \frac{1}{4 e^2}Y_{\alpha\beta} \, Y^{\alpha\beta}  - \frac{1}{4 e^2} \biggl(\frac{\overline{\lambda}}{\lambda}\biggr) \, Y_{\alpha\beta} \, \widetilde{\,\,Y\,\,}^{\alpha\beta} \biggr],\qquad e^2 = \frac{4\pi}{\lambda}.
\label{action2}
\end{equation}
The form of actions (\ref{action1}) and (\ref{action2}) is equivalent but, as we shall argue, 
the latter is more convenient than the former. 
According to Eq. (\ref{action2}) the gauge coupling increases 
when $\sqrt{\lambda}$ decreases and viceversa.  While the first part of the action (\ref{action2}) notoriously 
leads to a system of equation that is invariant under duality, the inclusion of the second term 
breaks both the duality symmetry and the Weyl symmetry. Equation (\ref{action2}) 
also suggests that the strength of anomalous interactions is measured by the ratio of $(\overline{\lambda}/\lambda)$. 
The simplest choice is to consider the case  
where $\overline{\lambda}$ and $\lambda$ are just proportional
(i.e. $\overline{\lambda} = \lambda_{0} \lambda$); this class of scenarios, in its simplicity, encompasses 
various possibilities examined in the previous literature and will be particularly suitable for the 
present ends. With these necessary specifications, from Eq. (\ref{action1}) the relevant evolution equations in the covariant form are given by:
\begin{equation}
\nabla_{\alpha} \biggl(\lambda \,Y^{\alpha\beta} \biggr) +  \nabla_{\alpha} \biggl(\overline{\lambda}\, \widetilde{\,Y\,}^{\alpha\beta} \biggr)=0, \qquad \nabla_{\alpha} \,\widetilde{\,Y\,}^{\alpha\beta} =0, \qquad \widetilde{\, Y\,}^{\alpha\beta} = \frac{1}{2} E^{\alpha\beta\rho\sigma} Y_{\rho\sigma}, 
\label{eq2}
\end{equation}
where  $Y_{\alpha\beta}$ is the gauge fields strength 
while $\widetilde{\,Y\,}^{\alpha\beta} $ is the dual field strength defined. We shall also introduce, as usual, 
 $E^{\alpha\beta\rho\sigma} = \epsilon^{\alpha\beta\rho\sigma}/\sqrt{- G}$ where $\epsilon^{\alpha\beta\rho\sigma}$ 
 is the Levi-Civita symbol in four-dimensions. The variation of Eq. (\ref{action1}) with respect to the metric leads to the total energy-momentum tensor does not depend on $\overline{\lambda}$:
\begin{equation}
{\mathcal T}_{\mu}^{\,\,\nu} = \frac{\lambda}{4 \pi} \biggl[ - Y_{\mu\, \alpha} \, Y^{\nu\alpha} + \frac{1}{4} \,Y_{\alpha\beta} Y^{\alpha\beta} \,\,\delta_{\mu}^{\nu} \biggr].
\label{eq4}
\end{equation}

\subsection{Comoving and physical fields}
It is practical to discuss the evolution equations and the components of the energy-momentum 
tensor not in terms of the physical fields but rather with the comoving (i.e. rescaled) 
fields.  In conformally flat background geometries $\overline{g}_{\mu\nu}= a^2(\tau) \eta_{\mu\nu}$ [where $a(\tau)$ is the scale factor and $\eta_{\mu\nu}$ 
is the Minkowski metric] the classical equations derived from Eq. (\ref{action1}) can be directly expressed 
in terms of the {\em comoving} fields (denoted hereunder by $\vec{E}$ and $\vec{B}$) which are 
related to the {\em physical} fields as:
\begin{equation}
\vec{\,E\,}  =  \, a^2 \, \sqrt{\lambda}\, \vec{\,E\,}^{(phys)}, \qquad \vec{\,B\,} = a^2 \, \sqrt{\lambda}\,\vec{\,B\,}^{(phys)}.
\label{ACT13}
\end{equation}
The components of the field strengths are directly expressible as $Y_{i\, 0} = - a^2 E_{i}^{(phys)}$,  $Y^{i\, j} = - \epsilon^{i\, j\, k} B_{k}^{(phys)}/a^2$ and similarly for the dual strength. While the physical fields are essential for the discussion of the actual magnetogenesis constraints, the explicit components of Eq. (\ref{eq2}) simplify greatly by using the comoving fields:
\begin{eqnarray}
 \vec{\nabla} \times \biggl(\sqrt{ \lambda} \, \vec{\,\,B\,\,}\biggr) = \partial_{\tau} \biggl( \sqrt{\lambda}  \vec{\,\,E\,\,}\biggr) + \biggl(\frac{\overline{\lambda}^{\prime}}{\sqrt{\lambda}}\biggr) \vec{B} +  \frac{\vec{\nabla} \, \overline{\lambda} \times \vec{\,\,E\,\,}}{\sqrt{\lambda}},\qquad \vec{\nabla}\cdot\biggl( \sqrt{\lambda}\, \vec{\,\,E\,\,} \biggr) = 
\frac{ \vec{\,\,B\,\,}\cdot \vec{\nabla} \lambda}{\sqrt{\lambda}},
\label{CE3}
\end{eqnarray}
where the prime denotes a derivation with respect to the conformal time coordinate 
$\tau$ while $\vec{\nabla}$ are the spatial gradients (we remind that, as usual, 
the conformal time coordinate $\tau$ is related to the cosmic time as $d\tau a(\tau) = d\,t$). 
Finally, from Eq. (\ref{eq2}) the explicit form of the Bianchi identity in terms 
of the comoving fields becomes:
\begin{equation}
\vec{\nabla} \times \biggl(\frac{ \vec{\,\,E\,\,}}{\sqrt{\lambda}}\biggr) + \partial_{\tau} \biggl(\frac{ \vec{\,\,B\,\,}}{\sqrt{\lambda}}\biggr) =0, \qquad \vec{\nabla}\cdot\biggl( \frac{ \vec{\,\,B\,\,}}{\sqrt{\lambda}}\biggr) =0.
\label{CE2}
\end{equation}
The various components of the energy-momentum tensor (\ref{eq4})
can be expressed either with the physical or with the comoving fields; using 
then $\vec{E}$ and $\vec{B}$ we have:
\begin{eqnarray}
{\mathcal T}_{0}^{\,\,0} = \rho_{B} + \rho_{E},\qquad {\mathcal T}_{0}^{\,\,i} = \frac{1}{4 \pi a^4} \bigl( \vec{\,E\,} \times \vec{\,B\,} \bigr)^{i},\qquad
{\mathcal T}_{i}^{\,\,j} = - (p_{E} + p_{B})\,\, \delta_{i}^{j} \,+ \Pi_{E\,\,i}^{\,\,\,j} \,+ \Pi_{B\,\,i}^{\,\,\,j}.
\label{ACT8}
\end{eqnarray}
In Eq. (\ref{ACT8}) we introduced the energy densities (i.e. $\rho_{B}= B^2/(8 \pi a^4)$ and $\rho_{E}= E^2/(8 \pi a^4)$), the pressures 
(i.e. $p_{B}= \rho_{B}/3$ and $p_{E}= \rho_{E}/3$) and the anisotropic stresses
\begin{eqnarray}
\Pi_{E\,\,i}^{\,\,\,j} &=& \frac{1}{4 \pi a^4} \biggl( E_{i} \, E^{j} - \frac{E^2}{3} \delta_{i}^{j} \biggr),\qquad 
\Pi_{B\,\,i}^{\,\,\,j} = \frac{1}{4 \pi a^4} \biggl( B_{i} \, B^{j} - \frac{B^2}{3} \delta_{i}^{j} \biggr), 
\label{ACT10}
\end{eqnarray}
of the hypermagnetic and hyperelectric fields; note also that $E^2 = \vec{\,E\,}\cdot \vec{\,E\,}$ and $B^2 = \vec{\,B\,}\cdot \vec{\,B\,}$. If the components of the energy-momentum tensor 
are expressed in terms of the comoving fields the redshift factor is standard and the 
dependence upon $\lambda$ is included in the definition of $\vec{E}$ and $\vec{B}$. 

\subsection{Quantum mechanical initial conditions} 
Because of the coupling between the two linear polarizations it is preferable to start with the appropriate action and with the related Hamiltonian. In spite of the different conventions and the more general treatment the results  of the present discussion do not differ from Ref. \cite{FIVEa0}. Since the Lorentz gauge condition is in general not preserved by a conformal rescaling, the Coulomb gauge  (i.e. $Y_{0} =0$ and $\vec{\nabla}\cdot \vec{Y} =0$) is always preferable \cite{TENa} so that  Eq. (\ref{action1}) becomes:
\begin{equation}
S_{Y} = \frac{1}{2}\int \,d\tau\, d^{3} x \, \biggl[ {\mathcal Y}_{i}^{\,\prime \,2} + {\mathcal F}^2 {\mathcal Y}_{i}^{\,2}  - 2 {\mathcal F}\, {\mathcal Y}_{i} \, {\mathcal Y}_{i}^{\,\prime} - \partial_{k} {\mathcal Y}_{i} \,\partial^{k} {\mathcal Y}_{k}+ \biggl(\frac{\overline{\lambda}^{\,\prime}}{\lambda}\biggr)
{\mathcal Y}_{i}\, \partial_{j} {\mathcal Y}_{k} \, \epsilon^{i \,j \,k}\biggr],
\label{actq}
\end{equation}
where $\vec{{\mathcal Y}} = \sqrt{\lambda/(4\pi)} \,\, \vec{Y}$ is the (rescaled) vector potential.  By dropping a total time derivative that does not contribute to the equations of motion, Eq. (\ref{actq})  follows from 
the action (\ref{action1}) by assuming that  $\sqrt{\lambda}$  only depends on the conformal time coordinate $\tau$. The  operators corresponding to the classical fields are promoted to the status of quantum operators by adding a caret  
on their expressions:
\begin{equation}
{\mathcal Y}_{i} \to \widehat{{\mathcal Y}}_{i}, \qquad \pi_{i} \to \widehat{\pi}_{i} = \widehat{{\mathcal Y}}_{i}^{\,\,\prime} - {\mathcal F} \, \widehat{{\mathcal Y}}_{i}, \qquad {\mathcal F} = \frac{\sqrt{\lambda}^{\,\prime}}{\sqrt{\lambda}},
\label{FOP1}
\end{equation}
where $\widehat{\pi}_{i}$ is the canonical momentum and ${\mathcal F}$ denotes throughout the rate of variation of $\sqrt{\lambda}$. In terms of the canonical fields and of the canonical momenta, the Hamiltonian operator is:
\begin{equation}
\widehat{H}_{Y}(\tau) = \frac{1}{2} \int d^3 x \biggl[ \widehat{\pi}_{i}^{2}  + 
{\mathcal F}\biggl( \widehat{\pi}_{i}\,\widehat{\mathcal Y}_{i} + \widehat{\mathcal Y}_{i}\,\widehat{\pi}_{i}\biggr)+
\partial_{i}\widehat{{\mathcal Y}}_{k} \,\,\partial^{i} \widehat{{\mathcal Y}}_{k} - \biggl(\frac{\overline{\lambda}^{\, \prime}}{ \lambda} \biggr)\, \widehat{{\mathcal Y}}_{i} \,\,\partial_{j} \widehat{{\mathcal Y}}_{k} \,\, \epsilon^{i \, j\, k}\biggr].
\label{hamq}
\end{equation}
Consequently the evolution equations of the field operators following form the Hamiltonian (\ref{hamq}) are (units $\hbar = c =1$ will be adopted):
\begin{eqnarray}
\widehat{\pi}_{i}^{\,\,\prime} &=& i\, \biggl[ \widehat{H}_{Y}, \widehat{\pi}_{i} \biggr] = - {\mathcal F} \, \widehat{\pi}_{i} + \nabla^2 \widehat{{\mathcal Y}}_{i} + \frac{\overline{\lambda}^{\,\prime}}{\lambda} \, \epsilon_{i\, j\, k} \partial^{j} \, \widehat{{\mathcal Y}}^{k},
\nonumber\\
\widehat{{\mathcal Y}}_{i}^{\,\,\prime} &=&   i\, \biggl[ \widehat{H}_{Y}, \widehat{{\mathcal Y}}_{i} \biggr] = \widehat{\pi}_{i} + {\mathcal F} \, \widehat{{\mathcal Y}}_{i}.
\label{can2}
\end{eqnarray}
The initial data of the field operators must obey canonical commutation relations 
at equal times:
\begin{equation}
\biggl[\widehat{{\mathcal Y}}_{i}(\vec{x}_{1}, \tau),\widehat{\pi}_{j}(\vec{x}_{2}, \tau)\biggr] = i \Delta_{ij}(\vec{x}_{1} - \vec{x}_{2}),\qquad 
\Delta_{ij}(\vec{x}_{1} - \vec{x}_{2}) = \int \frac{d^{3}k}{(2\pi)^3} e^{i \vec{k} \cdot (\vec{x}_{1} - \vec{x}_2)} p_{ij}(\hat{k}), 
\label{commrel}
\end{equation}
where $p_{ij}(\hat{k}) = (\delta_{ij} - \hat{k}_{i} \hat{k}_{j})$. The function $\Delta_{ij}(\vec{x}_{1} - \vec{x}_{2})$ is the transverse generalization of the Dirac delta function ensuring that  both the field operators and the canonical momenta are divergenceless. The two equations appearing in  (\ref{can2}) suggest the use of the circular basis where the right and left polarizations are defined, respectively, as $\hat{\varepsilon}^{(R)}= [\hat{e}^{(\oplus)} + i\, \hat{e}^{(\otimes)}]/\sqrt{2}$ and as $\hat{\varepsilon}^{(L)} = [\hat{e}^{(\oplus)} - i\, \hat{e}^{(\otimes)}]/\sqrt{2}$; as usual $\hat{e}^{\oplus}$ and $\hat{e}^{\otimes}$ denote two unit vectors mutually orthogonal and orthogonal to the direction of propagation $\hat{k}$ (i.e. $\hat{e}^{\oplus}\times \hat{e}^{\otimes} = \hat{k}$). From these definitions the following vector products are immediately obtained:
\begin{equation}
\hat{k} \times \hat{\varepsilon}^{(R)} = - i\, \hat{\varepsilon}^{(R)}, \qquad  \hat{k} \times \hat{\varepsilon}^{(L)} = i\, \hat{\varepsilon}^{(L)}, \qquad  \hat{\varepsilon}^{(R)\ast} =  \hat{\varepsilon}^{(L)}, \qquad  \hat{\varepsilon}^{(L)\ast} =  \hat{\varepsilon}^{(R)}.
\label{POL2}
\end{equation}
In the circular basis of Eq. (\ref{POL2})  the expansions of the field operators and and of the associated canonical momenta are:
\begin{eqnarray}
\widehat{{\mathcal Y}}_{i}(\vec{x},\tau) &=&  \sum_{\alpha= R,\, L} \, \,\int\frac{d^{3} k}{(2\pi)^{3/2}}\,\,
\biggl[ f_{k,\, \alpha}(\tau) \,\, \widehat{a}_{k,\,\alpha}\,\, \varepsilon^{(\alpha)}_{i}(\hat{k})\, \, e^{- i \vec{k} \cdot\vec{x}} +  f_{k,\, \alpha}^{\ast}(\tau) \,\, \widehat{a}^{\dagger}_{k,\alpha} \,\,\varepsilon^{(\alpha)\,\ast}_{i}(\hat{k})\,\,e^{ i \vec{k} \cdot\vec{x}}\biggr],
\label{expnew1}\\
\widehat{\pi}_{i}(\vec{x},\tau) &=&\sum_{\alpha=R,\, L}\,\, \int\frac{d^{3} k}{(2\pi)^{3/2}} \,\,
\biggl[ g_{k,\,\alpha}(\tau) \, \widehat{a}_{k,\alpha}  \,\,\varepsilon^{(\alpha)}_{i}(\hat{k}) \,\,e^{- i \vec{k} \cdot\vec{x}} +  g_{k,\alpha}^{\ast}(\tau) \,\, \widehat{a}^{\dagger}_{k,\alpha}\,\,\varepsilon^{(\alpha)\,\ast}_{i}(\hat{k})\,\,  e^{ i \vec{k} \cdot\vec{x}}\biggr], 
\label{expnew2}
\end{eqnarray}
where the creation and annihilation operators $\widehat{a}_{k,\,\alpha}$ and $\widehat{a}_{k,\,\alpha}^{\dagger}$ 
are directly defined in the circular basis and they obey the standard 
commutation relation $[\widehat{a}_{\vec{k}, \, \alpha}, \, \widehat{a}_{\vec{p}, \, \beta}] = \delta^{(3)}(\vec{k}- \vec{p})\, \delta_{\alpha\beta}$. The mode functions $f_{k,\, \alpha}(\tau)$ and $g_{k,\, \alpha}(\tau)$ will be often referred to as the 
{\em hypermagnetic} and {\em hyperelectric} mode functions respectively. Since the canonical commutation relations (\ref{commrel}) must be preserved by the time evolution the mode functions are normalized as: 
\begin{equation}
f_{k,\, \alpha} \, g^{\ast}_{k,\, \alpha} - f_{k,\, \alpha}^{\ast} \, g_{k,\, \alpha} =\,i,\qquad\qquad \alpha= R,\, L,
\label{MF2a}
\end{equation}
 so that Eq. (\ref{MF2a}) holds independently for each of the two circular polarizations. The actual evolution of the mode functions follows by inserting Eqs. (\ref{expnew1})--(\ref{expnew2}) into 
Eq. (\ref{can2}) and it is given by:
\begin{eqnarray}
f_{k,\, L}^{\,\prime} &=& g_{k,\,L} + {\mathcal F} f_{k,\, L},\qquad g_{k,\,L}^{\,\prime} = - k^2 \, f_{k,\, L} - {\mathcal F} \,g_{k,\,L}  + \,  \biggl(\frac{\overline{\lambda}^{\,\prime}}{\lambda}\biggr)\, k \, f_{k,\, L},
\label{MF8}\\
f_{k,\, R}^{\,\prime} &=& g_{k,\,R} + {\mathcal F} f_{k,\, R},\qquad
g_{k,\,R}^{\,\prime} = - k^2 \, f_{k,\, R} - {\mathcal F} \,g_{k,\,R}  - \,  \biggl(\frac{\overline{\lambda}^{\,\prime}}{\lambda}\biggr)\, k \, f_{k,\, R}.
\label{MF9}
\end{eqnarray}
Equations (\ref{MF8}) and (\ref{MF9}) differ by a sign in the last term at the right hand side. In the limit $\overline{\lambda} \to 0$ the circularly polarized mode functions must coincide up to an overall phase:
\begin{equation}
\lim_{\overline{\lambda} \to 0} f_{k,\,R} = \lim_{\overline{\lambda} \to 0} f_{k,\,L} = e^{-i \pi/4} f_{k}, \qquad \lim_{\overline{\lambda} \to 0} g_{k,\,R} = \lim_{\overline{\lambda} \to 0}  g_{k,\,L} = e^{i \pi/4} g_{k},
\label{MF10}
\end{equation} 
where $f_{k}$ and $g_{k}$ obey  Eqs. (\ref{MF8})--(\ref{MF9}) in the limit $\overline{\lambda}\to 0$; this 
observation is useful in order to check, a posteriori, the correctness of certain analytic 
continuations. Some details on this topic can be found in appendix \ref{APPA}.

\renewcommand{\theequation}{3.\arabic{equation}}
\setcounter{equation}{0}
\section{Power spectra, energy densities and gyrotropies}
\label{sec3}
The hyperelectric and hypermagnetic mode functions introduced above actually determine the corresponding 
quantum fields; in fact from Eqs. (\ref{expnew1})--(\ref{expnew2}) 
the hyperelectric and hypermagnetic fields are:
\begin{eqnarray}
\widehat{E}_{i}(\vec{x},\tau) &=& - \widehat{\pi}_{i}(\vec{x},\tau) = - \sqrt{\lambda} \biggl(\frac{\widehat{{\mathcal Y}}_{i}}{\sqrt{\lambda}}\biggr)^{\, \prime} 
\nonumber\\
&=& -   \sum_{\alpha= R,\,L} \, \,\int\frac{d^{3} k}{(2\pi)^{3/2}}\,\,
\biggl[ g_{k,\,\alpha}(\tau) \, \widehat{a}_{k,\alpha} \,\,  \varepsilon^{(\alpha)}_{i}(\hat{k})\,\,e^{- i \vec{k} \cdot\vec{x}} + 
\mathrm{h.c.}\biggr],
\label{PS1}\\
\widehat{B}_{k}(\vec{x}, \tau) &=& \epsilon_{i\,j \, k} \,\partial_{i}\, \widehat{{\mathcal Y}}_{j}
\nonumber\\
&=&  - \frac{i \, \,\epsilon_{i\,j\,k}}{(2\pi)^{3/2}}\,   \sum_{\alpha= R,\,L}\,\,\int\, d^{3} k\,\, k_{j} \,\,
\biggl[ f_{k,\, \alpha}(\tau) \, \widehat{a}_{k,\,\alpha}\, \,\,  \varepsilon^{(\alpha)}_{i}(\hat{k})\, e^{- i \vec{k} \cdot\vec{x}} - \mathrm{h.c.}\biggr],
\label{PS2}
\end{eqnarray}
where ``h.c.'' denotes the Hermitian conjugate of the preceding expression in each of the squared 
brackets. The Fourier representation of the operators $\widehat{E}_{i}(\vec{x},\tau)$ and $\widehat{B}_{k}(\vec{x},\tau)$
is: 
\begin{eqnarray}
\widehat{E}_{i}(\vec{q},\tau) &=& - \,\sum_{\alpha= R,\, L}\biggl[ \varepsilon_{i}^{(\alpha)}(\hat{q}) \, g_{q,\,\alpha} \widehat{a}_{\vec{q},\,\alpha} +\varepsilon_{i}^{(\alpha)\ast}(-\hat{q}) \, g_{q,\,\alpha}^{\ast} \widehat{a}_{-\vec{q},\,\alpha}^{\dagger}\biggr],
\label{PS3}\\
\widehat{B}_{k}(\vec{p},\tau) &=& - \, i\, \epsilon_{i\,j\,k} \,\sum_{\alpha= R,\,L}\biggl[ p_{i} \, \varepsilon_{j}^{(\alpha)}(\hat{p}) \, f_{p,\,\alpha} \widehat{a}_{\vec{p},\,\alpha} + p_{i} \, \varepsilon_{j}^{(\alpha)\ast}(-\hat{p})\, f_{p,\,\alpha}^{\ast} \widehat{a}_{-\vec{p},\,\alpha}^{\dagger} \biggr].
\label{PS4}
\end{eqnarray}
\subsection{Gyrotropic contributions}
From Eqs. (\ref{PS3})--(\ref{PS4}) we can compute the expectation values defining the 
corresponding two-point functions and the corresponding power spectra. The two-point functions 
will consists of a symmetric contribution and of an antisymmetric part that must vanish 
in the limit $\overline{\lambda} \to 0$:
\begin{eqnarray}
&& \langle \widehat{E}_{i}(\vec{k},\tau)\, \widehat{E}_{j}(\vec{p},\tau) \rangle = \frac{ 2 \pi^2 }{k^3} \biggl[\, P_{E}(k,\tau) \, p_{ij}(\hat{k}) 
+ P_{E}^{(G)}(k,\tau)\, \, i\, \epsilon_{i\, j\, \ell} \, \hat{k}^{\ell}\biggr] \, \delta^{(3)}(\vec{p} + \vec{k}),
\label{PS5}\\
&& \langle \widehat{B}_{i}(\vec{k},\tau)\, \widehat{B}_{j}(\vec{p},\tau) \rangle = \frac{ 2 \pi^2 }{k^3} \biggl[\, P_{B}(k,\tau) \, p_{ij}(\hat{k}) 
+ P_{B}^{(G)}(k,\tau)\, \,i\, \epsilon_{i\, j\, \ell} \, \hat{k}^{\ell}\biggr] \, \delta^{(3)}(\vec{p} + \vec{k}).
\label{PS6}
\end{eqnarray}
 In Eqs. (\ref{PS5})--(\ref{PS6}) $P_{E}(k,\tau)$ and $P_{B}(k,\tau)$ denote the hyperelectric and the hypermagnetic power spectra
while $P_{E}^{(G)}(k,\tau)$ and $P_{B}^{(G)}(k,\tau)$ are the corresponding gyrotropic contributions:
\begin{eqnarray}
P_{E}(k,\tau) &=& \frac{k^{3}}{4 \pi^2} \biggl[ \bigl| g_{k,\,L}\bigr|^2 + \bigl| g_{k,\,R}\bigr|^2 \biggr], \qquad
P_{B}(k,\tau) = \frac{k^{5}}{4 \pi^2} \biggl[ \bigl| f_{k,\,L}\bigr|^2 + \bigl| f_{k,\,R}\bigr|^2 \biggr],
\label{PS8}\\
P_{E}^{(G)}(k,\tau) &=& \frac{k^{3}}{4 \pi^2} \biggl[  \bigl| g_{k,\,L}\bigr|^2 - \bigl| g_{k,\,R}\bigr|^2 \biggr],\qquad
P_{B}^{(G)}(k,\tau) =  \frac{k^{5}}{4 \pi^2} \biggl[ \bigl| f_{k,\,L}\bigr|^2 - \bigl| f_{k,\,R}\bigr|^2\biggr].
\label{PS10}
\end{eqnarray}
Note that the gyrotropic spectra vanish when the hypermagnetic and hypermagnetic mode functions 
associated with the $L$ and $R$ polarizations differ only by an irrelevant phase; according to 
Eqs. (\ref{MF8})--(\ref{MF9}) this happens when $\overline{\lambda} \to 0$.
For the derivation of Eqs. (\ref{PS5})--(\ref{PS6}) we remind the following pair of identities 
\begin{equation}
2 \varepsilon_{i}^{(R)}(\hat{k}) \varepsilon_{j}^{(L)}(\hat{k}) = [ p_{ij}(\hat{k}) - i \, \epsilon_{i j \ell} \, \hat{k}^{\ell}], \qquad\qquad 
 2 \varepsilon_{i}^{(L)}(\hat{k}) \varepsilon_{j}^{(R)}(\hat{k}) = [ p_{ij}(\hat{k}) + i \, \epsilon_{i j \ell} \, \hat{k}^{\ell}],
 \label{PS6a}
 \end{equation}
 where $p_{ij}(\hat{k})$ is the traceless projector and $\epsilon_{i j \ell}$ is the Levi-Civita symbol in three dimensions.
The identities (\ref{PS6a}) can be directly obtained from the definitions of the $L$ and $R$ polarization 
(see Eq. (\ref{POL2}) and discussion therein); note that one of the two identities in Eq. (\ref{PS6a}) follow from the 
other (and vice versa) by complex conjugation.  
The power spectra $P_{B}(k,\tau)$ and $P_{E}(k,\tau)$ determine the averaged values of the various components of the energy-momentum  tensor given of Eq. (\ref{ACT8}) and, in particular, of the total energy density:
\begin{equation}
\langle \widehat{\rho}_{Y}(\vec{x},\tau) \rangle = \frac{1}{4 \pi a^4} \int \frac{d \, k}{k} \biggl[ P_{B}(k,\tau) + P_{E}(k,\tau) \biggr], \qquad\qquad
\widehat{\rho}_{Y}(\vec{x},\tau) = \widehat{\rho}_{E}(\vec{x},\tau) + \widehat{\rho}_{B}(\vec{x},\tau).
\label{PS10a}
\end{equation}
From Eq. (\ref{PS10a}) the explicit expression of the spectral energy density in critical units 
is therefore:
\begin{equation}
\Omega_{Y}(k,\tau) = \frac{d \langle \widehat{\rho}_{Y}\rangle}{d \ln{k}} = \frac{2}{3 H^2 M_{P}^2 a^{4}} \biggl[ P_{B}(k,\tau) + P_{E}(k,\tau) \biggr].
\label{PS10b}
\end{equation}
The gyrotropic spectra   $P_{B}^{(G)}(k,\tau)$ and $P^{(G)}_{E}(k,\tau)$ are relevant for baryogenesis (see section \ref{sec6} and discussion 
therein).  As already mentioned in the introduction, the notion of magnetic gyrotropy goes back to the analyses of Ref. \cite{zeld1} 
suggesting that the only possible pseudoscalar quadratic in $\vec{B}$ must be of the form $\vec{B}\cdot \vec{\nabla} \times \vec{B}$;  this quantity is called magnetic gyrotropy in analogy with the  kinetic gyrotropy $\vec{v} \cdot (\vec{\nabla} \times \vec{v})$ 
which naturally appears in the mean-field dynamo theory (see also \cite{zeld2,kaza}). If we introduce the 
hyperelectric and of the hypermagnetic gyrotropies as:
\begin{equation}
{\mathcal G}^{(E)}(\vec{x}, \tau) =  \vec{E} \cdot \vec{\nabla} \times \vec{E} , 
\qquad\qquad {\mathcal G}^{(B)}(\vec{x}, \tau) =  \vec{B} \cdot \vec{\nabla} \times \vec{B} , 
\label{PS10c}
\end{equation}
the corresponding quantum averages are determined by 
by $P^{(G)}_{E}(k,\tau)$ and $P^{(G)}_{B}(k,\tau)$ and they follow from Eqs. (\ref{PS5}) and (\ref{PS6}): 
will imply 
\begin{equation}
\langle \widehat{{\mathcal G}}^{(B)}(\vec{x},\tau) \rangle = 2 \int d\, k \, \,P_{B}^{(G)}(k,\tau),\qquad \qquad \langle \widehat{{\mathcal G}}^{(E)}(\vec{x},\tau) \rangle = 2 \int d\, k \, \,P_{E}^{(G)}(k,\tau).
\label{PS10d}
\end{equation}
As noted in Ref. \cite{FIVEa0} the systematic use of the hypermagnetic gyrotropy 
has some advantages in comparison with the case of the Chern-Simons number density (i.e. $n_{CS} \propto \vec{Y}\cdot\vec{B}$) which is however not gauge-invariant. This is lack of gauge-invariance  is  not crucial since it is well known that the difference of the Chern-Simons number density 
at different times [e.g. $\Delta n_{CS} = n_{CS}(\tau) - n_{CS}(0)$] is gauge-invariant and the same observation 
holds in the case of the Chern-Simons number density computed in the non-Abelian case \cite{CKN1}. What we are 
saying here is that, at a fixed time, $n_{CS}$ must be computed in the Coulomb gauge while locally in time  
the hypermagnetic gyrotropy is immediately gauge-invariant. 
We also note here that the time derivative of the Chern-Simons number is 
be proportional to $\vec{E}\cdot\vec{B}$ but if the (chiral) conductivity $\sigma$ is finite we will have 
that 
\begin{equation}
\langle \vec{E} \cdot \vec{B} \rangle = \frac{1}{\sigma} \langle \widehat{{\mathcal G}}_{B}(\vec{x},\tau) \rangle = \frac{2}{\sigma} \int d\, k \, \,P_{B}^{(G)}(k,\tau).
\label{PS10e}
\end{equation}
The phenomenological implications of Eqs. (\ref{PS10d})--(\ref{PS10e}) will be specifically discussed 
in section \ref{sec6} but before getting to the phenomenological implications 
it is necessary to deduce an accurate estimate 
of the gyrotropic spectra caused by the perturbative variation of the gauge coupling.

\subsection{Power spectra and duality transformations}
\label{subs32}
Before getting to the main point of the analysis it is useful to comment on the 
fate of the duality symmetry in the presence of pseudoscalar interactions. The first obvious observation is that 
Eqs. (\ref{CE3})--(\ref{CE2}) are not invariant under duality.
 This means, in practice, that they do not keep their 
form under the inversion of $\sqrt{\lambda}$ together with the simultaneous exchange of the hypermagnetic and 
hyperelectric fields according to: 
\begin{equation}
\sqrt{\lambda} \to \frac{1}{\sqrt{\lambda}}, \qquad \vec{\,\,E\,\,} \to \vec{\,\,B\,\,}, \qquad  \vec{\,\,B\,\,} \to - \,\vec{\,\,E\,\,}.
\label{CE4}
\end{equation}
However as long as the pseudoscalar interactions could be neglected (i.e. $\overline{\lambda} \ll \lambda$), the  
transformation (\ref{CE4}) leaves invariant Eqs. (\ref{CE3}) and (\ref{CE2}). 

At the quantum level the same reasoning 
could be repeated for the appropriate mode functions. In particular, in the limit $\overline{\lambda} \to 0$ Eqs. (\ref{MF8})--(\ref{MF9}) are also invariant under the following duality transformations\footnote{We recall that, by definition of ${\mathcal F}$, for an inversion of $\sqrt{\lambda}$,  ${\mathcal F} \to - {\mathcal F}$. }:
\begin{eqnarray}
\sqrt{\lambda} \to 1/\sqrt{\lambda}\qquad &\Rightarrow&  f_{k,\,L} \to g_{k,\,L}/k,\qquad g_{k,\,L} \to - k \,f_{k,\,L}, 
\nonumber\\
 &\Rightarrow& f_{k,\,R} \to g_{k,\,R}/k,\qquad g_{k,\,R} \to - k \,f_{k,\,R},
\nonumber\\
&\Rightarrow& P_{B}(k,\tau) \to P_{E}(k,\tau), \qquad P_{E}(k,\tau) \to P_{B}(k,\tau).
\label{MF10a}
\end{eqnarray}
Concerning the transformations of Eq. (\ref{MF10a}) two comments are in order:
\begin{itemize}
\item{} the above transformations do not involve the gyrotropic spectra simply because they 
vanish in the limit $\overline{\lambda} \to 0$;
\item{} when $\overline{\lambda} \neq 0$ it would be tempting to conclude that there exist 
some sort of symmetry between the gyrotropic spectra; indeed when 
\begin{equation}
 f_{k,\,L} \to g_{k,\,L}/k,\qquad g_{k,\,L} \to - k \,f_{k,\,L}, \qquad f_{k,\,R} \to g_{k,\,R}/k,\qquad g_{k,\,R} \to - k \,f_{k,\,R},
\label{MF10c}
\end{equation}
the gyrotropic spectra of Eqs. (\ref{PS8})--(\ref{PS10}) are exchanged as:
\begin{equation}
P^{(G)}_{B}(k,\tau) \to P^{(G)}_{E}(k,\tau), \qquad \qquad P^{(G)}_{E}(k,\tau) \to P^{(G)}_{B}(k,\tau).
\label{MF10b}
\end{equation}
However this is not a symmetry since Eqs. (\ref{MF8})--(\ref{MF9}) are not left 
invariant by the replacements (\ref{MF10c}) when $\overline{\lambda} \neq 0$.
\end{itemize}
In what follows (i.e. sections \ref{sec4} and \ref{sec5})
we shall compute the gauge power spectra in two manifestly dual situations with the purpose 
of deriving a set  of constraints on the strength of anomalous interactions 
during the quasi-de Sitter stage. Even if the primary objective of the forthcoming analysis 
is motivated by the consistency of our scenario, by comparing the gauge spectra obtained from 
a pair of dual evolutions it is possible to discuss more reliably the fate of the duality symmetry. 
A partial solution to the problem posed by Eqs. (\ref{MF10c})--(\ref{MF10b}) can be found at the end of section \ref{sec5} 
and, more precisely, in subsection \ref{subs54}.

\renewcommand{\theequation}{4.\arabic{equation}}
\setcounter{equation}{0}
\section{The case of increasing gauge coupling and the related bounds}
\label{sec4}
The production of the hypermagnetic gyrotropy can be discussed within a direct approach 
by specifying a model together with a set of pseudoscalar couplings. The approach followed 
here will instead reverse this logic by assuming a certain evolution 
of the gauge coupling with the purpose of deducing  a number of specific constraints 
for the strength of the anomalous interactions. From the physical viewpoint the 
evolution of the gauge coupling may be very complicated and various collateral 
 details may be added. Barring for all these possibilities it is however quite plausible 
 to address a pair of dual evolutions where the gauge coupling either increases or decreases 
 during inflation and then flattens out at late time. In the present section 
the first logical possibility will be addressed by considering the case where the gauge coupling increases during 
the inflationary phase and then flattens out during the post-inflationary evolution.
The same analysis shall be basically repeated  section \ref{sec5}, at a faster pace, 
in the case of a decreasing gauge coupling. If the gauge coupling first increases during a quasi-de Sitter 
stage of expansion and then flattens out after inflation the evolution 
of $\sqrt{\lambda}$ can be expressed as:
\begin{eqnarray}
\sqrt{\lambda} &=& \sqrt{\lambda_{1}} \biggl(-\frac{\tau}{\tau_{1}}\biggr)^{\gamma}, \qquad \tau \leq - \tau_{1}, 
\label{TWO1}\\
\sqrt{\lambda} &=& \sqrt{\lambda_{1}} \biggl[\frac{\gamma}{\delta} \biggl( \frac{\tau}{\tau_{1}} + 1\biggr) +1 \biggr]^{- \delta}, \qquad \tau \geq - \tau_{1}.
\label{FIVE2}
\end{eqnarray}
The explicit form of Eqs. (\ref{TWO1}) and (\ref{FIVE2}) is dictated by the continuity of $\sqrt{\lambda}$ and 
of $\sqrt{\lambda}^{\,\prime}$. The physical range of the parameters $\gamma$ and $\delta$ is therefore given by: 
\begin{equation} 
\gamma > 0, \qquad \mathrm{and}\qquad 0 \leq  \delta \ll  \gamma.
\label{FIVE3}
\end{equation}
The second condition in Eq. (\ref{FIVE3}) means that we are interested in the situation where the gauge coupling 
flattens out  {\em after} the end of the inflationary stage. The limit $\delta \to 0$ must be treated 
with some care: it is perfectly well defined at the end of the calculation but not before. 
In other words if we blindly take the limit $\delta \to 0$ and then compute the mode functions 
we run into potential discontinuities since the evolution of the mode functions (in their 
decoupled form) contain second time derivative of $\sqrt{\lambda}$. If $\delta \to 0$ in Eq, (\ref{FIVE2}) 
the first derivative of $\sqrt{\lambda}$ is not continuous while the second might have a discontinuity.
A posteriori, however, the limit $\delta \to 0$ is unambiguous and it just represents a practical way of 
estimating the physical situation defined by Eq. (\ref{FIVE3}) where $\delta \ll \gamma$.
For $\tau\leq - \tau_{1}$ we shall also assume that the geometry follows a quasi-de Sitter stage of expansion where\footnote{Because of a slightly different calligraphic style, the slow-roll parameter $\epsilon$ defined in Eq. (\ref{FIVE4}) cannot be confused 
with the circular polarizations $\hat{\varepsilon}^{(\alpha)}$ introduced in Eqs. (\ref{POL2}) and (\ref{expnew1})--(\ref{expnew2}).}
\begin{equation}
{\mathcal H} = a H = -\frac{1}{( 1- \epsilon) \tau}, \qquad \epsilon= - \frac{\dot{H}}{H^2} \ll 1, \qquad \tau\leq - \tau_{1},
\label{FIVE4}
\end{equation}
where the overdot denotes, as usual, a derivation with respect to the cosmic time coordinate $t$; $H = \dot{a}/a$ is the 
standard Hubble rate. For $\tau \geq - \tau_{1}$ the background expands in a standard decelerated manner. For the sake of concreteness for $\tau \geq - \tau_{1}$ the scale factor will evolve linearly with the conformal time coordinate $\tau$, i.e.  $a_{rad}(\tau) =[\tau + (\beta+1) \tau_{1}]/\tau_{1}$ where $\beta \simeq 1/(1 - \epsilon)$ is fixed by the continuity of the scale factor and of its derivative across the inflationary boundary $\tau= -\tau_1$. It is useful to mention, in this context, that the reheating 
 mechanism will be largely immaterial for the considerations developed here and it will anyway only affect 
 the highest mode of the spectrum that will be anyway ${\mathcal O}(a_{1} H_{1})$.

\subsection{Mode functions and their normalization the quasi-de Sitter stage}
The explicit solutions of the mode functions in the case of Eqs. (\ref{TWO1}) and (\ref{FIVE2})--(\ref{FIVE3}) 
 have been relegated  to the appendices \ref{APPA} and \ref{APPB} which will be often cited throughout 
 this section. We shall now focus on  the derivation of the power spectra and 
 on the related bounds involving the strength of the anomalous interactions: 
 this will be the most relevant aspect at least for the present ends. 
If Eq. (\ref{TWO1}) is inserted into Eqs. (\ref{MF8}) 
and (\ref{MF9}) we obtain the explicit expressions for the evolution of the mode functions 
during the inflationary phase. The pair of decoupled equations for the $L$ and $R$ mode functions obtained in this way has been 
reported in Eq. (\ref{IN2}) but it can also be written as:
\begin{equation}
\frac{d^2 f_{k,\, L}}{d z^2 } + \biggl[ - \frac{1}{4} + \frac{\zeta}{z} - \frac{\mu^2 - 1/4}{z^2} \biggr] f_{k,\, L}=0,\qquad 
\frac{d^2 f_{k,\, R}}{d z^2 } + \biggl[ - \frac{1}{4} - \frac{\zeta}{z} - \frac{\mu^2 - 1/4}{z^2} \biggr] f_{k,\, R}=0.
\label{IN6}
\end{equation}
Equation (\ref{IN6}) follows from Eq. (\ref{IN2}) of appendix \ref{APPA} by 
introducing the following rescaled quantities: 
\begin{equation}
z = 2 \, i \, k \,\tau, \qquad \zeta= i\, \lambda_{0} \, \gamma = i\, \overline{\zeta},\qquad \mu = | \gamma - 1/2 |,
\label{IN4}
\end{equation}
where  $\zeta$ is a purely imaginary quantity while $\overline{\zeta} = \lambda_{0} \gamma$ is real\footnote{The same notations will be employed for all the other quantities that 
are purely imaginary and that will appear in the forthcoming sections, e.g.  $\eta = i\, \overline{\eta}$, $\theta= i \overline{\theta}$  and so on and so forth.}
and it goes to zero in the limit $\lambda_{0} \to 0$.  From Eq. (\ref{IN6}) and (\ref{IN4}) we see that 
the parameter space of the model can be safely discussed in the $(\gamma, \, \overline{\zeta})$ plane. 
With the notations of Eq. (\ref{IN4}) the explicit expression of Eq. (\ref{IN6}) coincide with the canonical form of the Whittaker's 
equation (see e.g. Ref. \cite{abr2} and the related discussion in appendix \ref{APPA}).  The explicit solutions of Eq. (\ref{IN6}) with the correct boundary conditions for $\tau \ll - \tau_{1}$ are:
\begin{equation}
 f_{k\, R}(z) = C_{R}(k,\overline{\zeta}) \, W_{- \zeta,\, \mu}(z), \qquad \qquad f_{k\, L}(z) = C_{L}(k, \overline{\zeta}) \, W_{\zeta,\, \mu}(z).
 \label{IN8}
 \end{equation} 
The conformal time coordinate $\tau$ is 
always negative during the quasi-de Sitter stage of expansion and this observation impacts on the correct derivation 
of the corresponding asymptotic limits of the Whittaker's functions.  The normalization factors $C_{R}(k,\overline{\zeta})$ and $C_{L}(k,\overline{\zeta}) $ are then determined by requiring 
that for $\tau \ll - \tau_{1}$ the Wronskians of Eq. (\ref{MF2a}) are correctly normalized for the $L$-waves and for the $R$-waves:
\begin{equation} 
C_{R}(k,\overline{\zeta}) = \frac{e^{-i \pi/4 + \pi \overline{\zeta}/2}}{\sqrt{2 k}} ,\qquad\qquad
C_{L}(k,\overline{\zeta}) = \frac{e^{i \pi/4 - \pi \overline{\zeta}/2}}{\sqrt{2 k}}.
\label{IN11a}
\end{equation}
The phases $e^{\pm i\pi/4}$  appearing in Eq. (\ref{IN11a}) are required for the correct limit of the mode functions  when $\lambda_{0} \to 0$. When $\lambda_{0}\to 0$ we also have that $\overline{\zeta}\to 0$ and, in this limit, Eq.  (\ref{IN8}) must give back exactly the solutions obtained in the absence of anomalous interactions:
\begin{equation}
\lim_{\overline{\zeta} \to 0} f_{k,\,L}(\tau) = e^{i \pi/4} f_{k}(\tau), \qquad \lim_{\overline{\zeta} \to 0} f_{k,\,L}(\tau) = e^{- i \pi/4} f_{k}(\tau), 
\label{IN12}
\end{equation}
where $f_{k}(\tau) = f_{k,\,\otimes}(\tau) = f_{k,\,\oplus}(\tau)$ is the common value of the mode function for each of the two {\em linear polarizations}\footnote{We recall, in this respect, that with the present choice of circular modes (see Eq. (\ref{POL2}) and discussion therein) the relation between the circularly polarized mode functions and their 
linearly polarized counterpart is simply given by $f_{k,\,L} = (f_{k,\,\oplus} + i\, f_{k,\,\otimes})/\sqrt{2}$ 
and by  $f_{k,\,R} = (f_{k,\,\oplus} - i\, f_{k,\,\otimes})/\sqrt{2}$.}:
\begin{equation}
f_{k}(\tau) = \frac{{\mathcal N}_{\mu}}{\sqrt{ 2 k}}\, \sqrt{ - k \tau} \,\,H_{\mu}^{(1)}(-k\tau), \qquad {\mathcal N}_{\mu} = \sqrt{\frac{\pi}{2}} e^{i  \,\pi(2 \mu+1/4)}, 
\label{IN13}
\end{equation}
where $H_{\mu}^{(1)}(-k\tau)$ are the Hankel functions of the first kind \cite{abr2}. The Hankel limit 
of the Whittaker's functions has been swiftly discussed in appendix \ref{APPA} [see, 
in particular, Eqs. (\ref{EXP1}), (\ref{EXP2}) and (\ref{EXP3})]. Finally, 
inserting  Eq. (\ref{IN8}) into Eq. (\ref{IN3}) we obtain the hyperelectric mode functions
\begin{eqnarray}
&&g_{k\, L}(z) = 2 \, i\, k\, C_{L}(k,\overline{\zeta}) \biggl[ \frac{z - 2 (\zeta + \gamma)}{2 z} W_{\zeta,\, \mu}(z) - \frac{W_{\zeta+1,\, \mu}(z)}{z} \biggr],
\label{IN10}\\
&& g_{k\, R}(z) = 2 \, i\, k\, C_{R}(k, \overline{\zeta}) \biggl[ \frac{z + 2 (\zeta - \gamma)}{2 z} W_{-\zeta,\, \mu}(z) - \frac{W_{1-\zeta,\, \mu}(z)}{z} \biggr],
\label{IN11}
\end{eqnarray}
It can be can be verified, by direct substitution, that  Eqs. (\ref{IN8}), (\ref{IN10}) and (\ref{IN11}) satisfy the Wronskian normalization conditions of Eq. (\ref{MF2a}) for each of the two circular modes.

\subsection{Hypermagnetic and hyperelectric power spectra during the quasi-de Sitter stage}
Inserting Eqs. (\ref{IN8}), (\ref{IN10}) and (\ref{IN11})  into the 
general expressions of  Eqs. (\ref{PS8})--(\ref{PS10}) we obtain the explicit expressions of the hyperelectric 
and of the hypermagnetic power spectra valid for $\tau \leq - \tau_{1}$.
The most relevant limit of the power spectra is the one 
for which $k < a\, H$ since, in this range, the relevant wavelengths are larger than the effective horizon 
associated with the variation of the gauge coupling. In the conventional description of large-scale cosmological perturbations \cite{wein1,wein2,primer} a given wavelength is said to to exit the Hubble radius at some typical conformal time 
$\tau_{ex}$ during an inflationary stage of expansion and it is said to reenter at $\tau_{re}$, when the Universe still expands but in a decelerated manner.  An equivalent way of describing the same regime is to say that a given mode 
is beyond the horizon:  by a mode being beyond the horizon we only mean that the physical wavenumber 
is much less than the expansion rate and this does not necessarily 
have anything to do with causality \cite{wein2}. The same terminologies will also be employed 
hereunder with the important caveat that what matters, in our case, is the effective horizon associated 
with the variation of the gauge coupling. In other words the physical wavenumbers of the hyperelectric and hypermagnetic fields 
can be much smaller than the rate of variation of the gauge coupling (i.e. ${\mathcal F}$ in the present notations)
which now plays the role of the effective horizon. With these specifications the hypermagnetic power spectra are:
\begin{eqnarray}
P_{B}(k,\tau) &=& a^4 H^4 \, D(|\gamma -1/2|) \, Q_{B}(\overline{\zeta},\gamma) \, \biggl(\frac{k}{a H}\biggr)^{5 - 2 |\gamma -1/2|},
\label{PSA1}\\
P_{B}^{(G)}(k,\tau) &=& a^4 H^4 \, D(|\gamma -1/2|) Q^{(G)}_{B}(\overline{\zeta},\gamma) \, \biggl(\frac{k}{a H}\biggr)^{5 - 2 |\gamma -1/2|}.
\label{PSA2}
\end{eqnarray}
Throughout the whole discussion the function $D(x)$ will be used with the exactly same meaning and it is defined as:
\begin{equation}
D( x) = 2^{2 x -3} \frac{\Gamma^2(x)}{\pi^3},
\label{PSA2a}
\end{equation}
 where $\Gamma(x)$ denotes the conventional Gamma function \cite{abr2}. The remaining two auxiliary functions appearing in Eqs. (\ref{PSA1})--(\ref{PSA2}) are instead:
\begin{eqnarray}
Q_{B}(\overline{\zeta},\gamma) &=& \frac{1}{2} \biggl[ \frac{e^{- \pi \overline{\zeta}} \,\Gamma^2(|\gamma -1/2| +1/2)}{\bigl|\Gamma(1/2 - i \overline{\zeta} + |\gamma -1/2|)\bigr|^2} + \frac{e^{\pi \overline{\zeta}} \,\Gamma^2(|\gamma -1/2| +1/2)}{\bigl|\Gamma(1/2 + i \overline{\zeta} + |\gamma -1/2|)\bigr|^2} \biggr],
\nonumber\\
Q^{(G)}_{B}(\overline{\zeta},\gamma)  &=& \frac{1}{2} \biggl[ \frac{e^{- \pi \overline{\zeta}} \,\Gamma^2(|\gamma -1/2| +1/2)}{\bigl|\Gamma(1/2 - i \overline{\zeta} + |\gamma -1/2|)\bigr|^2} - \frac{e^{\pi \overline{\zeta}} \,\Gamma^2(|\gamma -1/2| +1/2)}{\bigl|\Gamma(1/2 + i \overline{\zeta} + |\gamma -1/2|)\bigr|^2} \biggr],
\label{AUX3}
\end{eqnarray}
where $Q_{B}(\overline{\zeta},\gamma) $ and $Q^{(G)}_{B}(\overline{\zeta},\gamma)$  only differ by a  crucial sign.
\begin{figure}[!ht]
\centering
\includegraphics[height=7.3cm]{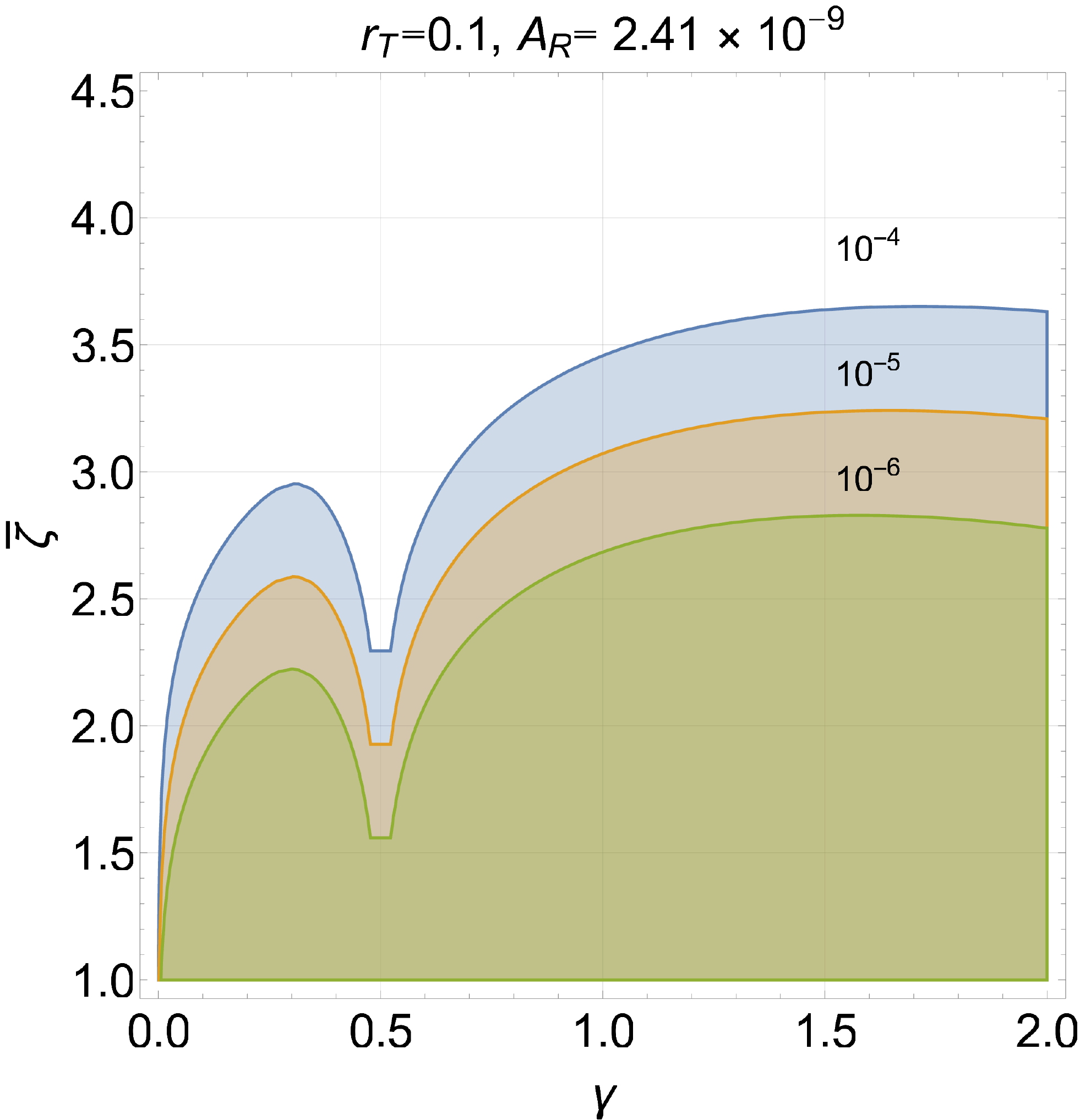}
\includegraphics[height=7.3cm]{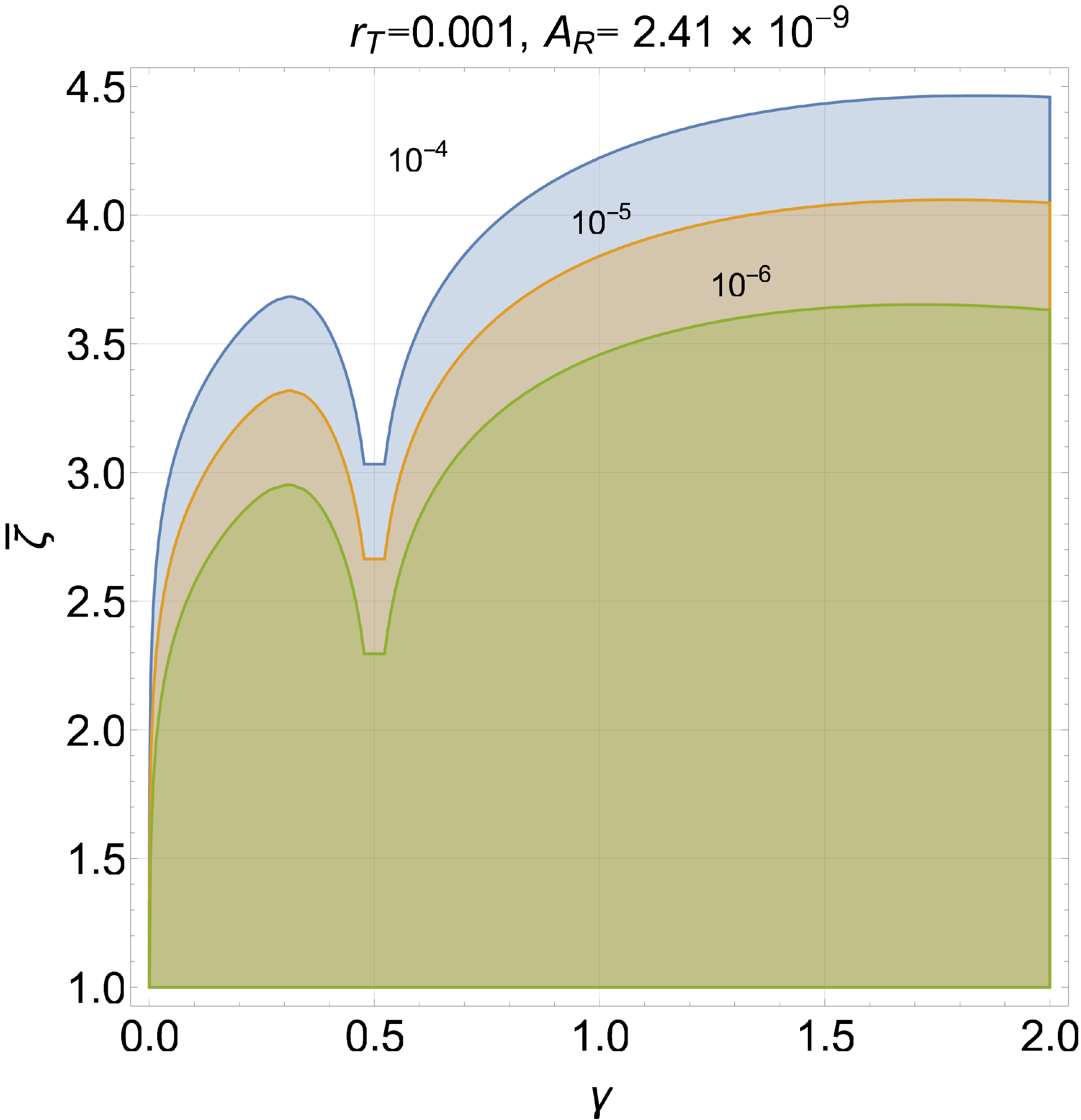}
\caption[a]{The critical density bounds during the quasi-de Sitter stage of expansion are illustrated in the $(\gamma, \, \overline{\zeta})$ plane. The difference between the left and right plots comes from a different value 
of the tensor to scalar ratio $r_{T}$ entering the estimate of $(H/M_{P})$}
\label{FFF1}      
\end{figure}
With the same notations used in Eq. (\ref{PSA1}) and (\ref{PSA2}) the hyperelectric spectra are instead given by:
\begin{eqnarray}
P_{E}(k,\tau) &=& a^4 H^4 \,D(\gamma +1/2) \, Q_{E}(\overline{\zeta},\gamma) \, \biggl(\frac{k}{a H}\biggr)^{4 - 2 \gamma},
\label{PSA3}\\
P^{(G)}_{E}(k,\tau) &=& a^4 H^4 \,D(\gamma +1/2) \, Q^{(G)}_{E}(\overline{\zeta},\gamma) \, \biggl(\frac{k}{a H}\biggr)^{4 - 2 \gamma},
\label{PSA4}
\end{eqnarray}
where $D(x)$ has been already introduced in Eq. (\ref{PSA2a}) while $Q_{E}(\overline{\zeta},\gamma)$ and $Q^{(G)}_{E}(\overline{\zeta},\gamma)$ are defined as:
\begin{equation}
Q_{E}(\overline{\zeta},\gamma) = \frac{1}{2} \biggl[ \frac{e^{- \pi \overline{\zeta}} \,\Gamma^2(\gamma)}{\bigl|\Gamma(\gamma - i \overline{\zeta} )\bigr|^2} + \frac{e^{\pi \overline{\zeta}} \,\Gamma^2(\gamma)}{\bigl|\Gamma(\gamma + i \overline{\zeta} )\bigr|^2} \biggr],\qquad
Q_{E}^{(G)}(\overline{\zeta}, \gamma) = \frac{1}{2} \biggl[ \frac{e^{- \pi \overline{\zeta}} \,\Gamma^2(\gamma)}{\bigl|\Gamma(\gamma - i \overline{\zeta} )\bigr|^2} - \frac{e^{\pi \overline{\zeta}} \,\Gamma^2(\gamma)}{\bigl|\Gamma(\gamma + i \overline{\zeta} )\bigr|^2} \biggr].
\label{AUX5}
\end{equation}
Since $\overline{\zeta} = \lambda_{0} \gamma$ (and, as required in Eq. (\ref{FIVE3}), $\gamma>0$) the  limits for $\overline{\zeta} \to 0$ of Eqs. (\ref{AUX3}) and (\ref{AUX5}) correspond to the situation where the anomalous interactions are absent (i.e. $\lambda_{0} \to 0$):
\begin{equation}
\lim_{\overline{\zeta}\to 0} Q_{X}(\overline{\zeta},\gamma) =1, \qquad \lim_{\overline{\zeta}\to 0} Q^{(G)}_{X}(\overline{\zeta},\gamma) =0,\qquad \mathrm{for}\quad X= E,\,\,B.
\label{AUX6}
\end{equation}
When the relevant wavelengths exceed the effective horizon associated 
with the evolution of the gauge coupling the ratio of the gyrotropic spectrum to its non-gyrotropic 
counterpart assumes a particularly simple form:
\begin{equation}
\lim_{(- k\tau)\ll 1} \frac{P^{(G)}_{B}(k,\tau)}{P_{B}(k,\tau)} = \lim_{(- k\tau)\ll 1} \frac{P^{(G)}_{E}(k,\tau)}{P_{E}(k,\tau)} = - \tanh{\pi \, \overline{\zeta}}, 
\label{ratio1}
\end{equation}
implying that for $\overline{\zeta} \gg 1$ the two components are of the same order while in the limit $\pi \overline{\zeta} < 1$ 
the gyrotropic contribution is always subleading.

\subsection{Bound on the strength of anomalous interactions}
The result of Eq. (\ref{ratio1}) suggests that the limits of Eqs. (\ref{PSA1})--(\ref{PSA2}) and 
(\ref{PSA3})--(\ref{PSA4}) for $\overline{\zeta} \to 0$  exactly coincide with the results 
valid in the absence of anomalous interactions (see e.g. \cite{EIGHTc} and references therein):
\begin{eqnarray}
\lim_{\overline{\zeta}\to 0} \,\,P_{B}(k,\tau) &=& a^4 H^4 \, D(|\gamma -1/2|) \, \biggl(\frac{k}{a H}\biggr)^{5 - | 2 \gamma -1|},
\label{AUX7}\\
\lim_{\overline{\zeta}\to 0} \,\,P_{E}(k,\tau) &=& a^4 H^4 \,D(\gamma +1/2) \, \biggl(\frac{k}{a H}\biggr)^{4 - 2 \gamma},
\label{AUX8}\\
\lim_{\overline{\zeta}\to 0} \,\,P^{(G)}_{B}(k,\tau) &=& \lim_{\overline{\zeta}\to 0} \,\,P^{(G)}_{E}(k,\tau) =0.
\label{AUX9}
\end{eqnarray}
This observation is relevant for the derivation of the general bounds on the parameter 
space of the model: since we want to make sure that the obtained constraints will have general validity 
we shall first consider the limit $\overline{\zeta} \to 0$.  Equations (\ref{AUX7}) and (\ref{AUX8}) assume a different form in 
two complementary 
situations, namely the cases\footnote{The case $\gamma =1/2$ 
should be separately considered but this discussion will be skipped for the sake of conciseness.} $\gamma >1/2$ and 
 $ 0 < \gamma < 1/2$. If $\gamma > 1/2$ the hypermagnetic spectrum $P_{B}(k,\tau)$ is  
scale-invariant for $\gamma = 3$; but for this value $\gamma$ the hyperelectric spectrum $P_{E}(k,\tau)$ diverges in the limit 
$(-k\tau) \ll 1$.  When $\gamma = 2$ the hyperelectric spectrum is scale-invariant but the corresponding hypermagnetic spectrum is  sharply increasing with $|k\, \tau|$. If $ 0 < \gamma < 1/2$
both the hyperelectric and the hypermagnetic spectra are increasing. We therefore conclude that the physical 
range\footnote{In the past it has been argued that this range is not 
relevant for magnetogenesis. It is however not sufficient that the gauge power spectra during inflation 
are sharply increasing with $k$ to conclude that the same spectra will be small at large-scales 
at later times; as we shall demonstrate this statement depends on the subsequent dynamical evolution and any conclusion should be 
drawn on the basis of the late-time spectra to be discussed later on in this section.} of $\gamma$ pinned down by the explicit expressions (\ref{AUX7})--(\ref{AUX8}) is given by $0< \gamma \leq 2$. 

The constraints on the strength of the 
anomalous interactions in the $(\gamma, \overline{\zeta})$ will now follow from the 
bounds on the spectral energy density during the 
quasi-de Sitter stage of expansion. Thanks to Eqs. (\ref{PSA1})--(\ref{PSA2}) and (\ref{PSA3})--(\ref{PSA4}) the general form of the spectral energy density given in Eq. (\ref{PS10b}) becomes 
\begin{equation}
\Omega_{Y}(k,\tau) = \frac{2}{3} \biggl(\frac{H}{M_{P}}\biggr)^2 \biggl[ D(|\gamma -1/2|) Q_{B}(\overline{\zeta},\gamma) \biggl(\frac{k}{ a H} \biggr)^{5 - | 2 \gamma -1|}
+ D(\gamma + 1/2) Q_{E}(\overline{\zeta},\gamma) \biggl(\frac{k}{ a H} \biggr)^{4 - 2 \gamma} \biggr], 
\label{EN1}
\end{equation}
and must always remain much smaller than $1$ during the whole quasi-de Sitter stage of expansion. Using Eqs. (\ref{AUX7})--(\ref{AUX8})  
 the limit of Eq. (\ref{EN1}) for $\overline{\zeta} \ll 1$  is:
\begin{equation}
\Omega_{Y}(k,\tau) = \frac{2}{3} \biggl(\frac{H}{M_{P}}\biggr)^2 \biggl[ D(|\gamma -1/2|) \biggl(\frac{k}{ a H} \biggr)^{5 - | 2 \gamma -1|}
+ D(\gamma + 1/2)  \biggl(\frac{k}{ a H} \biggr)^{4 - 2 \gamma} \biggr].
\label{EN2}
\end{equation}
From Eq. (\ref{EN2}) we then conclude that for $\overline{\zeta} \ll 1$ the spectral energy density is  subcritical\footnote{Note, in fact, that $(H/M_{P})$ can always 
be estimated as $\sqrt{\pi\,\epsilon\, {\mathcal A}_{{\mathcal R}}}$ where ${\mathcal A}_{{\mathcal R}} = {\mathcal O}(2.41) \times 10^{-9}$ is the amplitude of the scalar power spectrum evaluated at the pivot scale $k_{p} =0.002 \, \mathrm{Mpc}^{-1}$. }
(i.e. $\Omega_{Y}(k,\tau) \ll 1$) provided $ \gamma \leq 2$. If $\overline{\zeta}\geq 1$ we cannot avoid a more numerical discussion which is summarized in Fig. \ref{FFF1} where the two dimensional contours in the plane $(\gamma, \, \overline{\zeta})$ are illustrated in  various cases. For the sake of illustration in Fig. \ref{FFF1} the regions with the different shadings 
correspond to the maximal values of $\Omega_{Y}(k,\tau)$ indicated in the 
figure (i.e. $10^{-6}$, $10^{-5}$ and $10^{-4}$ from bottom to top). Note finally that in Fig. \ref{FFF1} we assumed the validity of the
consistency relations stipulating that $\epsilon = r_{T}/16$ where $r_{T}$ denotes the 
tensor to scalar ratio. We therefore conclude that as long as $\overline{\zeta} = {\mathcal O}(\gamma)$ 
the spectral energy density is subcritical.

\subsection{Late time evolution of the mode functions} 
For $\tau \geq -\tau_{1}$  the rate of variation of $\sqrt{\lambda}$ and the 
other relevant variables follow from Eq. (\ref{FIVE2}):
\begin{eqnarray}
{\mathcal F} &=& - \frac{\delta}{y(\tau)}, \qquad y(\tau) = \tau + \tau_{1} [ q(\gamma, \delta) +1 ], \qquad q(\gamma, \delta) = \delta/\gamma
\label{IN21ga}\\
\xi &=& i \overline{\xi}, \qquad \overline{\xi} = \lambda_{0} \delta, \qquad w = 2 i\, k\, y, \qquad \nu = (\delta + 1/2).
\label{IN21g}
\end{eqnarray}
Given that $\xi$ is also a purely imaginary quantity, $\overline{\xi}$ has the same meaning of $\overline{\zeta}$ [see Eq. (\ref{IN4})].  Furthermore since 
$\delta \ll \gamma$ [see Eq. (\ref{FIVE3})] we will also have that $\overline{\xi} \ll \overline{\zeta}$.
The relation of $y(\tau)$ and $w(\tau)$ to  the conformal time coordinate 
is always linear but it also contains a constant piece (multiplied by  $q(\gamma, \delta) = \delta/\gamma$) 
which is relevant for the accurate continuity of the various expressions.  With these precisions, for  $\tau \geq - \tau_{1}$  the canonical forms of the Whittaker's 
equations for the $L$ and $R$ polarizations are:
\begin{eqnarray}
&& \frac{d^2 f_{k,\, L}}{d w^2} + \biggl[ - \frac{1}{4} - \frac{\xi}{w} - \frac{\nu^2 -1/4}{w^2} \biggr] f_{k,\, L}=0,\qquad 
 \frac{d^2 f_{k,\, R}}{d w^2} + \biggl[ - \frac{1}{4} + \frac{\xi}{w} - \frac{\nu^2 -1/4}{w^2} \biggr] f_{k,\, R}=0,
\label{IN21d}\\
&& g_{k,\, L} = 2\, i\, k\,\biggl[\frac{d f_{k,\, L}}{d w} - \frac{\delta}{w}  f_{k,\, L}\biggr], \qquad g_{k,\, R} = 2\, i\, k\, \biggl[\frac{d f_{k,\, R}}{d w} - \frac{\delta}{w}  f_{k,\, R}\biggr].
\label{IN21f}
\end{eqnarray}
The late-time form of the mode functions now follows from the continuity of the background and of the geometry.
For the present ends it is particularly convenient to express $f_{k\, X}(\tau)$ and $g_{k\, X}(\tau)$ for $\tau > -\tau_{1}$ 
in terms of the corresponding mode functions computed for  $\tau \leq - \tau_{1}$:
\begin{eqnarray}
f_{k\, X}(\tau) &=&  A_{f\,f}^{(X)}(w_{1}, \, w,\, \delta) \overline{f}_{k,\, X} +  A_{f\,g}^{(X)}(w_{1}, \, w,\, \delta) \, \frac{\overline{g}_{k,\,X}}{k},
\label{MATRIX1}\\
g_{k\, X}(\tau) &=&  A_{f\,f}^{(X)}(w_{1}, \, w,\, \delta) k\, \overline{f}_{k,\, X} +  A_{g\,g}^{(X)}(w_{1}, \, w,\, \delta) \, \overline{g}_{k,\,X},
\label{MATRIX2}
\end{eqnarray}
where $X = L,\, R$ since Eqs. (\ref{MATRIX1})--(\ref{MATRIX2}) hold independently for the left- and for the right-movers.
Note that the functions $\overline{f}_{k,\,X}$ and $\overline{g}_{k,\,X}$  are just the mode functions valid for $\tau \leq - \tau_{1}$ evaluated for $\tau = -\tau_{1}$, i.e.\footnote{For instance from Eq. (\ref{IN8}) $\overline{f}_{k,\,L} = C_{R}(k,\overline{\zeta}) \, W_{- \zeta,\, \mu}(- 2 i k \tau_{1})$, $\overline{f}_{k,\, L} = C_{L}(k, \overline{\zeta}) \, W_{\zeta,\, \mu}(- 2 i\, k\tau_{1})$ and similarly for the hyperelectric mode functions of Eqs. (\ref{IN10})--(\ref{IN11}).} 
\begin{equation}
\overline{f}_{k,\,X}\equiv f^{(inf)}_{k,\,X}(-\tau_{1}), \qquad \mathrm{and} \qquad \overline{g}_{k,\,X}\equiv g^{(inf)}_{k,\,X}(-\tau_{1}), \qquad \mathrm{for}\quad X= R,\, L.
\end{equation}
The variable $w_{1}$ appearing in Eqs. (\ref{MATRIX1})--(\ref{MATRIX2}) is, by definition, $w1= w(-\tau_{1})$. Hence, from the definition of $y(\tau)$ given in Eq. (\ref{IN21ga}) we will also have that $w_{1} =  2 i\, k\, \tau_{1} q$. The various terms appearing in Eqs. (\ref{MATRIX1})--(\ref{MATRIX2}) can also be ordered in an appropriate mixing matrix: 
\begin{equation}
{\mathcal M}^{(X)}(w_{1},\, w,\, \delta) =  \left(\matrix{ A^{(X)}_{f\, f}(w_{1},\, w,\, \delta)
& A^{(X)}_{f\,g}(w_{1},\, w,\, \delta)&\cr
A^{(X)}_{g\,f}(w_{1},\,w,\, \delta) & A^{(X)}_{g\,g}(w_{1},\,w,\, \delta)&\cr}\right).
\label{MATRIX3}
\end{equation}
Since the Wronskian normalization has been fixed in the limit $\tau \to - \infty$  (and it is preserved by the 
continuity and differentiability of the evolution) the determinant  of ${\mathcal M}^{(X)}(w_{1},\, w,\, \delta)$ must be equal to $1$:
\begin{equation}
A_{f\,f}^{(X)}(w_{1}, \, w,\, \delta) A^{(X)}_{g\,g}(w_{1},\,w,\, \delta) -A^{(X)}_{f\,g}(w_{1},\, w,\, \delta) A^{(X)}_{g\,f}(w_{1},\,w,\, \delta) =1.
\label{MATRIX4}
\end{equation}
The  explicit forms of the various entries of the matrix (\ref{MATRIX3}) are  
determined from the linearly independent solutions of Eqs. (\ref{IN21d}) and (\ref{IN21f}) 
listed in appendix \ref{APPB} [i.e. Eqs. (\ref{IN21h}) and (\ref{IN21m})--(\ref{IN21n})]:  
\begin{eqnarray}
A^{(L)}_{f\, f}(w_{1}, w, \delta) &=& -\frac{e^{\pi  \overline{\xi} }}{2w_1} \biggl\{(w_1-2  \delta - 2\xi )W_{-\xi ,\nu }\left(-w_1\right) W_{\xi ,\nu }(w) + 2 W_{1-\xi ,\nu }(-w_1) W_{\xi ,\nu }(w)
\nonumber\\
&+& W_{-\xi ,\nu }(-w) \biggl[\left(2  \delta -2 \xi +w_1\right) W_{\xi ,\nu }\left(w_1\right)-2 W_{\xi +1,\nu }(w_1)\biggr]\biggr\},
 \label{IN22L}\\
 A^{(L)}_{f\, g}( w_{1}, w, \delta) &=& \frac{i}{2} e^{\pi  \overline{\xi} } \biggl[W_{-\xi ,\nu }\left(-w_1\right) W_{\xi ,\nu }(w)-W_{-\xi ,\nu }(-w) W_{\xi ,\nu }\left(w_1\right)\biggr],  
\label{IN23L}\\
A^{(L)}_{g\, f}(w_{1}, w, \delta)  &=& \frac{i \, e^{\pi \overline{\xi}}}{2 w\, w_{1}} 
\biggl\{\biggl[\left(2 \delta +2 \xi -w_1\right) W_{-\xi ,\nu }(-w_1) - 2 W_{1-\xi ,\nu }(-w_1)\biggr]  \biggl[(2 \delta -2 \xi +w) W_{\xi,\nu }(w)
\nonumber\\
&-& 2 W_{\xi +1,\nu }(w)\biggr] -  \biggl[(2 \delta + 2\xi -w) W_{-\xi ,\nu }(-w)-2 W_{1-\xi ,\nu }(-w)\biggr] \biggl[\left(2 \delta -2 \xi +w_1\right)
   W_{\xi ,\nu }(w_1)
 \nonumber\\  
&-& 2 W_{\xi +1,\nu }(w_1)\biggr]\biggr\},
\label{IN24L}\\
 A^{(L)}_{g\, g}(w_{1}, w, \delta)  &=& -\frac{e^{\pi  \overline{\xi}}}{2w} \biggl\{W_{\xi ,\nu }\left(w_1\right) \biggl[(w-2 \delta - 2\xi ) W_{-\xi ,\nu }(-w)+2 W_{1-\xi ,\nu }(-w)\biggr]
\nonumber\\
&+& W_{-\xi ,\nu}\left(-w_1\right) \biggl[(2 \delta -2 \xi +w) W_{\xi ,\nu }(w)-2 W_{\xi +1,\nu }(w)\biggr]\biggr\}.
\label{IN25L}
\end{eqnarray}
In Eqs. (\ref{IN22L})--(\ref{IN23L}) and (\ref{IN24L})--(\ref{IN25L}) the dependence 
upon $\delta$ appears, implicitly, in $\nu(\delta)$ [because of Eq. (\ref{IN21g})]  and also in $\xi = i \overline{\xi}$ [since $\overline{\xi} = \lambda_{0} \delta$].  The entries of the companion matrix  valid for the $R$-polarization (i.e. ${\mathcal M}^{(R)}(w_{1},\, w,\, \delta)$) have a form similar to ${\mathcal M}^{(L)}(w_{1},\, w,\, \delta)$
and they are obtained from the expressions reported in appendix \ref{APPB} (see, 
in particular, Eqs. (\ref{IN21ha}) and (\ref{IN21ma})--(\ref{IN21na})).  Just to state the rule of thumb we now compare the result of Eq. (\ref{IN23L})
with the analog expression of $A^{(R)}_{f\, g}( w_{1}, w, \delta)$:
\begin{equation}
A^{(R)}_{f\, g}( w_{1}, w, \delta) = \frac{i}{2} e^{-\pi  \overline{\xi} } \biggl[W_{-\xi ,\nu }\left(w_1\right) W_{\xi ,\nu }(-w)-W_{-\xi ,\nu }(w) W_{\xi ,\nu }\left(-w_1\right)\biggr].  
\label{COMP1a}
\end{equation}
Therefore, to obtain the 
explicit form of ${\mathcal M}^{(R)}(w_{1},\, w,\, \delta)$ once the entries of ${\mathcal M}^{(L)}(w_{1},\, w,\, \delta)$ are known it is sufficient to flip the signs of $w_{1}$ and $w$ in Eqs. (\ref{IN22L})--(\ref{IN23L}) and (\ref{IN24L})--(\ref{IN25L}) and to flip the sign of $\overline{\xi}$ in the normalization appearing in the different entries.  It is finally interesting to analyze the Hankel limit of the mixing matrices, namely  the limit $\overline{\xi} \to 0$ (see also Eqs. (\ref{EXP0})--(\ref{EXP2}) and discussion therein). Since $\overline{\xi} = \lambda_{0} \, \delta$, the limit $\overline{\xi} \to 0$  can be realized in two complementary (but physically distinct) cases. The first interesting limit is $\lambda_{0} \to 0$ (while $\delta \neq 0$); in this case 
the anomalous contribution vanishes but the gauge coupling 
is still increasing for $\tau \geq - \tau_{1}$. In this case the matrix elements of ${\mathcal M}^{(R)}(w_{1},\, w,\, \delta)$ and of 
${\mathcal M}^{(L)}(w_{1},\, w,\, \delta)$ coincide and are given by:
\begin{equation}
\lim_{\lambda_{0} \to 0} \,{\mathcal M}^{(R)}(w_{1},\, w,\, \delta)  = \lim_{\lambda_{0} \to 0} \, {\mathcal M}^{(L)}(w_{1},\, w,\, \delta) = \left(\matrix{ A_{f\, f}(\tau_{1},\, \tau,\, \delta)
& A_{f\,g}(\tau_{1},\, \tau,\, \delta)&\cr A_{g\,f}(\tau_{1},\, \tau,\, \delta) & A_{g\,g}(\tau_{1},\, \tau,\, \delta)&\cr}\right).
\label{IN25N}
\end{equation}
where  the explicit expressions of the various entries appearing in the common limit of Eq. (\ref{IN25N}) 
are:
\begin{eqnarray}
A_{f\, f}(k,\tau, \tau_{1}) &=& \frac{\pi}{2} \sqrt{q x_{1}} \sqrt{ k y} \biggl[ Y_{\nu -1}( q x_{1}) J_{\nu}(k y) - J_{\nu-1}(q x_{1}) Y_{\nu}(k y) \biggr],
\nonumber\\
A_{f\, g}(k, \tau, \tau_{1}) &=& \frac{\pi}{2} \sqrt{q x_{1}} \sqrt{ k y} \biggl[ J_{\nu}( q x_{1}) Y_{\nu}(k y) - Y_{\nu}(q x_{1}) J_{\nu}(k y) \biggr],
\nonumber\\
A_{g\, f}(k, \tau, \tau_{1}) &=& \frac{\pi}{2} \sqrt{q x_{1}} \sqrt{ k y} \biggl[ Y_{\nu - 1}( q x_{1}) J_{\nu -1}(k y) - J_{\nu-1}(q x_{1}) Y_{\nu-1}(k y) \biggr],
\nonumber\\
A_{g\, g}(k, \tau, \tau_{1}) &=& \frac{\pi}{2} \sqrt{q x_{1}} \sqrt{ k y} \biggl[ J_{\nu}( q x_{1}) Y_{\nu-1}(k y) - Y_{\nu}(q x_{1}) J_{\nu-1}(k y) \biggr],
\label{AAAin1}
\end{eqnarray}
where $ x_{1} = k \tau_{1}$ while $y$ has the same definition given before (see prior to Eq. (\ref{IN21g})); note 
that the matrix elements (\ref{AAAin1}) still obey Eq. (\ref{IN25N}).

The second way in which the limit $\overline{\xi} \ll 1$ can be physically 
realized suggests that $\delta \to 0$ (while $\lambda_{0} \neq 0$). In this case the anomalous contributions
remain for $\tau \geq - \tau_{1}$ but the gauge coupling flattens out, at least approximately.
By keeping fixed $x = k \tau $ and $x_{1} = k \tau_{1}$ the common limit of ${\mathcal M}^{(L)}(w_{1},\, w,\, \delta)$ 
and of ${\mathcal M}^{(R)}(w_{1},\, w,\, \delta)$ is now even simpler:
\begin{eqnarray}
&&A_{f\, f}(k,\tau, \tau_{1}) = \cos{[k (\tau+ \tau_{1})]} + {\mathcal O}(\delta),\qquad 
A_{f\, g}(k,\tau, \tau_{1}) = \sin{[k (\tau+ \tau_{1})]} + {\mathcal O}(\delta),
\nonumber\\
&& A_{g\, f}(k,\tau, \tau_{1}) = -\sin{[k (\tau+ \tau_{1})]} + {\mathcal O}(\delta),\qquad 
A_{g\, f}(k,\tau, \tau_{1}) = \cos{[k (\tau+ \tau_{1})]} + {\mathcal O}(\delta).
\label{AAAin2}
\end{eqnarray}
The finiteness 
of the result (\ref{AAAin2}) demonstrates that the continuity  
of the mode functions is an essential requirement for the correctness of the asymptotic results.
 
\subsection{Late-time hypermagnetic and hyperelectric power spectra}
The matrices ${\mathcal M}^{(X)}(w_{1},\, w,\, \delta)$ introduced in 
 Eq. (\ref{MATRIX3})  effectively mix the hypermagnetic and the hyperelectric mode functions computed at the end of inflation. 
 There is therefore no reason to expect that the late-time hypermagnetic power 
 spectra will have to coincide (either exactly or approximately) with the hypermagnetic spectra at the end of inflation, as it is sometimes assumed. In their full form the late-time spectra for $\tau \geq - \tau_{1}$  will be given by:
\begin{eqnarray}
P_{B}(k,\tau) &=& \frac{k^{5}}{4 \pi^2}\sum_{X= L,\,R} \biggl|A_{f\, f}^{(X)}(w_{1},\, w,\, \delta)\,\, \overline{f}_{k,\,X} + A_{f\, g}^{(X)}(w_{1},\, w,\, \delta) \,\, \frac{\overline{g}_{k,\,X}}{k}\biggr|^2,
\label{MGC3}\\
P_{E}(k,\tau) &=& \frac{k^{3}}{4 \pi^2} \,\sum_{X= L,\,R} \biggl| A_{g\, f}^{(X)}(w_{1},\, w,\, \delta) \,k \,\overline{f}_{k,\,X} + A_{g\, g}^{(X)}(w_{1},\, w,\, \delta)\,\, \overline{g}_{k,\,X}\,\,\biggr|^2.
\label{MGC4}
\end{eqnarray}
Since the hyperelectric and hypermagnetic mode functions at the end of inflation are not of the same order, also 
the contributions entering Eqs. (\ref{MGC3})--(\ref{MGC4}) are not all comparable. For instance in the case of  the hypermagnetic power spectrum of Eq. (\ref{MGC3}) the  first term inside the square modulus dominates against the second term:
\begin{equation}
\biggl| A^{(X)}_{f\, g}(w_{1}, w, \delta)\, \frac{\overline{g}_{k,\,X}}{k} \biggr| \gg \biggl| A_{f\, f}^{(X)}( w_{1}, w, \delta) \, \overline{f}_{k,\,X} \biggr|,\qquad \mathrm{for} \quad X = R,\,\, L.
\label{MGC5}
\end{equation}
The simplest way of proving the inequality (\ref{MGC5}) is to consider
 the physically relevant regime  where the modes 
 are still larger than the effective horizon for $\tau \geq -\tau_{1}$ (i.e. $k \tau_{1} < k \tau \ll 1$)
 while the gauge coupling either flattens out or even freezes;  according to Eq. (\ref{FIVE3})  this regime 
corresponds to the situation where\footnote{The first requirement in Eq. (\ref{MGC6}) is the most relevant 
and it implies that  $\delta$ must range between $0$ and $1/2$; in this interval we then have $\nu = 1/2 -\delta$.
In the opposite regime (where $\delta$ is comparable with $\gamma$)
 the gauge coupling does not flatten out for $\tau \geq - \tau_{1}$. } 
\begin{equation}
0 < \delta \ll \gamma\qquad \Rightarrow 0< \delta <1/2\qquad \Rightarrow  \nu = 1/2 - \delta.
\label{MGC6}
\end{equation}
Using Eq. (\ref{MGC6}) in the limit  $  k\tau_{1} < k\tau \ll 1$ together with the explicit expressions 
of the mode functions we obtain that Eq. (\ref{MGC5}) is verified in spite of the value of $\lambda_{0}$:
this happens since for $\delta \to 0$ the limit $\overline{\xi} \to 0$ is verified for any $\lambda_{0}= {\mathcal O}(1)$. Equation (\ref{MGC5}) is complemented by the analog inequality (actually motivated by duality as we shall see at the end of section \ref{sec5}) valid 
in the hyperelectric case, namely: 
 \begin{equation}
  \biggl| A_{g\, f}^{(X)}(w_{1},\, w,\, \delta) \,k \,\overline{f}_{k,\,X}  \biggr| \ll \biggl| A_{g\, g}^{(X)}(w_{1},\, w,\, \delta) \overline{g}_{k,\,X}\biggr|,\qquad \mathrm{for} \quad X = R,\,\, L.
\label{MGC7}
\end{equation}
If the variables $w$ and $w_{1}$ are traded for $x = k \tau$ and for  $x_{1} = k\tau_{1}$ the general expression of the hypermagnetic and hyperelectric power spectra can be written as:
\begin{eqnarray}
P_{B}(k,\tau) &=& \frac{k^5}{2\pi^2} \biggl[ \biggl| A^{(L)}_{f\, g}( x_{1}, \,x, \delta)\, \frac{\overline{g}_{k,\, L}}{k}\biggr|^2 +
\biggl| A^{(R)}_{f\, g}( x_{1}, \,x, \delta)\, \frac{\overline{g}_{k,\, R}}{k}\biggr|^2 \biggr],
\label{MGC8}\\
P_{E}(k,\tau) &=& \frac{k^3}{2\pi^2} \biggl[ \biggl| A^{(L)}_{g\, g}( x_{1}, \,x, \delta)\, \overline{g}_{k,\, L}\biggr|^2 +
\biggl| A^{(R)}_{f\, g}( x_{1}, \,x, \delta)\, \overline{g}_{k,\, R} \biggr|^2 \biggr].
\label{MGC9}
\end{eqnarray}
Equations (\ref{MGC8}) and (\ref{MGC9}) are actually very interesting since they show that, {\em when the gauge coupling first increases and then flattens out,  the late-time power spectra are determined by the hyperelectric power spectra at the end of inflation}.
For a fixed value of $x_{1}$ and $x$ we have that $A^{(L)}_{f\, g}( x_{1}, \,x, \delta)$ and 
$ A^{(R)}_{f\, g}( x_{1}, \,x, \delta)$ differ at most by terms ${\mathcal O}(\delta)$ in the limit $\delta \to 0$ 
\begin{equation}
\lim_{\delta \ll 1}  \biggl[A^{(R)}_{f\, g}( x_{1}, \,x, \delta) - A^{(L)}_{f\, g}( x_{1}, \,x, \delta)\biggr] = {\mathcal O}(\delta).
\label{MGC10}
\end{equation}
Since the reasonable  values of $\delta$ do not exceed ${\mathcal O}(0.1)$,  the power spectra can be estimated, in this range, as: 
 \begin{eqnarray}
P_{B}(k,\tau) &=& a_{1}^{4} \, H_{1}^4 \, D(\gamma + 1/2) \, x_{1}^{4 - 2 \gamma } \,
F_{B}(x_{1}, x, \delta, \overline{\zeta},\gamma),
\nonumber\\
F_{B}(x_{1}, x, \delta, \overline{\zeta},\gamma) &=& \biggl|A^{(X)}_{f\, g}( x_{1}, \,x, \delta)\biggr|^2 Q_{E}(\overline{\zeta},\gamma) = [\sin{(x+ x_{1})} + {\mathcal O}(\delta) ]^2 \, Q_{E}(\overline{\zeta},\gamma).
\label{MGC11}
\end{eqnarray}
The limit $\delta \to 0$ can be explicitly compared with the estimates  obtained when, for instance, $\delta = {\mathcal O}(10^{-2})$ and this 
is what will also be specifically illustrated in section \ref{sec6}. With the same approach leading to Eq. (\ref{MGC11}) the hyperelectric power spectrum becomes
\begin{eqnarray}
P_{E}(k,\tau) &=& a_{1}^{4} \, H_{1}^4 \, D(\gamma + 1/2) \, x_{1}^{4 - 2 \gamma } \, F_{E}(x_{1}, x, \delta, \overline{\zeta},\gamma),
\nonumber\\
F_{E}(x_{1}, x, \delta, \overline{\zeta},\gamma) &=&\biggl|A^{(X)}_{g\, g}( x_{1}, \,x, \delta)\biggr|^2 Q_{E}(\overline{\zeta},\gamma) =  [\cos{(x+ x_{1})} + {\mathcal O}(\delta) ]^2 \, Q_{E}(\overline{\zeta},\gamma).
\label{MGC12}
\end{eqnarray}
Equations  (\ref{MGC11})--(\ref{MGC12})  are therefore a consequence  of Eqs. (\ref{MGC8})--(\ref{MGC9}) and they 
explicitly show that the late-time  hypermagnetic spectra are determined by  their hyperelecric counterpart at the end of the inflationary phase;  this is ultimately the reason why $Q_{E}(\overline{\zeta},\gamma)$ appears in Eq. (\ref{MGC11}). The amplitudes of the two spectra are however quite different: while $P_{B}(k,\tau)$ 
exhibits standing oscillations going as $\sin^2{(x+ x_{1})}$ the hyperelectric oscillations 
go instead as $P_{E}(k,\tau) \propto \cos^2{(x+ x_{1})}$. This means that 
while $P_{B}(k,\tau) \ll P_{E}(k,\tau)$ when $k \tau \ll 1$, as the relevant modes cross the Hubble 
radius $P_{B}(k,\tau_{k}) \simeq P_{E}(k,\tau_{k})$ where $\tau_{k}$ is defined by the condition 
$k \,\tau_{k} = {\mathcal O}(1)$.  The same strategy employed to estimate 
the power spectra can be used with the corresponding gyrotropies. We shall omit this discussion for the 
sake of conciseness and mention the final  results for 
$P_{B}^{(G)}(k,\tau)$ and $P_{E}^{(G)}(k,\tau)$:
\begin{eqnarray}
P_{B}^{(G)}(k,\tau) &=& a_{1}^{4} \, H_{1}^4 \, D(\gamma + 1/2) \, x_{1}^{4 - 2 \gamma } \, Q^{(G)}_{B}(\overline{\zeta},\gamma)\,  \sin^2{(x+ x_{1})},
\label{MGCC13}\\
P_{E}^{(G)}(k,\tau) &=& a_{1}^{4} \, H_{1}^4 \, D(\gamma + 1/2) \, x_{1}^{4 - 2 \gamma } \, Q^{(G)}_{E}(\overline{\zeta},\gamma)\,  \cos^2{(x+ x_{1})}.
\label{MGCC14}
\end{eqnarray}
We note in passing that the standing waves appearing in Eqs. (\ref{MGCC13}) and (\ref{MGCC14}) are the gauge analog of the Sakharov oscillations \cite{SAK1,SAK2}.   

\renewcommand{\theequation}{5.\arabic{equation}}
\setcounter{equation}{0}
\section{The case of decreasing gauge coupling and the related bounds}
\label{sec5}
While in section \ref{sec4} we examined a phase of increasing gauge coupling, 
in this section we shall examine the dual situation where the evolution 
of $\sqrt{\lambda}$ is given by:
\begin{eqnarray}
\sqrt{\lambda} &=& \sqrt{\lambda_{1}} \biggl( - \frac{\tau}{\tau_{1}} \biggr)^{-\widetilde{\,\gamma\,}} , \qquad \tau < - \tau_{1},
\label{ADD1}\\
\sqrt{\lambda} &=& \sqrt{\lambda_{1}} \biggl[\frac{\widetilde{\,\gamma\,}}{\widetilde{\,\delta\,}} \biggl( \frac{\tau}{\tau_{1}} + 1\biggr) +1 \biggr]^{\widetilde{\,\delta\,}}, \qquad \tau \geq - \tau_{1}.
\label{ADD2} 
\end{eqnarray}
If we compare Eqs. (\ref{ADD1})--(\ref{ADD2}) with Eqs. (\ref{TWO1})--(\ref{FIVE2}) we see that
 {\em the quantities with tilde will always correspond to the situation where 
the gauge coupling decreases}. This notation simplifies the comparison of the power spectra in the dual cases, as we shall see at the end of this section. 
The physical region of the parameters $\widetilde{\gamma}$ and $\widetilde{\delta}$ appearing in Eqs. (\ref{ADD1})--(\ref{ADD2}) 
 is:
\begin{equation} 
\widetilde{\,\gamma\,} > 0, \qquad  \widetilde{\,\delta\,} \geq 0, \qquad \mathrm{and}\qquad 0< \widetilde{\,\delta\,} \ll  \widetilde{\,\gamma\,}.
\label{ADD3}
\end{equation}

We want to stress that dual profiles  of Eqs. (\ref{TWO1})--(\ref{FIVE2}) and (\ref{ADD1})--(\ref{ADD2}) are not physically equivalent. 
 Equation (\ref{ADD1}) implies that at the onset of the inflationary phase the gauge coupling may be very large\footnote{ In Eq. (\ref{COMP1}) the 
conformal time has been traded for the scale factors by using Eq. (\ref{FIVE4})  in the limit $\epsilon \ll 1$.
The ratio of the scale factors during a quasi-de Sitter stage 
can be related to the  number of $e$-folds $N$ as $(a_{i}/a_{f})^{\widetilde{\,\gamma\,}} = e^{ - N \,\widetilde{\,\gamma\,}}\ll 1$. If we estimate the total number of $e$-folds as $N = {\mathcal O}(60)$ (or larger)
we have that  $\sqrt{\lambda_{i}}$ will be ${\mathcal O}(10^{-60})$ (or smaller). }:
\begin{equation}
\sqrt{\lambda_{i}} = \sqrt{\lambda_{1}} \biggl(\frac{a_{i}}{a_{f}}\biggr)^{\widetilde{\,\gamma\,}} \ll 1\qquad\qquad  \Rightarrow\qquad \qquad
e(\tau_{i}) = \frac{\sqrt{4\pi}}{\sqrt{\lambda_{i}}} \gg 1,
\label{COMP1}
\end{equation}
where, by definition, $\lambda_{i} = \lambda(- \tau_{i})$. Equation (\ref{COMP1}) implies that the evolution of the gauge coupling starts from a 
non-perturbative regime unless $\sqrt{\lambda_{1}}$ is 
extremely large: only in this way we could possibly have $\sqrt{\lambda_{i}} = {\mathcal O}(1)$. Whenever 
$\sqrt{\lambda_{1}} \gg 1$ the gauge coupling will be extremely minute at the end of inflation 
and this is at odds with the fact that during the decelerated stage of expansion 
we would like to have $e^2 = {\mathcal O}(10^{-2})$ but not much smaller. 
Furthermore, as we shall discuss in section \ref{sec6}, the physical power spectra are suppressed as $\lambda_{1}^{-1}$ and this 
will make their contribution marginal for the phenomenological implications. A possibility 
suggested previously has been that the $\sqrt{\lambda}$ increases during inflation, decreases 
sharply during reheating, and then flattens out again. This suggestion has been originally put forward 
in Refs. \cite{mg1a} (see also \cite{mg1b}) and it would imply a sudden decrease of $\sqrt{\lambda}$ in an extremely short time. This 
hopeful possibility will not be necessary in the present context. 

\subsection{Mode functions and their normalization during the quasi-de Sitter stage}
During the inflationary phase the explicit expressions of the evolution of the mode functions 
are given by Eq. (\ref{DEC2}) (see also the related discussion in the appendix \ref{APPA}). Equations (\ref{MF8}) 
and (\ref{MF9}) become:
\begin{equation}
 \frac{d^2 f_{k,\, L}}{d z^2 } + \biggl[ - \frac{1}{4} - \frac{\eta}{z} - \frac{\widetilde{\,\mu\,}^2 - 1/4}{z^2} \biggr] f_{k,\, L}=0, \qquad 
\frac{d^2 f_{k,\, R}}{d z^2 } + \biggl[ - \frac{1}{4} + \frac{\eta}{z} - \frac{\widetilde{\,\mu\,}^2 - 1/4}{z^2} \biggr] f_{k,\, R}=0,
\label{DEC6}
\end{equation}
where the variables appearing in Eq. (\ref{DEC6}) are now defined as: 
\begin{equation}
z = 2 \, i \, k \,\tau,\qquad \widetilde{\,\mu\,} = \widetilde{\,\gamma\,} + 1/2, \qquad \eta= i\, \overline{\eta},\qquad \overline{\eta}= \lambda_{0} \, \widetilde{\,\gamma\,}.
\label{DEC4}
\end{equation}
Equations (\ref{DEC6})--(\ref{DEC4}) are  written in the canonical Whittaker form (see also Eq. (\ref{EW1})).
By comparing Eqs. (\ref{DEC6})--(\ref{DEC4}) with  Eqs. (\ref{IN6})--(\ref{IN4}) it is useful to stress
the differences but also the possible symmetries especially for what concerns the sign
of the anomalous contributions. From Eqs. (\ref{DEC6})--(\ref{DEC4}) the solutions for $f_{k\, L}(z)$ and $f_{k\, R}(z)$ are:
\begin{equation}
f_{k\, L}(z) = \widetilde{\,C\,}_{L}(k, \overline{\eta}) \, W_{-\eta,\, \widetilde{\mu}}(z), \qquad \qquad  f_{k\, R}(z) = \widetilde{\,C\,}_{R}(k,\overline{\eta}) \, W_{\eta,\, \widetilde{\mu}}(z),
 \label{DEC8}
 \end{equation} 
where the  factors $\widetilde{\,C\,}_{R}(k,\overline{\eta})$ and $\widetilde{\,C\,}_{L}(k,\overline{\eta}) $ are fixed 
from the Wronskian normalization condition (\ref{MF2a}) in the limit $\tau \ll - \tau_{1} $:
\begin{equation} 
\widetilde{\,C\,}_{R}(k,\overline{\eta}) = \frac{e^{-i \pi/4 - \pi \overline{\eta}/2}}{\sqrt{2 k}} ,\qquad\qquad
\widetilde{\,C\,}_{L}(k,\overline{\eta}) = \frac{e^{i \pi/4 + \pi \overline{\eta}/2}}{\sqrt{2 k}}.
\label{DEC9}
\end{equation}
As before the results of appendix \ref{APPA} can be used together with 
 Eq. (\ref{DEC8}) so that the explicit form of the hyperelectric mode functions is:
\begin{eqnarray}
&&  g_{k\, R}(z) = 2 \, i\, k\, \widetilde{\,C\,}_{R}(k, \overline{\eta}) \biggl[ \frac{z + 2 (\widetilde{\gamma} -\eta)}{2 z} W_{\eta,\, \widetilde{\mu}}(z) - \frac{W_{1+\eta,\, \widetilde{\mu}}(z)}{z} \biggr],
\label{DEC11}\\
&& g_{k\, L}(z) = 2 \, i\, k\, \widetilde{\,C\,}_{L}(k, \overline{\eta}) \biggl[ \frac{z + 2 (\eta +\widetilde{\gamma})}{2 z} W_{-\eta,\, \widetilde{\mu}}(z) - \frac{W_{1- \eta,\, \widetilde{\mu}}(z)}{z} \biggr].
\label{DEC12}
\end{eqnarray}

\subsection{Hypermagnetic and hyperelectric power spectra during the quasi-de Sitter stage}
After inserting the normalized solutions of Eqs. (\ref{DEC8})--(\ref{DEC9}) into Eqs. (\ref{PS8})--(\ref{PS10}) 
the explicit form of the hypermagnetic power spectra when the typical wavelengths are larger 
than the effective horizon is:
\begin{eqnarray}
\widetilde{\,P\,}_{B}(k,\tau)&=&  a^4 \, H^4 \, D(\widetilde{\,\gamma\,}+1/2) \, \biggl(\frac{k}{a H}\biggr)^{4 - 2 \widetilde{\,\gamma\,}} \, \widetilde{\,Q\,}_{B}(\overline{\eta}, \widetilde{\,\gamma\,}),
\label{PSB1}\\
\widetilde{\,P\,}^{(G)}_{B}(k,\tau)&=& a^4 \, H^4 \, D(\widetilde{\,\gamma\,}+1/2) \, \biggl(\frac{k}{a H}\biggr)^{4 - 2 \widetilde{\,\gamma\,}} \,
\widetilde{\,Q\,}^{(G)}_{B}(\overline{\eta}, \widetilde{\,\gamma\,}),
\label{PSB2}
\end{eqnarray}
where $D(x)$ has been already defined in Eq. (\ref{PSA2a}) and it has here exactly the same meaning in terms of the new argument; $\widetilde{\,Q\,}_{B}(\overline{\eta}, \widetilde{\gamma})$ and $\widetilde{\,Q\,}^{(G)}_{B}(\overline{\eta}, \widetilde{\gamma})$ are now given by: 
\begin{equation}
\widetilde{\,Q\,}_{B}(\overline{\eta}, \widetilde{\,\gamma\,}) = \frac{1}{2}\biggl[\frac{e^{\pi \overline{\eta}}\,\,\Gamma^2(\widetilde{\,\gamma\,})}{\bigl| \Gamma( \widetilde{\,\gamma\,} - i \overline{\eta})\bigr|^2} + \frac{e^{-\pi \overline{\eta}}\,\,\Gamma^2(\widetilde{\,\gamma\,})}{\bigl| \Gamma( \widetilde{\,\gamma\,} + i \overline{\eta})\bigr|^2} \biggr],
\qquad 
\widetilde{\,Q\,}^{(G)}_{B}(\overline{\eta}, \widetilde{\,\gamma\,}) = \frac{1}{2}\biggl[\frac{e^{\pi \overline{\eta}}\,\,\Gamma^2(\widetilde{\,\gamma\,})}{\bigl| \Gamma(\widetilde{\,\gamma\,} - i \overline{\eta})\bigr|^2} - \frac{e^{-\pi \overline{\eta}}\,\,\Gamma^2(\widetilde{\,\gamma\,})}{\bigl| \Gamma(\widetilde{\,\gamma\,}+ i \overline{\eta})\bigr|^2} \biggr].
\label{PSB4}
\end{equation}
The limit  $\overline{\eta}\to 0$  mirrors the case $\overline{\zeta}\to 0$ discussed in Eq. (\ref{AUX6}):
\begin{equation}
\lim_{\overline{\eta}\to 0} \widetilde{\,Q\,}_{B}^{(G)}(\overline{\eta}, \widetilde{\,\gamma\,})=1, \qquad \lim_{\overline{\eta}\to 0} \widetilde{\,Q\,}_{B}^{(G)}(\overline{\eta}, \widetilde{\,\gamma\,}) =0.
\label{PSB4a}
\end{equation}
The  asymptotic expression of the hyperelectric mode functions 
(see Eqs. (\ref{DEC15})--(\ref{DEC16})) is slightly more cumbersome and, to avoid  confusions, the limits $\overline{\eta} \ll 1$ and $\overline{\eta}\gg 1$  will be separately 
discussed. If $\overline{\eta} \ll 1$  the hyperelectric spectra are 
\begin{equation}
\widetilde{\,P\,}_{E}(k,\tau) = a^4 \, H^4 D(|\widetilde{\,\gamma\,} -1/2|) \, \biggl(\frac{k}{a H}\biggr)^{ 5 - | 2 \widetilde{\,\gamma\,} -1|},\qquad \widetilde{\,P\,}_{E}^{(G)}(k,\tau) =0.
\label{PSB5}
\end{equation}
All in all in the limit $ \overline{\eta} \to 0$ the gauge power spectra computed in the case of decreasing coupling are:
 \begin{eqnarray}
\lim_{\overline{\eta}\to 0} \,\,\widetilde{\,P\,}_{B}(k,\tau) &=& a^4 H^4 \, \,D(\widetilde{\,\gamma\,} +1/2) \, \biggl(\frac{k}{a H}\biggr)^{4 - 2 \widetilde{\gamma}}, 
\label{AUX7D}\\
\lim_{\overline{\eta}\to 0} \,\,\widetilde{\,P\,}_{E}(k,\tau) &=& a^4 H^4 \,\,D(|\widetilde{\,\gamma\,} -1/2|) \, \biggl(\frac{k}{a H}\biggr)^{5 - | 2 \widetilde{\gamma} -1|},
\label{AUX8D}\\
\lim_{\overline{\eta}\to 0} \,\,\widetilde{\,P\,}^{(G)}_{B}(k,\tau) &=& \lim_{\overline{\eta}\to 0} \,\,\widetilde{\,P\,}^{(G)}_{E}(k,\tau) =0.
\label{AUX9D}
\end{eqnarray}
In the opposite limit  $\overline{\eta} \gg 1$ we have instead:
\begin{eqnarray}
\widetilde{\,P\,}_{E}(k,\tau) &=& a^4 \, H^4 D(\widetilde{\,\gamma\,}+1/2) \, \biggl(\frac{k}{a H}\biggr)^{ 4 - 2 \widetilde{\,\gamma\,}} \widetilde{\,Q\,}_{E}(\overline{\eta}, \widetilde{\gamma}), 
\label{PSB6a}\\
\widetilde{\,P\,}^{(G)}_{E}(k,\tau) &=& a^4 \, H^4 D(\widetilde{\,\gamma\,}+1/2) \, \biggl(\frac{k}{a H}\biggr)^{ 4 - 2 \widetilde{\,\gamma\,}} \widetilde{\,Q\,}^{(G)}_{E}(\overline{\eta}, \widetilde{\,\gamma\,}), 
\label{PSB6b}
\end{eqnarray}
where $\widetilde{\,Q\,}_{E}(\overline{\eta}, \widetilde{\,\gamma\,})$ and $\widetilde{\,Q\,}^{(G)}_{E}(\overline{\eta}, \widetilde{\,\gamma\,})$
are now defined as:
\begin{eqnarray}
 \widetilde{\,Q\,}_{E}(\overline{\eta}, \widetilde{\,\gamma\,}) = \frac{\overline{\eta}^2}{2} \biggl[ \frac{e^{\pi \overline{\eta}}\,\Gamma^2( \widetilde{\,\gamma\,})}{\bigl| \Gamma(\widetilde{\,\gamma\,} + i \overline{\eta})\bigr|^2} + \frac{e^{-\pi \overline{\eta}}\,\Gamma^2(\widetilde{\,\gamma\,})}{\bigl| \Gamma(\widetilde{\,\gamma\,} - i \overline{\eta})\bigr|^2 }\biggr], \qquad 
  \widetilde{\,Q\,}^{(G)}_{E}(\overline{\eta}, \widetilde{\gamma}) = \frac{\overline{\eta}^2}{2} \biggl[ \frac{e^{\pi \overline{\eta}}\,\Gamma^2( \widetilde{\gamma})}{\bigl| \Gamma( \widetilde{\gamma} + i \overline{\eta})\bigr|^2} - \frac{e^{-\pi \overline{\eta}}\,\Gamma^2( \widetilde{\gamma})}{\bigl| \Gamma(  \widetilde{\gamma} - i \overline{\eta})\bigr|^2 }\biggr].
\label{PSB7}
 \end{eqnarray}
 When the relevant wavelengths exceed the effective horizon associated 
with the evolution of the gauge coupling the ratio of the gyrotropic spectrum to its non-gyrotropic 
counterpart is common to the hyperelectric and to the hypermagnetic case:
\begin{equation}
\lim_{(- k\tau)\ll 1} \frac{\widetilde{\,P\,}^{(G)}_{B}(k,\tau)}{\widetilde{\,P\,}_{B}(k,\tau)} = \lim_{(- k\tau)\ll 1} \frac{\widetilde{\,P\,}^{(G)}_{E}(k,\tau)}{\widetilde{\,P\,}_{E}(k,\tau)} = \tanh{\pi \, \overline{\eta}}. 
\label{ratio2}
\end{equation}
If  Eqs. (\ref{AUX7D})--(\ref{AUX8D}) are inserted into Eq. (\ref{PS10b}) the spectral energy density is:
 \begin{equation}
\Omega_{Y}(k,\tau) = \frac{2}{3} \biggl(\frac{H}{M_{P}}\biggr)^2 \biggl[ D(\widetilde{\gamma} +1/2) \biggl(\frac{k}{ a H} \biggr)^{4 - 2 \widetilde{\gamma}}
+ D(|\widetilde{\gamma} - 1/2|)  \biggl(\frac{k}{ a H} \biggr)^{5 - | 2 \widetilde{\gamma}- 1|} \biggr],
\label{EN1A}
\end{equation}
and it is valid in the limit $\overline{\eta} \to 0$. In  Eq. (\ref{EN1A})  the flat hypermagnetic spectrum corresponds to the case $\widetilde{\gamma} =2$ and the 
relevant region of the parameter space corresponds therefore to the interval  $0 < \widetilde{\gamma} \leq 2$ since for $\widetilde{\gamma}> 2$ the spectral energy density $\Omega_{Y}(k,\tau)$ cannot be subcritical. 
\begin{figure}[!ht]
\centering
\includegraphics[height=7.3cm]{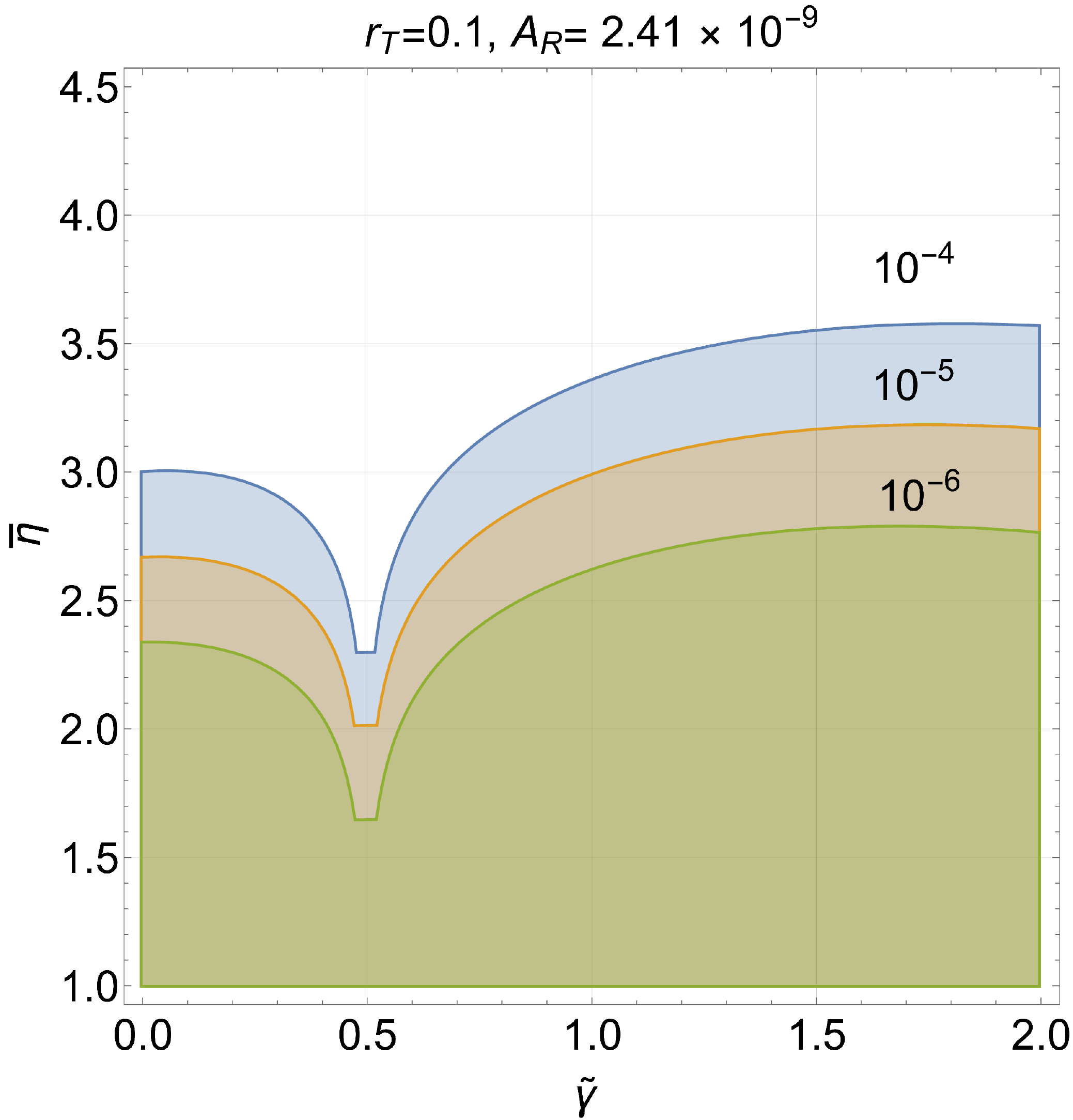}
\includegraphics[height=7.3cm]{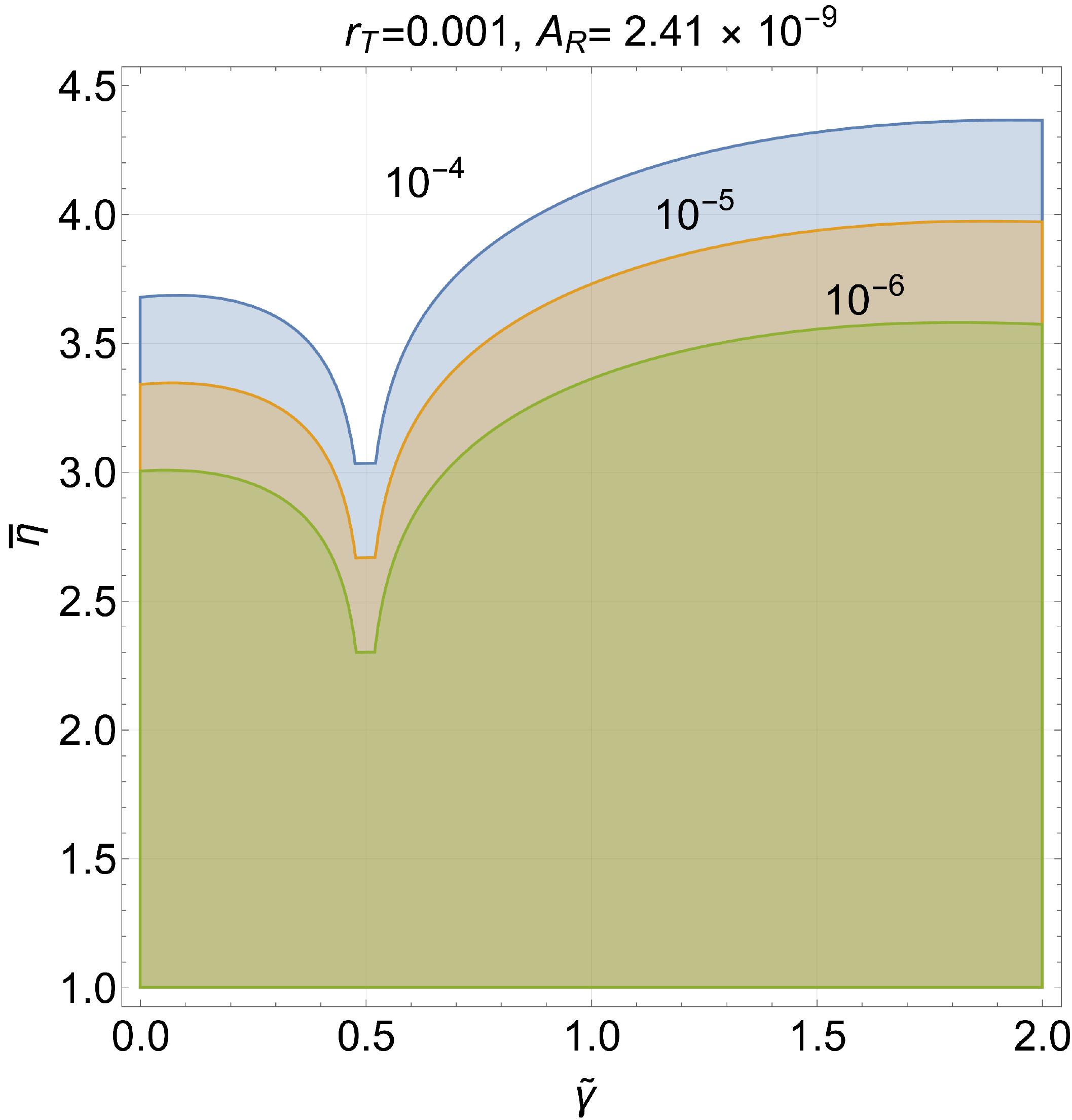}
\caption[a]{The critical density bounds during the quasi-de Sitter stage of expansion are illustrated in the $(\widetilde{\gamma}, \, \overline{\eta})$ plane. As in  difference between the left and right plots comes from a different value 
of the tensor to scalar ratio $r_{T}$ entering the estimate of $(H/M_{P})$.}
\label{FFF2}      
\end{figure}
If $\overline{\eta}\geq 1$ we can look directly at Fig. \ref{FFF2} where we illustrate the two dimensional contours in the plane $(\widetilde{\gamma}, \, \overline{\eta})$.  As in Fig. \ref{FFF1} the 
various shadings correspond to the values of $\Omega_{Y}(k,\tau)$ indicated in the plots (i.e. $10^{-6}$, $10^{-5}$ and $10^{-4}$ from bottom to top). In spite of the differences between 
Figs. \ref{FFF1} and \ref{FFF2} we can therefore conclude that for $0<\widetilde{\gamma} \leq 2$ and $\overline{\eta} >1$ the physical region of the parameters corresponds to $\overline{\eta} = {\mathcal O}(\widetilde{\gamma})$.

\subsection{Late-time mode functions and power spectra}
For $\tau \geq - \tau_{1}$ the rate of variation of $\sqrt{\lambda}$ becomes, in this case ${\mathcal F} =  \widetilde{\delta}/y(\tau)$ where  $y(\tau) = \tau + \tau_{1} [ q(\widetilde{\gamma}, \widetilde{\delta}) +1]$ 
has the same definition given before but now $q(\widetilde{\gamma}, \widetilde{\delta}) = \widetilde{\gamma}/\widetilde{\delta}$. We can then define the appropriate set of rescaled variables, i.e. 
\begin{equation}
\theta = i \,\overline{\theta}, \qquad \overline{\theta} = \lambda_{0} \widetilde{\delta}, \qquad w = 2 i\, k\, y, \qquad \nu = |\widetilde{\delta}  -1/2|,
\label{EN3C}
\end{equation}
and then obtain the explicit evolution of the mode functions. Since these steps reproduce, with due differences, 
the discussion already presented in section \ref{sec4} the details will be omitted. The interested 
reader may however find useful the explicit form of the linearly independent solutions 
for the evolution of the mode functions. These results have been relegated to the appendix \ref{APPB} 
 but they are instrumental in deriving 
the elements of the mixing matrix in the case of the  $(\widetilde{\gamma}, \widetilde{\delta})$ transition\footnote{See in particular Eqs. (\ref{IN21hDEC})--(\ref{IN21nDEC}) in the case of the $L$-polarization and Eqs. (\ref{IN21haDEC})--(\ref{IN21naDEC}) in the case of the $R$-polarization. }:
\begin{equation}
\widetilde{\,{\mathcal M}\,}^{(X)}(w_{1},\, w,\,\widetilde{\delta}) =  \left(\matrix{ \widetilde{A}^{(X)}_{f\, f}(w_{1},\, w,\, \widetilde{\delta})
& \widetilde{A}^{(X)}_{f\,g}(w_{1},\, w,\, \widetilde{\delta})&\cr
\widetilde{A}^{(X)}_{g\,f}(w_{1},\,w,\,\widetilde{\delta}) & \widetilde{A}^{(X)}_{g\,g}(w_{1},\,w,\, \widetilde{\delta})&\cr}\right),
\label{MATRIX3a}
\end{equation}
where, as before, $X= L,\,R$. Even though Eq. (\ref{MATRIX3a})  is 
formally analog to the matrix defined in Eq. (\ref{MATRIX3}),  the various entries are clearly different in the two situations. 
The explicit results valid in the case of Eq. (\ref{MATRIX3a}) can be found in Eqs. (\ref{ATFFL})--(\ref{ATGGL}) (see also the discussion thereafter). When the gauge coupling decreases the explicit form of the power spectra will be
\begin{eqnarray}
\widetilde{\,P\,}_{B}(k,\tau) &=& \frac{k^{5}}{4 \pi^2} \sum_{X= R,\,L}\biggl|\widetilde{A}_{f\, f}^{(X)}(x,\, x_{1},\, \delta) \overline{f}_{k,\,X} + \widetilde{A}_{f\, g}^{(X)}(x,\, x_{1},\, \delta) \frac{\overline{g}_{k,\,X}}{k}\biggr|^2,
\label{MGC13}\\
\widetilde{\,P\,}_{E}(k,\tau) &=& \frac{k^{3}}{4 \pi^2} \sum_{X= R,\,L}\biggl|\widetilde{A}_{g\, f}^{(X)}(x,\, x_{1},\, \delta)\, k\, \overline{f}_{k,\,X} + \widetilde{A}_{g\, g}^{(X)}(x,\, x_{1},\, \delta) \, \overline{g}_{k,\,X}\biggr|^2.
\label{MGC14}
\end{eqnarray}
In Eqs. (\ref{MGC13})--(\ref{MGC14}) $\overline{f}_{k,\,X} = f^{(inf)}_{k, \, X}(-\tau_{1})$ and $\overline{g}_{k,\,X} = g^{(inf)}_{k, \, X}(-\tau_{1})$ where the inflationary mode functions are the ones given in Eqs. (\ref{DEC8}) and (\ref{DEC9}).  
The explicit form of $\widetilde{\,{\mathcal M}\,}^{(X)}(w_{1},\, w,\,\widetilde{\delta}) $ together 
with the different expressions for $\overline{f}_{k,\,X}$ and $\overline{g}_{k,\,X}$
imply that the contributions to the gauge power spectra are now  in a different hierarchy:
\begin{eqnarray}
\biggl|\widetilde{A}_{f\, f}^{(X)}(x,\, x_{1},\, \delta) \overline{f}_{k,\,X} \biggr| \gg \biggl| \widetilde{A}_{f\, g}^{(X)}(x,\, x_{1},\, \delta) \frac{\overline{g}_{k,\,X}}{k} \biggr|, 
\label{MGC15}\\
\biggl|\widetilde{A}_{g\, f}^{(X)}(x,\, x_{1},\, \delta)\, k\, \overline{f}_{k,\,X} \biggr| \gg \biggl| \widetilde{A}_{g\, g}^{(X)}(x,\, x_{1},\, \delta) \frac{\overline{g}_{k,\,X}}{k} \biggr|.
\label{MGC16}
\end{eqnarray}
The inequalities of Eqs. (\ref{MGC15})--(\ref{MGC16}) should be compared with the analog results holding in the case 
of increasing coupling [see  Eqs. (\ref{MGC5})--(\ref{MGC7})]. The simplest way of proving the correctness 
of Eqs. (\ref{MGC15})--(\ref{MGC16}) is to consider the physical region of the parameters where 
$\widetilde{\delta} \ll 1$. In this limit it can be shown that the difference between the $L$ and $R$ 
coefficients for $\tau\geq - \tau_{1}$ will be  ${\mathcal O}(\widetilde{\delta})$, i.e.  
\begin{equation}
\lim_{\widetilde{\delta} \ll 1}  \biggl[\widetilde{A}^{(R)}_{f\, g}( x_{1}, \,x, \widetilde{\delta}) - \widetilde{A}^{(L)}_{f\, g}( x_{1}, \,x, \widetilde{\delta})\biggr] = {\mathcal O}(\widetilde{\delta}).
\label{MGC17}
\end{equation}
The electric and the magnetic power spectra will therefore be:
\begin{eqnarray}
\widetilde{\,P\,}_{B}(k,\tau) &=& \frac{k^{5}}{4 \pi^2} \biggl[ \biggl|\widetilde{A}_{f\, f}^{(L)}(x,\, x_{1},\, \delta) \overline{f}_{k,\,L} \biggr|^2 
+ \biggl| \widetilde{A}_{f\, f}^{(R)}(x,\, x_{1},\, \delta) \overline{f}_{k,\,R} \biggr|^2 \biggr],
\label{MGC18}\\
\widetilde{\,P\,}_{E}(k,\tau) &=& \frac{k^{3}}{4 \pi^2} \biggl[ \biggl|\widetilde{A}_{g\, f}^{(L)}(x,\, x_{1},\, \delta)\, k\, \overline{f}_{k,\,L} \biggr|^2  + \biggl|\widetilde{A}_{g\, f}^{(R)}(x,\, x_{1},\, \delta) \,k\,\overline{f}_{k,\,R}\biggr|^2 \biggr].
\label{MGC19}
\end{eqnarray}
Following the same considerations developed above we will therefore have that the explicit form of the power spectra 
will be:
 \begin{eqnarray}
\widetilde{\,P\,}_{B}(k,\tau) &=& a_{1}^{4} \, H_{1}^4 \, D(\gamma + 1/2) \, \biggl(\frac{k}{a_{1} \, H_{1}}\biggr)^{4 - 2 \widetilde{\gamma} } \, 
\widetilde{F}_{B}(x_{1}, x, \widetilde{\delta}, \overline{\eta},\widetilde{\gamma}),
\nonumber\\
\widetilde{F}_{B}(x_{1}, x, \delta,x_{1}, x, \widetilde{\delta}, \overline{\eta},\widetilde{\gamma}) &=& \biggl|\widetilde{A}^{(X)}_{f\, g}( x_{1}, \,x, \widetilde{\delta})\biggr|^2 \widetilde{\,Q\,}_{B}(\overline{\eta},\gamma).
\label{MGC20}
\end{eqnarray}
Similarly the hyperelectric spectrum turns out to be
\begin{eqnarray}
\widetilde{\,P\,}_{E}(k,\tau) &=& a_{1}^{4} \, H_{1}^4 \, D(\widetilde{\gamma} + 1/2) \, \biggl(\frac{k}{a_{1} \, H_{1}}\biggr)^{4 - 2 \widetilde{\gamma} } \, \widetilde{\,F\,}_{E}(x_{1}, x, \widetilde{\delta}, \overline{\eta},\widetilde{\gamma}),
\nonumber\\
\widetilde{\,F\,}_{E}(x_{1}, x, \widetilde{\delta}, \overline{\eta},\widetilde{\gamma}) &=&\biggl|A^{(X)}_{g\, g}( x_{1}, \,x, \widetilde{\delta})\biggr|^2\, \widetilde{\,Q\,}_{B}(\overline{\eta},\gamma).
\label{MGC21}
\end{eqnarray}
The same analysis can be discussed in the case of the gyrotropic spectra but it will  be omitted since it can be 
easily deduced by using the reported results. 

\subsection{Anomalous contributions and quasi-dual power spectra}
\label{subs54}
The results of section \ref{sec4} and of the present section have been obtained in two 
manifestly dual situations. We want now to compare the final form of the power 
spectra valid in the two cases.
Starting from Eqs. (\ref{AUX7D})--(\ref{AUX8D}) when $\overline{\eta} \to 0$, the power spectra shall be compared with the analog expressions coming from Eqs. (\ref{AUX7})--(\ref{AUX8}) valid when the gauge coupling 
increases in the absence of anomalous terms (i.e. for $\overline{\zeta} \to 0$). Given the explicit form 
 of $\sqrt{\lambda}$ [see e.g. Eqs. (\ref{TWO1}) and (\ref{ADD1})] the  duality transformation stipulates that $\sqrt{\lambda} \to 1/\sqrt{\lambda}$ (i.e. $\gamma\to \widetilde{\gamma}$). But, as expected, if $\gamma \to \widetilde{\gamma}$ the hypermagnetic power spectra computed when the gauge coupling increases 
turn into the hyperelectric power spectra obtained for decreasing gauge coupling and vice versa (i.e.
 $P_{B}(k,\tau) \to \widetilde{\,P\,}_{E}(k,\tau)$ and  $P_{E}(k,\tau) \to \widetilde{\,P\,}_{B}(k,\tau)$):
 \begin{equation}
\sqrt{\lambda} \to 1/\sqrt{\lambda}\qquad \Rightarrow\gamma \to \widetilde{\gamma}\qquad \Rightarrow P_{E}(k,\tau) \to \widetilde{\,P\,}_{B}(k,\tau), \qquad P_{B}(k,\tau) \to \widetilde{\,P\,}_{E}(k,\tau).
\label{DUAL1}
\end{equation}
Equation (\ref{DUAL1}) can also be written by explicitly recalling the equations where the corresponding 
results have been obtained, namely:
 \begin{equation}
\sqrt{\lambda} \to 1/\sqrt{\lambda}\qquad \Rightarrow\gamma \to \widetilde{\gamma}\qquad \Rightarrow (\ref{AUX8}) \to (\ref{AUX7D}), \qquad (\ref{AUX7}) \to (\ref{AUX8D}).
\label{DUAL1a}
\end{equation}
So far we just reinstated that when the anomalous interactions are absent 
 (i.e. $\overline{\zeta} \to 0$ in Eqs. (\ref{AUX7})--(\ref{AUX8}) and $\overline{\eta} \to 0$ in Eqs. (\ref{AUX7D})--(\ref{AUX8D}))
duality is a good symmetry of the problem and it exchanges the hyperelectric and the hypermagnetic power spectra.
The same result can be obtained, incidentally, for the late-time power spectra as long as the anomalous terms 
vanish (i.e. $\overline{\xi}\to 0$ and $\overline{\theta} \to 0$). 

Let us now move to the case where $\overline{\zeta}\neq 0$ and $\overline{\eta}\neq 0$  and let us compare 
 the hyperelectric power spectra of Eqs. (\ref{PSA3})--(\ref{PSA4}) with their hypermagnetic analog obtained in the case of decreasing gauge coupling (i.e. Eqs. (\ref{PSB1})--(\ref{PSB2})):  
\begin{eqnarray}
&& \sqrt{\lambda} \to 1/\sqrt{\lambda}\qquad \Rightarrow  \gamma \to \widetilde{\gamma} \qquad \Rightarrow\qquad  \overline{\zeta} \to \overline{\eta} \qquad \Rightarrow 
\nonumber\\
&& P_{E}(k,\tau) \to \widetilde{\,P\,}_{B}(k,\tau), \qquad P^{(G)}_{E}(k,\tau) \to - \,\widetilde{\,P\,}^{(G)}_{B}(k,\tau).
\label{DUAL2}
\end{eqnarray}
According to Eq. (\ref{DUAL2})  it is still true that a duality transformation 
exchanges the hyperelectric and the hypermagnetic power spectra even if the anomalous interactions   
are present. More precisely the comparison between Eqs. (\ref{DUAL1}) and (\ref{DUAL2}) shows that if $\gamma\to \widetilde{\gamma}$ (and viceversa) we have that $\zeta \to \eta$ and viceversa\footnote{This simply happens 
because $\overline{\zeta}= \lambda_{0} \, \gamma$ and $\overline{\eta} = \lambda_{0} \widetilde{\gamma}$; see, in this respect, the 
specific definitions of $\overline{\zeta}$ and $\overline{\eta}$ of Eqs. (\ref{IN4}) and (\ref{DEC4}).}
While in Eq. (\ref{DUAL1}) the gyrotropies vanish as a consequence of the absence of anomalous interactions, in Eq. (\ref{DUAL2}) the hyperelectric and the hypermagnetic gyrotropies are exchanged under duality 
but up to a sign; the reason for this result is that for $\gamma \to \widetilde{\gamma}$:
\begin{equation}
Q_{E}(\overline{\zeta},\gamma) \to \widetilde{\,Q\,}_{B}(\overline{\eta}, \widetilde{\gamma}), \qquad\qquad 
Q_{E}^{(G)}(\overline{\zeta},\gamma) \to - \widetilde{\,Q\,}^{(G)}_{B}(\overline{\eta}, \widetilde{\gamma}).
\label{DUAL3}
\end{equation}

According to Eq. (\ref{DUAL2}) is still true that, under duality, $P_{E}(k,\tau) \to \widetilde{\,P\,}_{B}(k,\tau)$, however,  by comparing Eqs. (\ref{PSA1})--(\ref{PSA2}) with Eqs. (\ref{PSB6a})--(\ref{PSB6b}) we conclude that 
the reciprocal transformation is not satisfied i.e.  $P_{B}(k,\tau) \not\rightarrow \widetilde{\,P\,}_{E}(k,\tau)$.
All in all we can say that, in the presence of anomalous interactions, the gauge power spectra 
for increasing and decreasing gauge couplings are connected as follows: 
\begin{eqnarray}
&& \sqrt{\lambda} \to 1/\sqrt{\lambda}\quad \Rightarrow \gamma \to \widetilde{\gamma} \qquad \Rightarrow \overline{\zeta} \to \overline{\eta}\qquad \Rightarrow
 \nonumber\\
&& P_{E}(k,\tau) \rightarrow \widetilde{\,P\,}_{B}(k,\tau),\qquad P^{(G)}_{E}(k,\tau) \to \widetilde{\,P\,}^{(G)}_{B}(k,\tau),
\nonumber\\
&& P_{B}(k,\tau) \not\rightarrow \widetilde{\,P\,}_{E}(k,\tau),\qquad P^{(G)}_{B}(k,\tau) \not\rightarrow \widetilde{\,P\,}^{(G)}_{E}(k,\tau).
\label{DUAL4}
 \end{eqnarray}
To be even more specific we remind of the explicit equations where the results appearing in Eq. (\ref{DUAL4}) have been obtained, namely:
\begin{eqnarray}
&& \sqrt{\lambda} \to 1/\sqrt{\lambda}\quad \Rightarrow \gamma \to \widetilde{\gamma} \qquad \Rightarrow \overline{\zeta} \to \overline{\eta}\qquad \Rightarrow
 \nonumber\\
&&(\ref{PSA3}) \rightarrow (\ref{PSB1}),\qquad (\ref{PSA4}) \to (\ref{PSB2}),
\nonumber\\
&& (\ref{PSA1}) \not\rightarrow (\ref{PSB6a}),\qquad (\ref{PSA2}) \not\rightarrow (\ref{PSB6b}).
\label{DUAL4a}
 \end{eqnarray}
The remnant of the duality symmetry expressed by Eqs. (\ref{DUAL4})--(\ref{DUAL4a}) is ultimately related to the late-time behaviour of the gauge power spectra. This result is useful not only at a technical level but also 
for the phenomenological applications since it shows that the late-time form of the power 
spectra does not depend on the approximation schemes but it follows, to some extent, 
from symmetry considerations.

\renewcommand{\theequation}{6.\arabic{equation}}
\setcounter{equation}{0}
\section{Baryogenesis and magnetogenesis requirements}
\label{sec6}
\subsection{Reentry all along the radiation-dominated phase}
The shortest wavelengths reentering the effective horizon throughout the radiation-dominated 
phase affect the BAU while the largest wavelengths reentering just before matter-radiation
equality will have to be compared with the magnetogenesis requirements. 
Indeed the largest $k$-modes (i.e. smallest wavelengths) reentering the effective horizon between the end of inflation and the 
electroweak epoch act as Abelian sources of the potentially anomalous currents \cite{FIVEa0,FIVEa1}
and ultimately produce the BAU. The magnetogenesis requirements are instead associated with the 
modes that reenter the Hubble radius 
just before matter radiation equality: these modes are comparable with the typical scale of the gravitational 
collapse of the protogalaxy which is ${\mathcal O}(\mathrm{Mpc})$.

In this analysis we shall assume that for   $T> T_{ew}$ the electroweak symmetry is restored while around $T\simeq T_{ew}$ the  ordinary magnetic fields are proportional 
to the hypermagnetic fields through the cosine of the Weinberg's angle $\theta_{W}$, i.e. $\cos{\theta_{W}} \, \vec{B}$.  Assuming, for the sake of concreteness, that the electroweak temperature $T_{ew}$  is ${\mathcal O}(100)$ GeV,  when $T < T_{ew}$ the $SU_{L}(2)\otimes U_{Y}(1)$ symmetry is broken down 
to $U_{\mathrm{em}}(1)$ and the produced gauge fields will survive in the plasma as ordinary magnetic fields evolving 
in an electrically neutral plasma \cite{mhd1,mhd2} (see also \cite{moffat,parker,zeldovich}). 
While below $T_{ew}$ the fermions couple to the gauge fields 
through the standard vector currents,  above $T_{ew}$ the coupling of the hyercharge to fermions is chiral 
and the baryonic currents are anomalous\footnote{  
The $SU_{L}(2)$ anomaly is typically responsible for $B$ and $L$ non-conservation via instantons and sphalerons; the $U_{Y}(1)$ anomaly might lead to the transformation of the infra-red modes of the hypercharge field into fermions \cite{CKN3,CKN4,CKN5}.}. This means that the hypermagnetic gyrotropy associated with the modes reentering prior to the phase transition must be released into fermions for $T< T_{ew}$. In this section we shall first pin down the regions where the gyrotropy is sufficiently intense to seed the BAU;  at the electroweak epoch while the backreaction constraints and the other magnetogenesis requirements are concurrently satisfied.  

\subsection{Hypermagnetic gyrotropy and baryogenesis}
\begin{figure}[!ht]
\centering
\includegraphics[height=7.3cm]{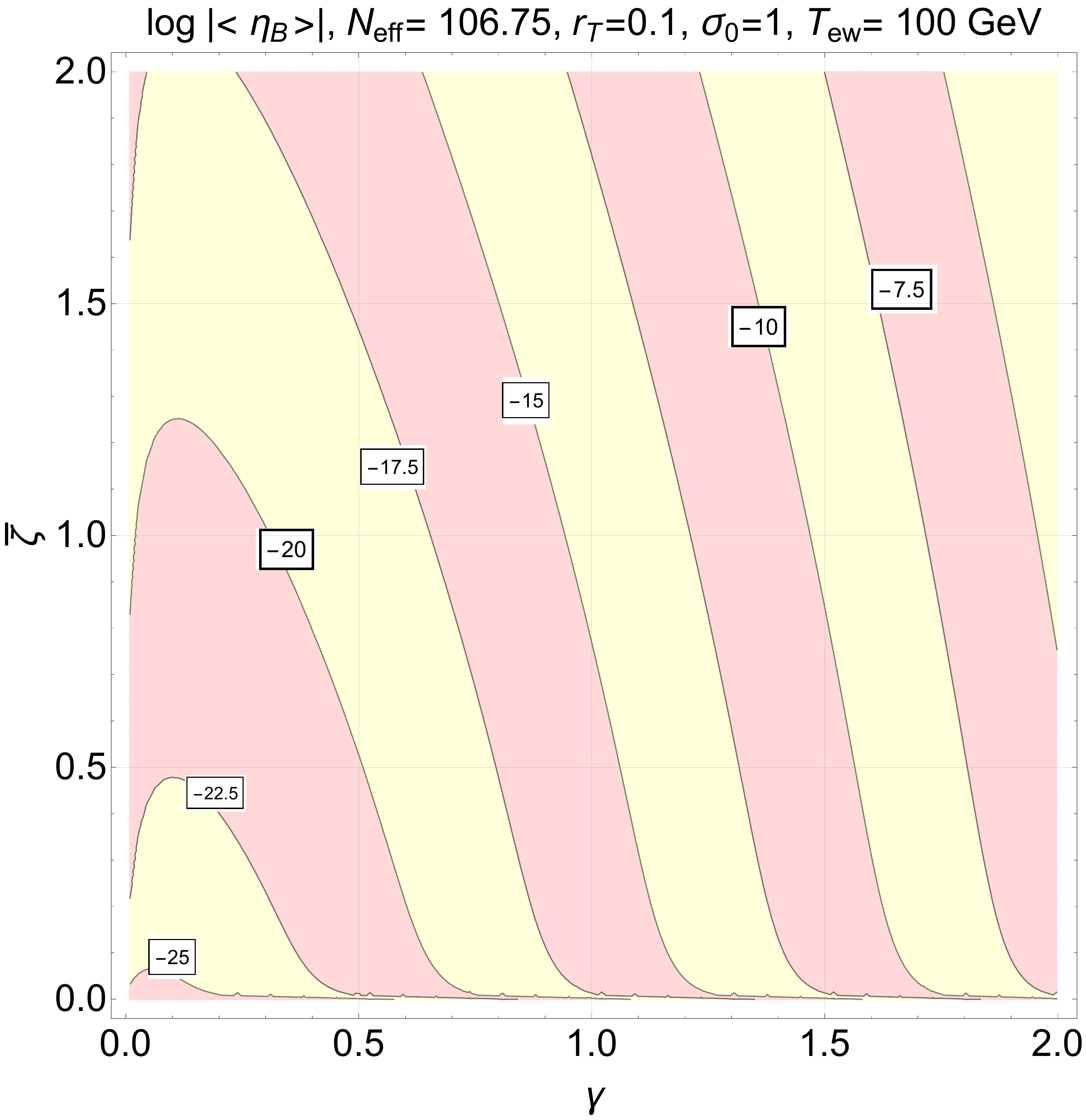}
\includegraphics[height=7.3cm]{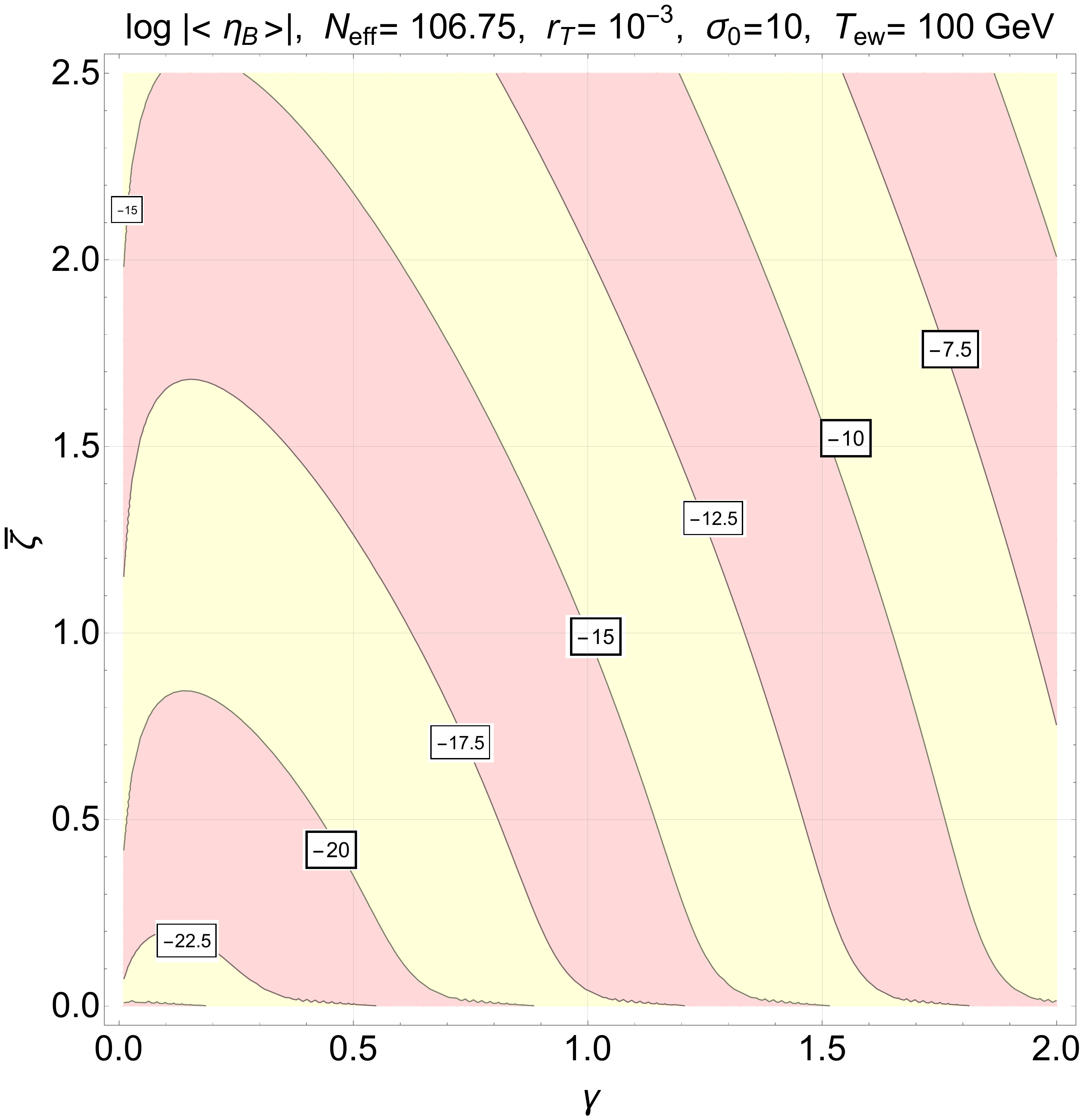}
\caption[a]{The contours of the common logarithm of the BAU are illustrated for a fiducial set of parameters assuming $T_{ew} =100 \,\mathrm{GeV}$; note that 
$r_{T}=0.1$ (left plot) and $r_{T} =0.001$ (right plot). }
\label{FFF3}      
\end{figure}
While the contribution of the hypermagnetic gyrotropy determines the comoving baryon to entropy ratio $\eta_{B} = n_{B}/\varsigma$, 
 the hyperelectric gyrotropy is washed out by the effect of the chiral conductivity.  Denoting with $N_{eff}$  the effective number of relativistic degrees of freedom at the electroweak epoch and recalling the notations of Eqs. (\ref{PS10c})--(\ref{PS10d})
 the expression of the BAU \cite{FIVEa0,FIVEa1}
is:
\begin{equation}
\eta_{B}(\vec{x},\tau) = \frac{n_{B}}{\varsigma} = \frac{ 3 \alpha' n_{f}}{8\pi \, H} \biggl(\frac{T}{\sigma}\biggl) \frac{{\mathcal G}^{(B)}(\vec{x}, \tau)}{{\mathcal H} \,a^4 \rho_{crit}},\qquad\qquad {\mathcal G}^{(B)}(\vec{x}, \tau) =  \vec{B} \cdot \vec{\nabla} \times \vec{B},
\label{BAU}
\end{equation}
where $\varsigma= 2 \pi^2 T^3 N_{eff}/45$ is the entropy density of the plasma, $\alpha^{\prime} = g'^2/(4\pi)$ (with $g'\simeq 0.3 $) is the $U(1)_{Y}$ coupling at the electroweak time and  $n_{f}$ is the number of fermionic generations. In what follows $N_{eff}$ shall be fixed to its standard model value (i.e. $N_{eff}= 106.75$).  Equation (\ref{BAU}) holds when the rate of the slowest reactions in the plasma (associated with the right-electrons) is larger than the dilution rate caused by the hypermagnetic field itself: at the phase transition the hypermagnetic gyrotropy  is converted back into fermions since the ordinary magnetic fields does not couple to fermions. Since all quantities in Eq. (\ref{BAU})  are comoving, $\langle \,\eta_{B}(\vec{x},\tau)\,\rangle$ for $\tau = \tau_{ew}$ is 
  determined by the averaged gyrotropy:
   \begin{equation} 
 \langle \, \eta_{B}(\vec{x}, \tau_{ew})\, \rangle = \frac{3 \, n_{f}\, \alpha^{\prime\, 2 }}{ 4 \, \pi\sigma_{0}\, a_{ew}^4\, \rho_{crit} {\mathcal H}_{ew}} \int_{0}^{k_{\sigma}} \, P_{B}^{(G)}(k,\tau) \, dk,
 \label{BAU1a}
 \end{equation}
 where $\sigma_{0}$ accounts for the theoretical uncertainty associated with the determination of the chiral conductivity  of the electroweak plasma according to $\sigma_{c} = \sigma_{0} T/\alpha^{\prime}$ \cite{cond1,cond2}.  The upper limit of integration in Eq. (\ref{BAU1a}) coincides with  diffusivity momentum and  it is useful to express $k_{\sigma}$ in units of ${\mathcal H}_{ew} = a_{ew} H_{ew}$ where 
 $H_{ew}^{-1} = {\mathcal O}(1) \, \mathrm{cm}$ is the Hubble radius at the electroweak time:
 \begin{equation}
 \frac{k_{\sigma}}{a_{ew} \, H_{ew}} = 3.5 \times 10^{8} \, \sqrt{\frac{\sigma_{0}}{\alpha^{\prime} \, N_{eff}}}\, \biggl(\frac{T_{ew}}{100\, \mathrm{GeV}} \biggr)^{-1/2}.
 \label{ksigma}
 \end{equation}
The typical diffusion scale exceeds the electroweak Hubble rate by approximately $8$ orders of magnitude and the corresponding range of wavelengths $\lambda_{\sigma}$ will roughly 
be smaller than ${\mathcal O}(10)$ nm (i.e. ${\mathcal O}(10^{-8}) \mathrm{cm}$) at $\tau_{ew}$. By definition
  $x_{1} = k/(a_{1} H_{1})$ and $\tau_{ew} \simeq \tau \gg \tau_{1}$; thus the explicit expression of Eq. (\ref{MGCC13}) 
becomes:
\begin{equation}
P_{B}^{(G)}(k,\tau) = a_{1}^4 \, H_{1}^4 \, D(\gamma +1/2) Q_{E}^{(G)}(\overline{\zeta}, \gamma)  \biggl(\frac{k}{a_{1} H_{1}}\biggr)^{ 4 - 2 \gamma} \sin^2{k\tau}.
\label{BAU1b}
\end{equation} 
It is relevant to remark that Eq. (\ref{BAU1b}) does not depend crucially on the approximation scheme 
but it is rather a direct consequence of the symmetries of the system (see in particular Eqs. (\ref{DUAL4})--(\ref{DUAL4a}) and discussion therein).
Recalling now the average of the  comoving gyrotropy computed in Eq. (\ref{PS10d}), the result of Eq. (\ref{BAU1b}) leads 
to the wanted estimate of $\langle \, \eta_{B}(\vec{x}, \tau_{ew})\, \rangle$:
\begin{eqnarray}
 \langle \, \eta_{B}(\vec{x}, \tau_{ew})\, \rangle &=& \frac{2 \, n_{f} \, \alpha^{\prime \, 2}}{\sigma_{0}} \, \biggl(\frac{H_{1}}{M_{P}}\biggr)^2   D(\gamma +1/2) Q_{E}^{(G)}(\overline{\zeta}, \gamma)  \biggl(\frac{a_{ew} \, H_{ew}}{a_{1} \, H_{1}}\biggr)^{ 4 - 2 \gamma} {\mathcal I}(\gamma, k_{\sigma}), 
 \label{BAU1c}\\
  {\mathcal I}(\gamma, k_{\sigma}) &=&  \int_{0}^{x_{\sigma}} \, \, x^{4 - 2 \gamma} \, \, \sin^2{x} \,\,d x, \qquad 
  x = \frac{k}{a_{ew} H_{ew}}, \qquad  x_{\sigma} = \frac{k_{\sigma}}{a_{ew} H_{ew}}.
\label{BAU1d} 
 \end{eqnarray}
 Since the evolution between $\tau_{1}$ and $\tau_{ew}$ is 
dominated by radiation, in Eq. (\ref{BAU1c}) we took into account that that $\rho_{crit} = 3 H_{ew}^2 \overline{M}_{P}= 3 H_{1}^2 \overline{M}_{P} (a_{1}/a_{ew})^4$ with $\overline{M}_{P} = M_{P}/\sqrt{8 \pi}$. For the same reason we will also have 
that $(a_{ew}/a_{1})(H_{ew}/H_{1}) = \sqrt{H_{ew}/H_{1}}$.

For the range 
of $k_{\sigma}$ entering Eq. (\ref{ksigma}) the integral $ {\mathcal I}(\gamma, k_{\sigma})$ extends between 
$0$ and about $10^{8}$;  as long as $\gamma \leq 2$,  the integral of
Eq. (\ref{BAU1d}) is estimated as
${\mathcal I}(\gamma, k_{\sigma}) = c_{\gamma} x_{\sigma}^{5 - 2 \gamma}$ where $c_{\gamma} = 1/[2 (5 - 2\gamma)]$. 
This result is very accurate for  $x_{\sigma} > 10^{3}$ and $ \gamma\leq 2$. For $ x_{\sigma} = 10^{3}$ and $\gamma=2$
 the ratio between the exact and the approximate results is $0.9$ while it is practically $1$ as long as 
$x_{\sigma} > 10^{3}$ for the whole range $0< \gamma \leq 2$. Taking into account this  estimate and the other explicit figures mentioned above,  the final expression of the BAU will be: 
\begin{eqnarray}
 \langle \, \eta_{B}(\vec{x}, \tau_{ew})\, \rangle &=& \frac{2 \, n_{f} \, \alpha^{\prime \, 2}}{\sigma_{0}} \, \biggl(\frac{H_{1}}{M_{P}}\biggr)^{\gamma}   \biggl(\frac{H_{ew}}{M_{P}}\biggr)^{ 2 -  \gamma} \biggl(\frac{k_{\sigma}}{a_{ew}\, H_{ew}}\biggr)^{5 - 2 \gamma} {\mathcal J}^{(G)}(\overline{\zeta},\gamma), 
 \label{BAU1e}\\
{\mathcal J}^{(G)}(\overline{\zeta},\gamma) &=& \frac{2^{2 \gamma -3} \Gamma^2(\gamma +1/2)} {\pi^3 ( 5 - 2 \gamma)}  Q_{E}^{(G)}(\overline{\zeta}, \gamma),
\label{BAU1f}
\end{eqnarray}
where,  the function ${\mathcal J}^{(G)}(\overline{\zeta}, \gamma)$  collects all the previously introduced 
numerical factors not related to the physical scales of the problem\footnote{In doing so we  made explicit the expressions of $D(\gamma +1/2)$ and of $c_{\gamma}= 1/[2 (5 - 2\gamma)]$ by recalling that, throughout this paper,  the function $D(x)$ has been defined as $D(x) = 2^{2 x -3} \Gamma^2(x)/\pi^3$.}.  
\begin{figure}[!ht]
\centering
\includegraphics[height=7.3cm]{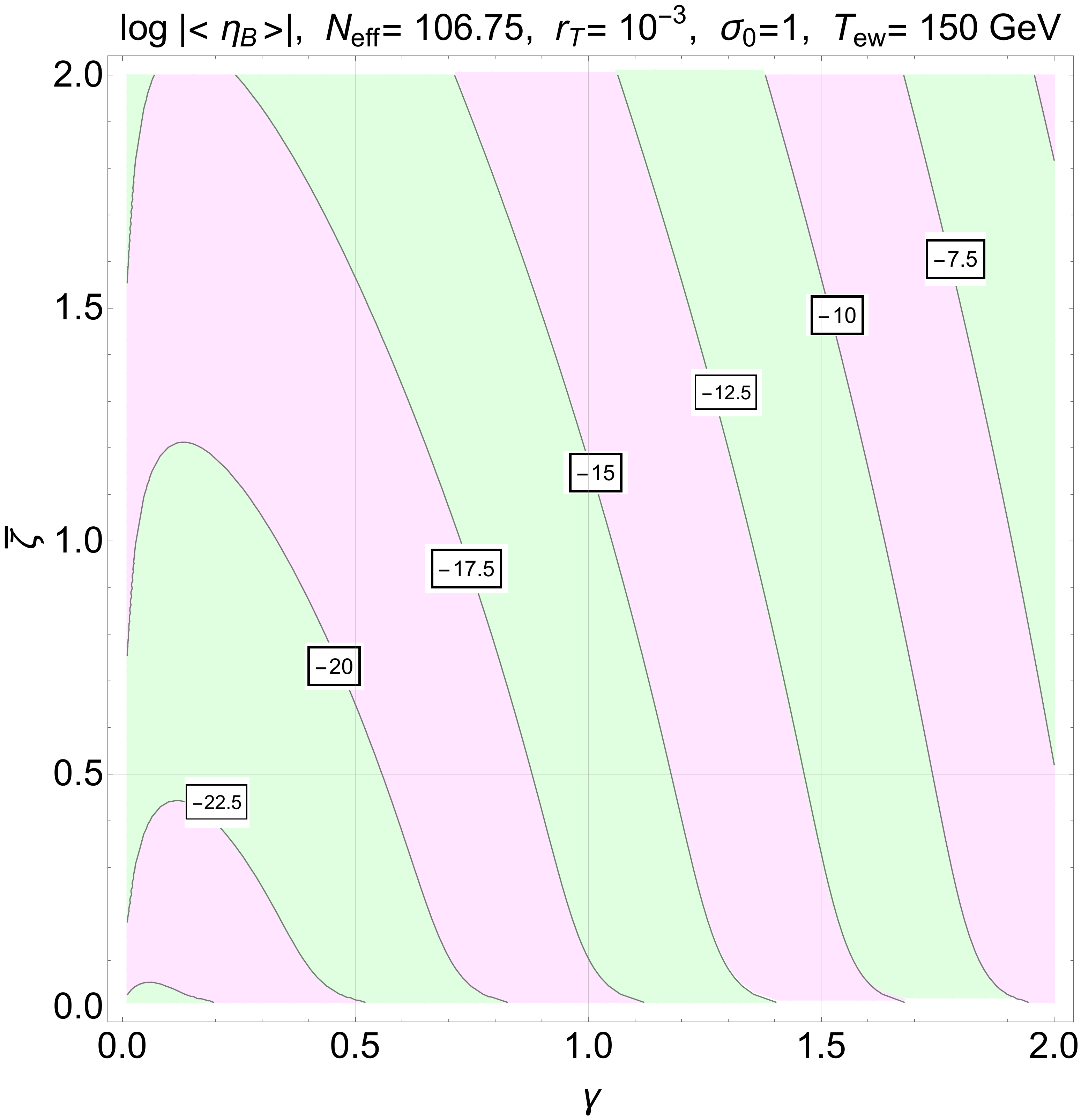}
\includegraphics[height=7.3cm]{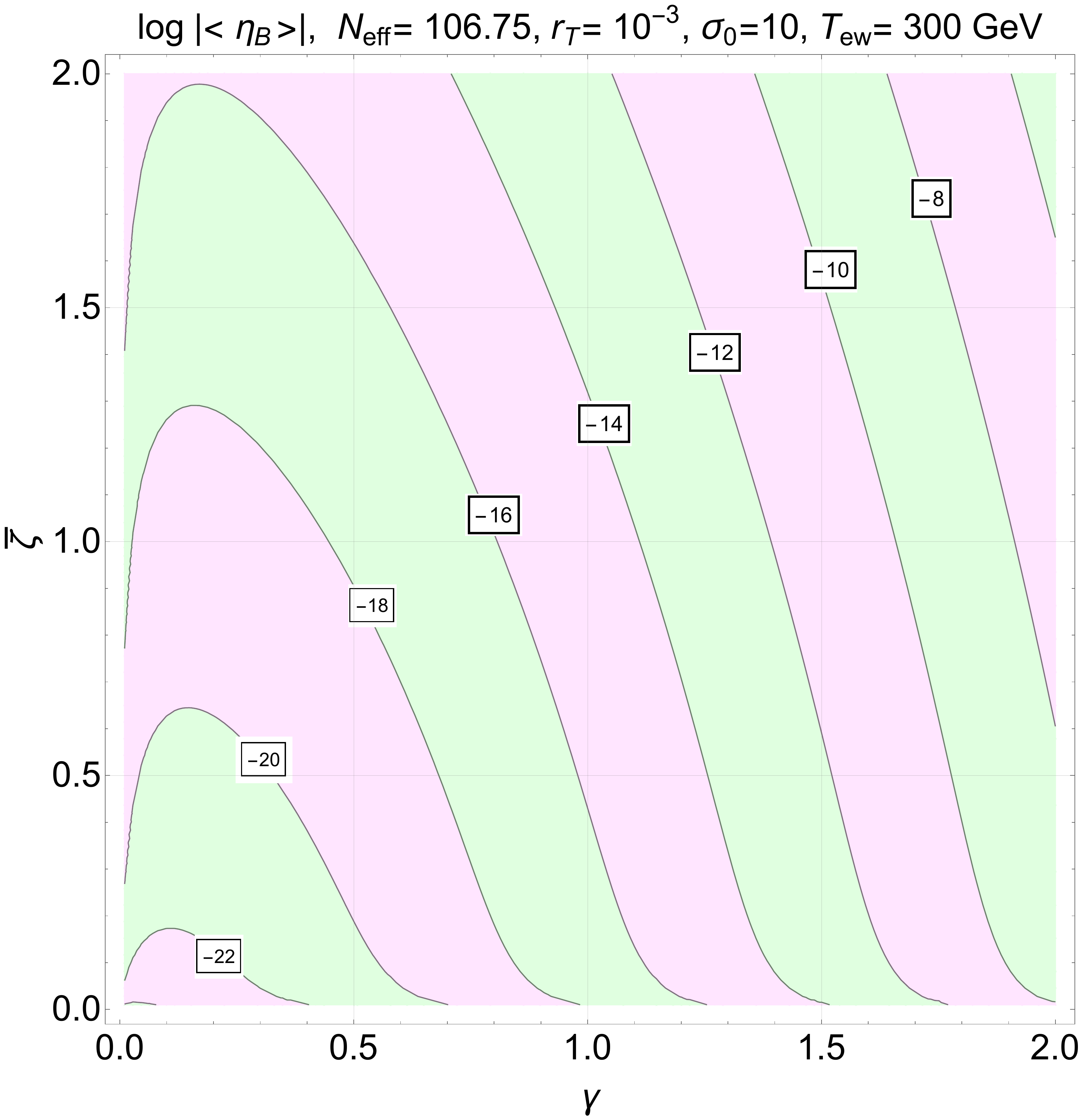}
\caption{We illustrate the contours of the common logarithm of the BAU for larger values of $T_{ew}$ (i.e. $T_{ew} =150\, \mathrm{GeV}$ and $T_{ew} =300\, \mathrm{GeV}$) and for two extreme values of $\sigma_{0}$. }
\label{FFF4}      
\end{figure}
The quantities appearing in Eq. (\ref{BAU1e}) can be easily related to the late-time cosmological parameters. In particular 
$H_{1}/M_{P} = \sqrt{\pi \, r_{T}\, {\mathcal A}_{{\mathcal R}}}/4$ where $r_{T}$ is the standard 
tensor to scalar ratio and ${\mathcal A}_{{\mathcal R}} = 2.41\times 10^{-9}$ is the amplitude 
of the scalar power spectrum evaluated at the pivot scale $k_{p} =0.002\, \mathrm{Mpc}^{-1}$. The other two ratios 
of scales appearing in Eq.  (\ref{BAU1e}) follow directly from Eq. (\ref{ksigma}) and from the explicit 
value of $T_{ew}$ and $N_{eff}$, i.e. $H_{ew}/M_{P} = \sqrt{4 \pi^3N_{eff}/45} (T_{ew}/M_{P})^2$.
The common logarithm of the BAU for a reference temperature of $100$ GeV is illustrated in Fig. \ref{FFF3} for different values of $r_{T}$. While the physical range of values\footnote{There have been recent attempts 
to make the bound even more stringent.  So for instance in Ref. \cite{RT1} the combination 
of different data sets implies $r_{T} <0.07$ while in Ref. \cite{RT2} values $r_{T} <0.064$ are quoted. 
These bounds are not based on a direct assessment of the $B$-mode polarization but rather on the effect of the tensor modes of the geometry on the remaining temperature and polarization 
anisotropies.}  of $r_{T}$ is below $10^{-2}$ \cite{RT1,RT2}, 
a drastic decrease in $r_{T}$ (e.g. $r_{T} = {\mathcal O}(10^{-4})$) has a mild impact on the numerical value of the BAU 
since $\eta_{B} \propto r_{T}^{\gamma/2}$. The interesting region of the parameter space is the one 
where $\eta_{B}$ is larger than $10^{-10}$: indeed we have to foresee the BAU might be diluted depending 
on the specific dynamics of the phase transition.  In Fig. \ref{FFF4} 
a different range of temperatures is illustrated but the features of the contours are fully compatible 
with the ones of Fig. \ref{FFF3}. It is finally useful to remark that $\sigma_{0}$ parametrizes, in practice, the indetermination 
in the estimate of $\sigma_{c}$  and this is why we shall assume that $ 1 \leq \sigma_{0} \leq 10$.
Both in Figs. \ref{FFF3} and \ref{FFF4} different values of $\sigma_{0}$ have been selected just for the sake of illustration. 
The baryon asymmetry produced on this way has a number of further consequences which have been partially 
investigated in the past \cite{sev1,sev2,sev3}. These discussions are beyond the scopes of the present discussion whose 
main purpose is to chart the parameter space of a potentially interesting scenario.

\subsection{Requirements from magnetogenesis}
The considerations associated with the BAU involved typical $k$-modes in the range
$a_{ew} \, H_{ew} \leq k < k_{\sigma}$. While the modes inside the Hubble radius at the 
electroweak time reentered right after inflation, for the effects related to magnetogenesis the relevant comoving scales are of 
the order of the Mpc at the epoch of the gravitational collapse of the protogalaxy,  the corresponding wavelengths 
reentered the effective horizon just prior to matter-radiation equality.
If$\tau_{k}=1/k$ denotes the reentry time of a generic wavelength, the ratio between $\tau_{k}$ and the time of matter-radiation equality $\tau_{eq}$ is\footnote{ In what follows $H_{0} 
= 100 \, h_{0} \, (\mathrm{km}/\mathrm{Mpc}) \, \mathrm{Hz}$ denotes the present value of the Hubble rate;
according to the standard notations  $\Omega_{X0}$ is the (present) critical fraction 
of a given species.  All the other standard cosmological notations 
will be assumed for the late-time parameters within the concordance paradigm (see e.g. \cite{wein1,wein2, primer}).}:
\begin{equation}
\frac{\tau_{k}}{\tau_{eq}} = 1.06 \times 10^{-6} \biggl(\frac{h_{0}^2 \Omega_{M0}}{0.1386} \biggr) 
\biggl(\frac{h_{0}^2 \Omega_{R 0}}{4.15\times 10^{-5}}\biggr)^{-1/2} \, \biggl( \frac{k}{\mathrm{Mpc}^{-1}}\biggr)^{-1}, 
\label{PH1}
\end{equation}
since $(\tau_{k}/\tau_{eq}) = \sqrt{2} (H_{0}/k) \Omega_{M0}/\Omega_{R0}$. 
Even if the reentry of the the Mpc scales occurred much later than the electroweak time,  Eq. (\ref{PH1}) shows that, within 
the concordance paradigm,  the wavenumbers $k = {\mathcal O}(\mathrm{Mpc}^{-1})$ still reentered 
prior to matter radiation equality.  We shall assume that the gauge coupling freezes at $\tau_{k}$ even if 
this is not strictly necessary. In fact for $\tau > \tau_{k}$ the conductivity dominates and the duality 
symmetry is further broken because of the presence of Ohmic currents,
\begin{figure}[!ht]
\centering
\includegraphics[height=7.3cm]{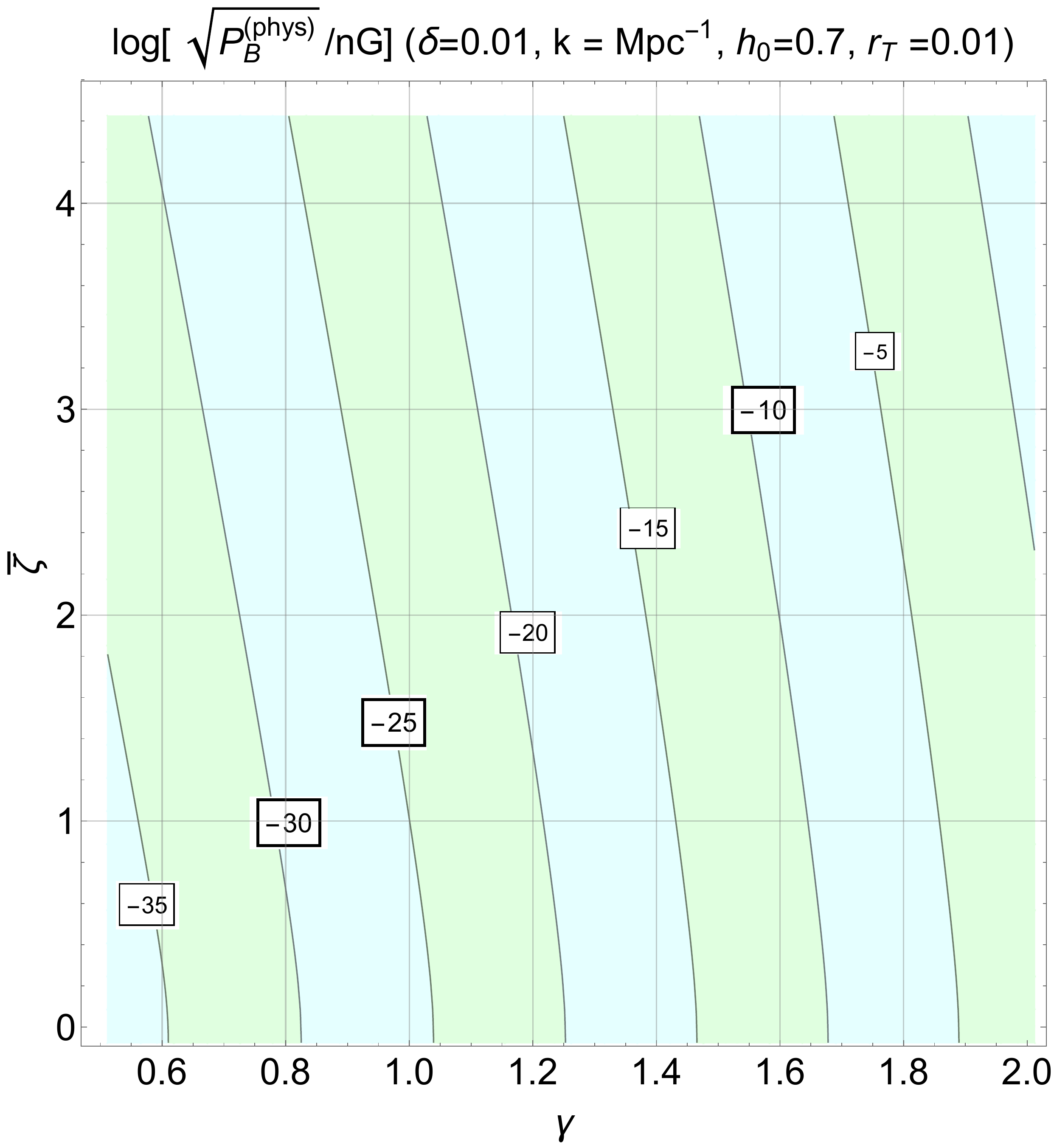}
\includegraphics[height=7.3cm]{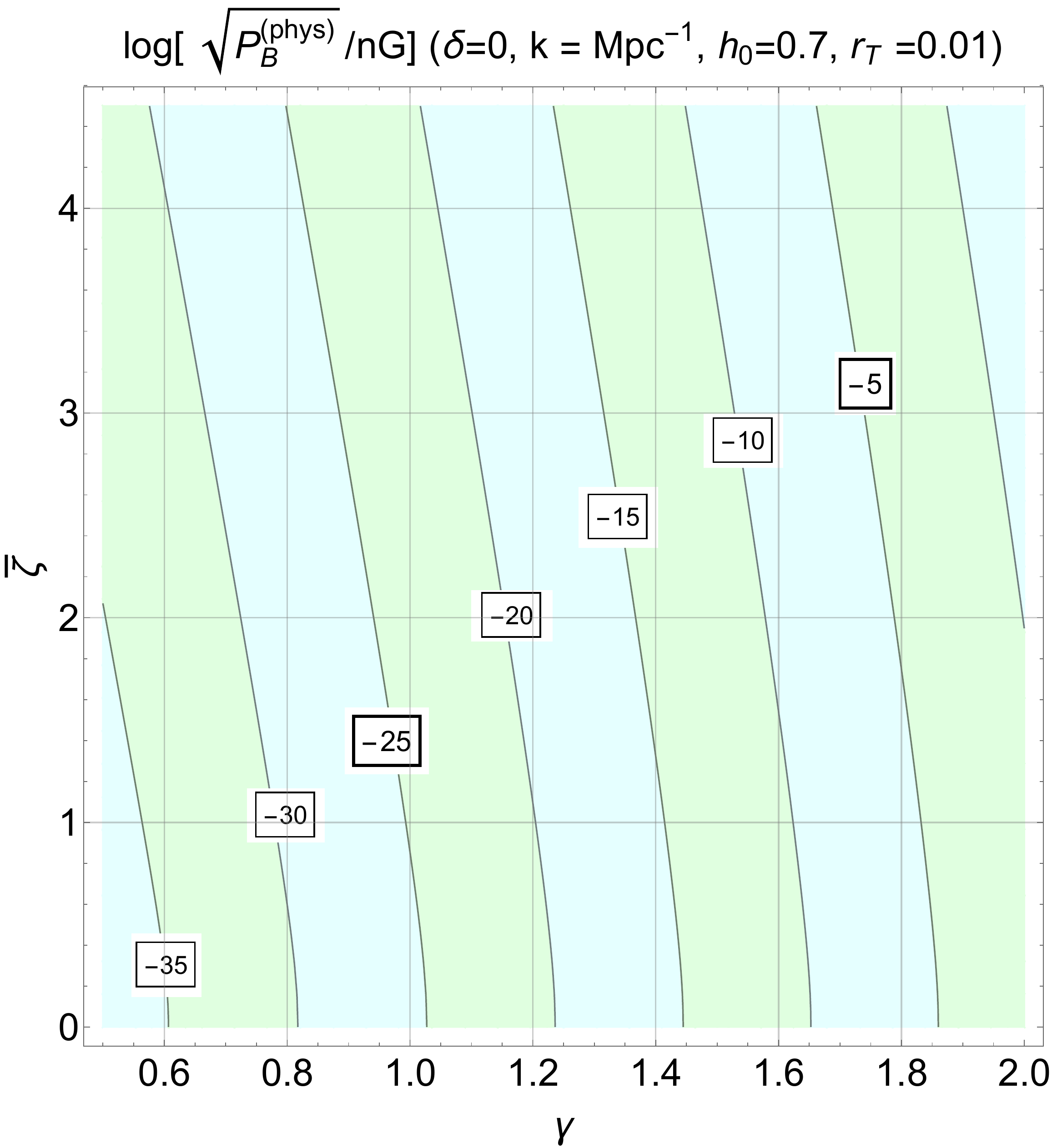}
\caption[a]{The parameter space is illustrated in the $(\gamma, \, \overline{\zeta})$ plane. The numbers appearing on the various contours correspond to the common logarithm of $\sqrt{P_{B}^{(phys)}}$ expressed in nG units.}
\label{FFF5}      
\end{figure}
so that the evolution of the magnetic and of the electric mode functions is very different: while 
the electric fields are suppressed by the finite value of the conductivity, the magnetic fields are not dissipated
at least for typical scales smaller than the magnetic diffusivity scale. In practice the mode functions 
for $\tau \geq \tau_{k}$ will be suppressed with respect to their values at $\tau_{k}$ as:
\begin{equation}
f_{k}(\tau) = f_{k}(\tau_{k}) e^{- k^2/k_{d}^2}, \qquad g_{k}(\tau) = (k/\sigma_{em}) g_{k}(\tau_{k}) e^{- k^2/k_{d}^2},\qquad 
k_{d}^{-2} = \int_{\tau_{k}}^{\tau} \,\,d \, z/\sigma_{em}(z),
\label{PH2}
\end{equation}
where $\sigma_{em}$ is the standard conductivity of the plasma (to be distinguished from its 
chiral counterpart $\sigma_{c}$ discussed above) and $k_{d}$ denotes the magnetic diffusivity 
momentum. The ratio $(k/k_{d})^2$ appearing in Eq. (\ref{PH2}) is actually extremely small in the 
phenomenologically interesting situation;  $\tau = \tau_{\mathrm{eq}}$ the ratio $(k/k_{d})^2$ turns out to be:
\begin{equation}
\biggl(\frac{k}{k_{d}}\biggr)^2 = \frac{4.75 \times 10^{-26}}{ \sqrt{2 \, h_{0}^2 \Omega_{M0} (z_{\mathrm{eq}}+1)}} \, \biggl(\frac{k}{\mathrm{Mpc}^{-1}} \biggr)^2,
\label{PH3}
\end{equation}
where $\Omega_{M0}$ is the present critical fraction in dusty matter and $z_{\mathrm{eq}} + 1 = a_{0}/a_{\mathrm{eq}}\simeq {\mathcal O}(3200)$ is the redshift of matter-radiation equality. The magnetic spectrum for 
$k < k_{d}$ is practically not affected by the conductivity while the electric power spectrum is
suppressed\footnote{Prior to decoupling the  electron-photon and electron-proton interactions tie the temperatures close together, the conductivity scales as $\sigma_{em} \sim 
\sqrt{T/m_{e}} T/\alpha_{em}$ where $m_{e}$ is the electromagnetic mass, $T$ is the temperature and $\alpha_{em}$ is the fine structure constant. For  instance, if we take $T= {\mathcal O}(\mathrm{eV})$ we get $k/\sigma = {\mathcal O}(10^{-30})$ for $k= {\mathcal O}(\mathrm{Mpc}^{-1})$. The electric power spectrum will then be suppressed, in comparison with the magnetic spectrum, by a factor ranging between $40$ and $60$ orders of magnitude.} by $k^2/\sigma_{em}^2\ll 1$.  

If a given quantity is not affected by the expansion of the background the 
comoving and physical fields could be used interchangeably. For instance in the case 
of the BAU $n_{B}$ (i.e. the baryon concentration) and $\varsigma$ (the entropy density) 
redshift in the same way so that $\eta_{B}$ could be computed directly in terms 
of the comoving fields. What matters for the 
magnetogenesis requirements are the instead physical power spectra prior to gravitational collapse since 
we now have to compare the values of the power spectra with a set of bounds valid at a given time.
Recalling Eq. (\ref{ACT13}) we have that the physical power spectrum is:
\begin{equation}
P_{B}^{(phys)}(k,\tau) = \frac{P_{B}(k,\tau)}{a^4 \lambda}\, \cos^2{\theta_{W}} , \qquad \Rightarrow \quad P_{B}^{(phys)}(k,\tau_{*}) = \frac{P_{B}(k,\tau_{k})}{a_{k}^4 \lambda_{k}}\biggl(\frac{a_{k}}{a_{*}}\biggr)^4\, \cos^2{\theta_{W}}.
\label{PH4}
\end{equation} 
The first expression of Eq. (\ref{PH4}) is just the definition of the physical power spectrum obtained directly 
from Eq. (\ref{ACT13}) by evaluating, after symmetry breaking, the two-point function of Eq. (\ref{PS6}) in terms of the physical fields. The  second expression appearing in Eq. (\ref{PH4}) is instead the physical power spectrum computed at a reference time 
$\tau_{*} > \tau_{k}$ under the further assumption that the gauge coupling 
is effectively frozen for $\tau > \tau_{k}$. 
\begin{figure}[!ht]
\centering
\includegraphics[height=7.3cm]{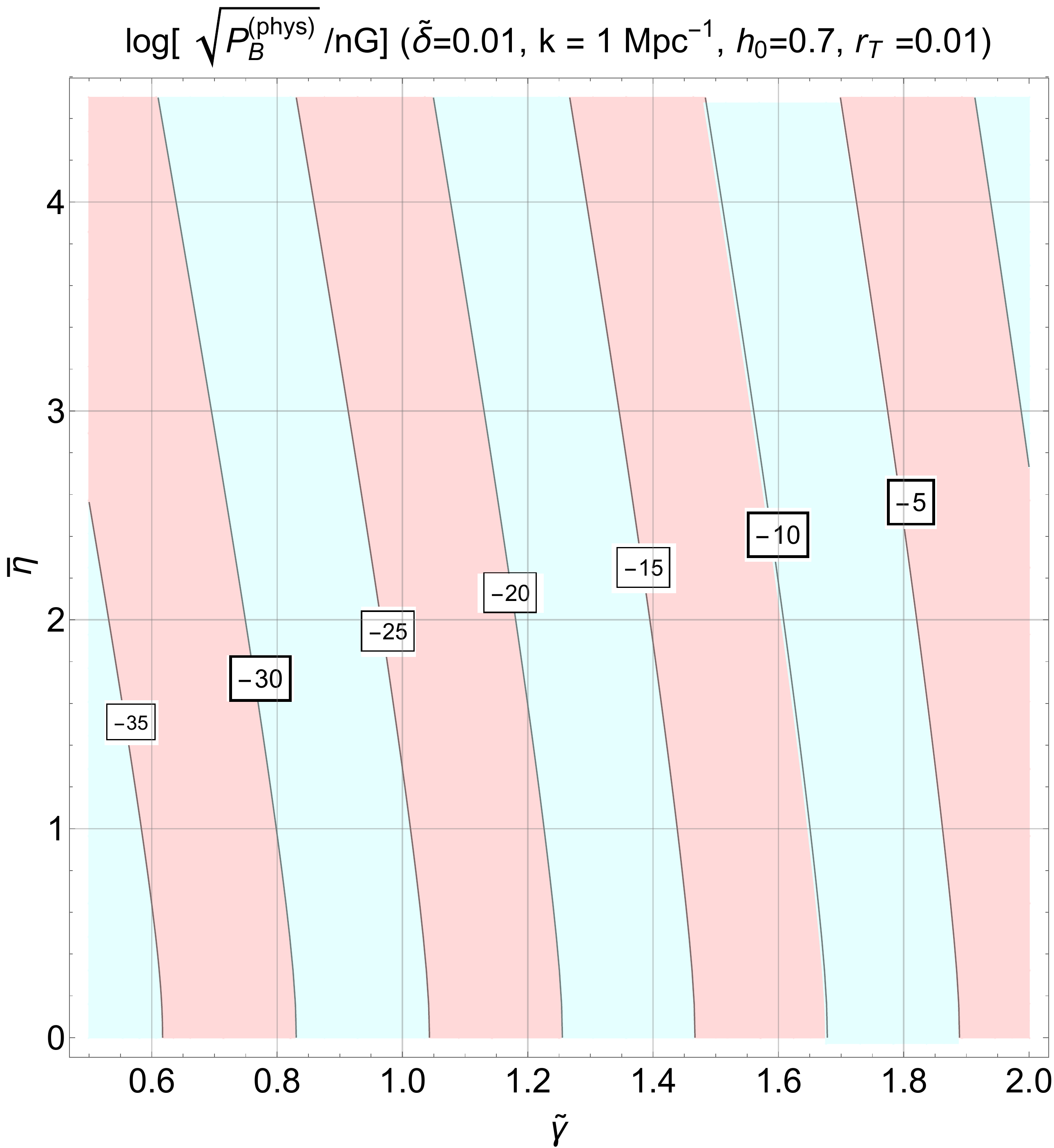}
\includegraphics[height=7.3cm]{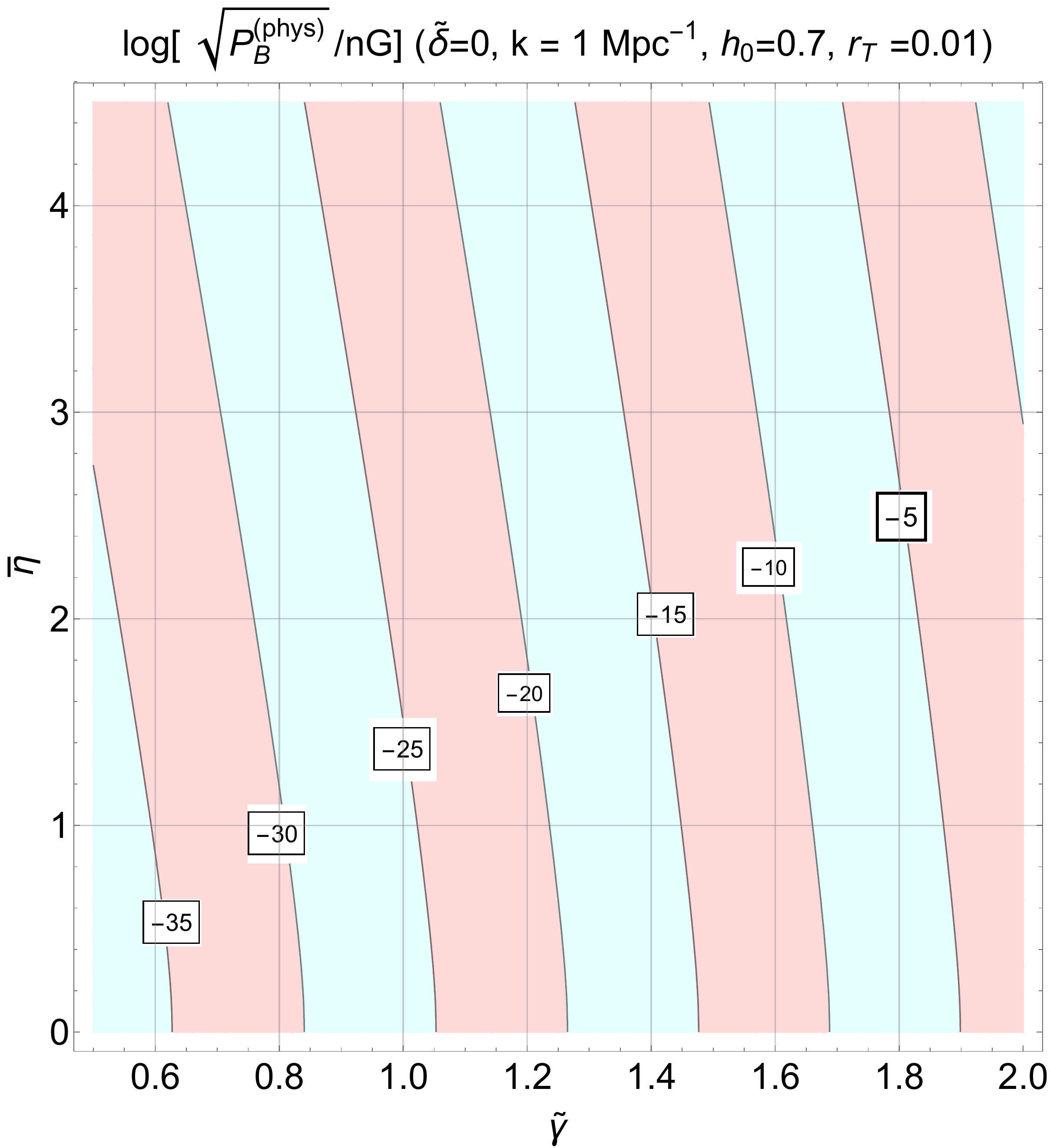}
\caption[a]{The parameter space is illustrated in the $(\widetilde{\gamma}, \, \overline{\eta})$ plane. As in Fig. \ref{FFF3} the numbers appearing on the various contours correspond to the common logarithm of $\sqrt{P_{B}^{(phys)}}$ expressed in nG units.}
\label{FFF6}      
\end{figure}
Even though the parameter 
space is easily charted by varying the relevant physical parameters,  it is useful, in a preliminary perspective,  to discuss the orders of magnitude of the problem by evaluating the physical counterpart of the comoving spectrum given in  Eq. (\ref{MGC11}).
 Inserting then Eq. (\ref{MGC11}) into Eq. (\ref{PH4}) we obtain\footnote{Note that ${\mathcal J}(\overline{\zeta},\gamma)$ is the 
counterpart of Eq. (\ref{BAU1f}) valid in the non-gyrotropic case.}:
 \begin{eqnarray}
 P_{B}^{(phys)}(k,\tau_{*}) &=& \frac{H_{1}^4}{\lambda_{1}} \biggl(\frac{a_{1}}{a_{k}}\biggr)^4 \, \biggl(\frac{a_{k}}{a_{*}}\biggr)^{4}  \biggl(\frac{k}{a_{1} H_{1}}\biggr)^{4 - 2 \gamma} {\mathcal J}(\overline{\zeta},\gamma)\cos^2{\theta_{W}} \,\,\sin^2{k\tau_{k}},
\nonumber\\
 {\mathcal J}(\overline{\zeta},\gamma) &=& Q_{E}(\overline{\zeta},\gamma) D(\gamma +1/2),
 \label{PH5}
\end{eqnarray}
where we considered, for the sake of simplicity, the situation where $\delta \ll \gamma$ so that the evolution of $\lambda$ 
could be effectively neglected for $\tau \geq - \tau_{1}$. The order of magnitude 
of the physical power spectrum does not have to coincide with 
the observed values the magnetic fields of the galaxies (or of the clusters).
The way large-scale magnetic fields originate is a rather broad subject 
that will not be discussed here in detail; various more specific discussions exist 
(see e. g. \cite{rev1,rev2,rev3}). In a conservative perspective  the magnetogenesis requirements 
 roughly demand that the magnetic fields at the time of the gravitational collapse of the protogalaxy should be approximately larger than a (minimal) power spectrum which can be estimated between ${\mathcal O}(10^{-32})\,\mathrm{nG}^2$ and ${\mathcal O}(10^{-22})\, \mathrm{nG}^2$:
\begin{equation}
\log{\biggl(\frac{\sqrt{P^{(phys)}_{B}(k,\tau_{*})}}{\mathrm{nG}} \biggr)} \geq - \omega, \qquad 11 < \omega < 16.
\label{PH6}
\end{equation}
The least demanding requirement  of Eq. (\ref{PH6}) (i.e. $\sqrt{P^{(phys)}_{B}(k,\tau_{*})} \geq 10^{-16}\,\mathrm{nG}$) follows by assuming 
that, after compressional amplification, every rotation of the galaxy increases the initial magnetic field of one $e$-fold. According to some this requirement is not completely realistic since it takes more than one $e$-fold  to increase the value of the magnetic field by one order of magnitude  and this is the rationale for the most demanding condition of Eq. (\ref{PH6}), i.e. $\sqrt{P^{(phys)}_{B}} \geq 10^{-11}\,\mathrm{nG}$. 
\begin{figure}[!ht]
\centering
\includegraphics[height=7.3cm]{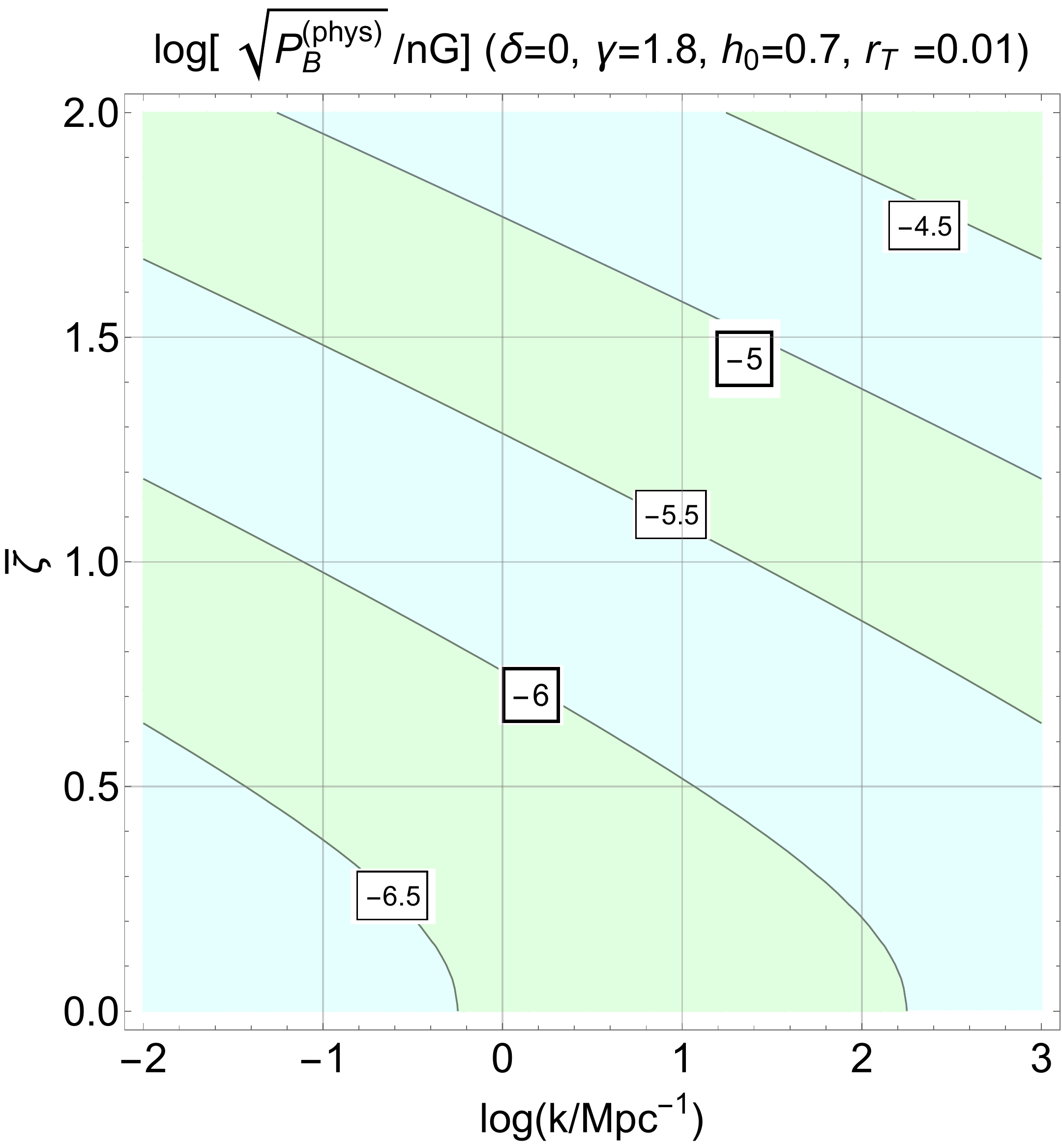}
\includegraphics[height=7.3cm]{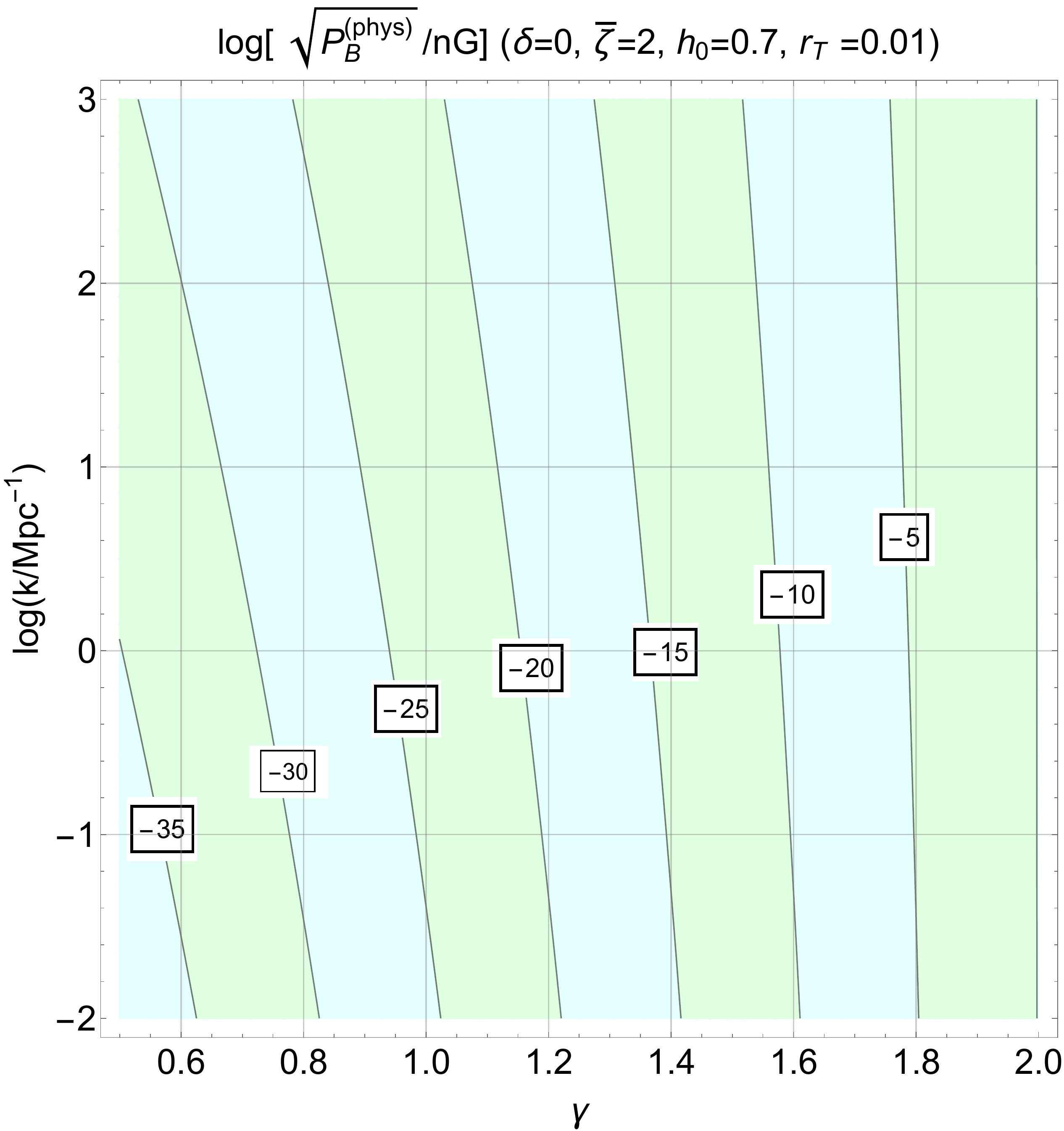}
\caption[a]{The parameter space is illustrated in the case of increasing gauge coupling and 
for different comoving scales. As in the previous plots the numbers appearing on the various contours correspond to the common logarithm of $\sqrt{P_{B}^{(phys)}}$ expressed in nG units.}
\label{FFF7}      
\end{figure}
It is now useful to estimate the overall amplitude of Eq. (\ref{PH5}) 
 and to compare the obtained result with the conditions of Eq. (\ref{PH6}). For this purpose let us 
 neglect, for a moment, the scale dependence of Eq. (\ref{PH5}) and focus on the overall amplitude; in this case 
 we will have that 
 \begin{equation}
\sqrt{P^{(phys)}_{B}(k,\tau_{*})} = \frac{H_{1}^2}{\lambda_{1}} \biggl(\frac{a_{1}}{a_{*}}\biggr)^2 \cos^2{\theta_{W}}  \simeq 
\frac{e_{1} \sqrt{\pi}}{32} \, r_{T}\, {\mathcal A}_{{\mathcal R}} \,M_{P}^2 \, \biggl(\frac{a_{1}}{a_{*}}\biggr)^2 \cos^2{\theta_{W}} ,
\label{PH7}
\end{equation}
where, as before, we traded $(H_{1}/M_{P})^2$ for $\pi r_{T} {\mathcal A}_{{\mathcal R}}/16$ and 
recalled the definition of the gauge coupling at $\tau_{1}$ (i.e. $e_{1} = \sqrt{4\pi/\lambda_{1}}$).
Equation (\ref{PH7}) should now be expressed in units of nG and, for this purpose, we recall that 
$M_{P}^2 = 2.15\times 10^{-66} \,\,\mathrm{nG}$. The final result for Eq. (\ref{PH7}) 
can therefore be written as:
 \begin{equation}
\sqrt{P^{(phys)}_{B}(k,\tau_{0})} = 4.89 \times 10^{-4} \biggl(\frac{r_{T}}{0.01}\biggr)^{1/2} \, 
\biggl(\frac{{\mathcal A}_{{\mathcal R}}}{2.41\times 10^{-9}}\biggr)^{1/2}\, \biggl( \frac{h_{0}^2 \Omega_{R0}}{4.15\times 
10^{-5}}\biggr)^{1/2} \,\,\, \mathrm{nG}.
\label{PH8}
\end{equation}
Even if the figures of Eq. (\ref{PH6}) seem even too large in comparison with Eq. (\ref{PH8}) 
a substantial reduction comes from the scale dependence that can be estimated by following exactly the same strategy 
outlined before. The relevant result is:
\begin{equation}
\frac{k}{a_{1}\, H_{1}} = 10^{-23.05}\,\, \biggl(\frac{k}{\mathrm{Mpc}^{-1}}\biggr)\, \biggl(\frac{r_{T}}{0.01}\biggr)^{-1/4}\,\,\biggl(\frac{h_{0}^2 \Omega_{R0}}{4.15\times 10^{-5}}\biggr)^{-1/4} \,\,\biggl(\frac{{\mathcal A}_{{\mathcal R}}}{2.41\times10^{-9}}\biggr)^{-1/4}.
\label{PH9}
\end{equation}
Last but not least the physical power spectrum will be further affected by ${\mathcal J}(\overline{\zeta},\gamma)$
which cannot be evaluated in its asymptotic limit since both $\overline{\zeta}$ and $\gamma$ 
are both quantities ${\mathcal O}(1)$; indeed the most interesting region of the parameter 
space (i.e. where the constraints of Figs. \ref{FFF1} and \ref{FFF2} are satisfied) occurs for relatively modest values 
of the anomalous interactions.  We shall therefore analyze the physical power spectra in numerical terms:
 the values of the common logarithm of the physical power spectra will be discussed 
 for the different values of the parameters. In Fig. \ref{FFF5}  and in all the forthcoming figures 
the labels denote $\sqrt{P_{B}^{(phys)}}$ in units on nG.
In Fig. \ref{FFF5} the physical power spectrum is illustrated in 
the $(\gamma, \overline{\zeta})$ plane: in both plots the gauge coupling increases 
during inflation and then flattens out by always remains smaller than $1$; however in the plot at the left $\delta =0.01$
while in the plot at the right $ \delta \to 0$. In the subsequent plots 
these two conceptually different physical situations will be often distinguished with the purpose of stressing that that the numerical differences are minimal. 

The critical density constraint discussed in Fig. \ref{FFF1} would imply that $\overline{\zeta} = {\mathcal O}(\gamma)$ 
and also, somehow, that $\zeta \leq 2$. By looking at the intercept on the $\gamma$ axis we see 
that the phenomenologically reasonable values of $\gamma$ correspond to spectra that are blue or, at most, quasi-flat but always slightly increasing with $k$. The effect of non-vanishing $\overline{\zeta}$ can be appreciated 
by comparing a five value of $\gamma$ for different values of $\overline{\zeta}$: the increase of $\overline{\zeta}$ from $0$ to $2$ or even $3$ is not crucial in order to satisfy the magnetogenesis constraints and it can be traded for an increase 
in $\gamma$. 
\begin{figure}[!ht]
\centering
\includegraphics[height=7.3cm]{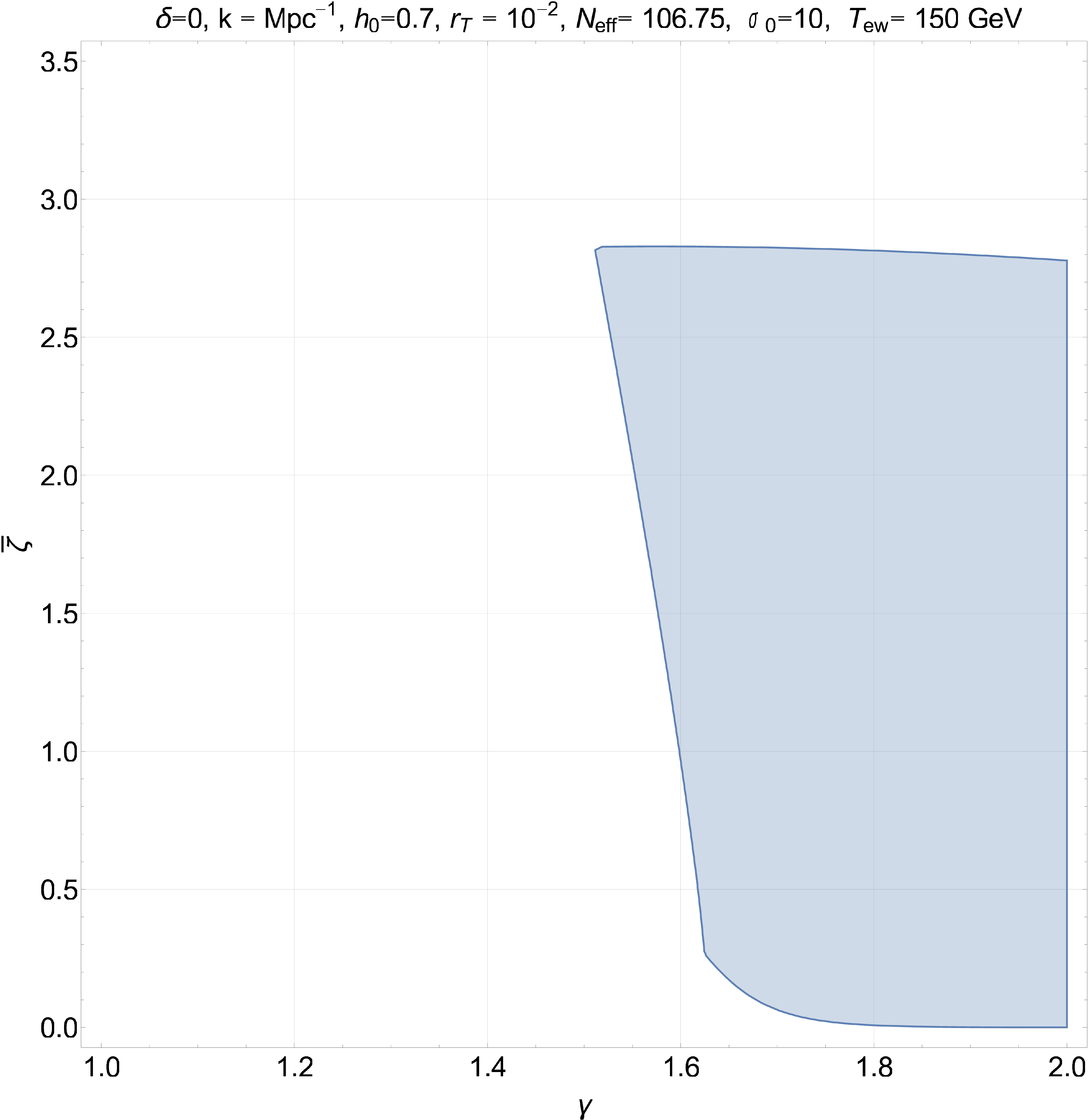}
\includegraphics[height=7.3cm]{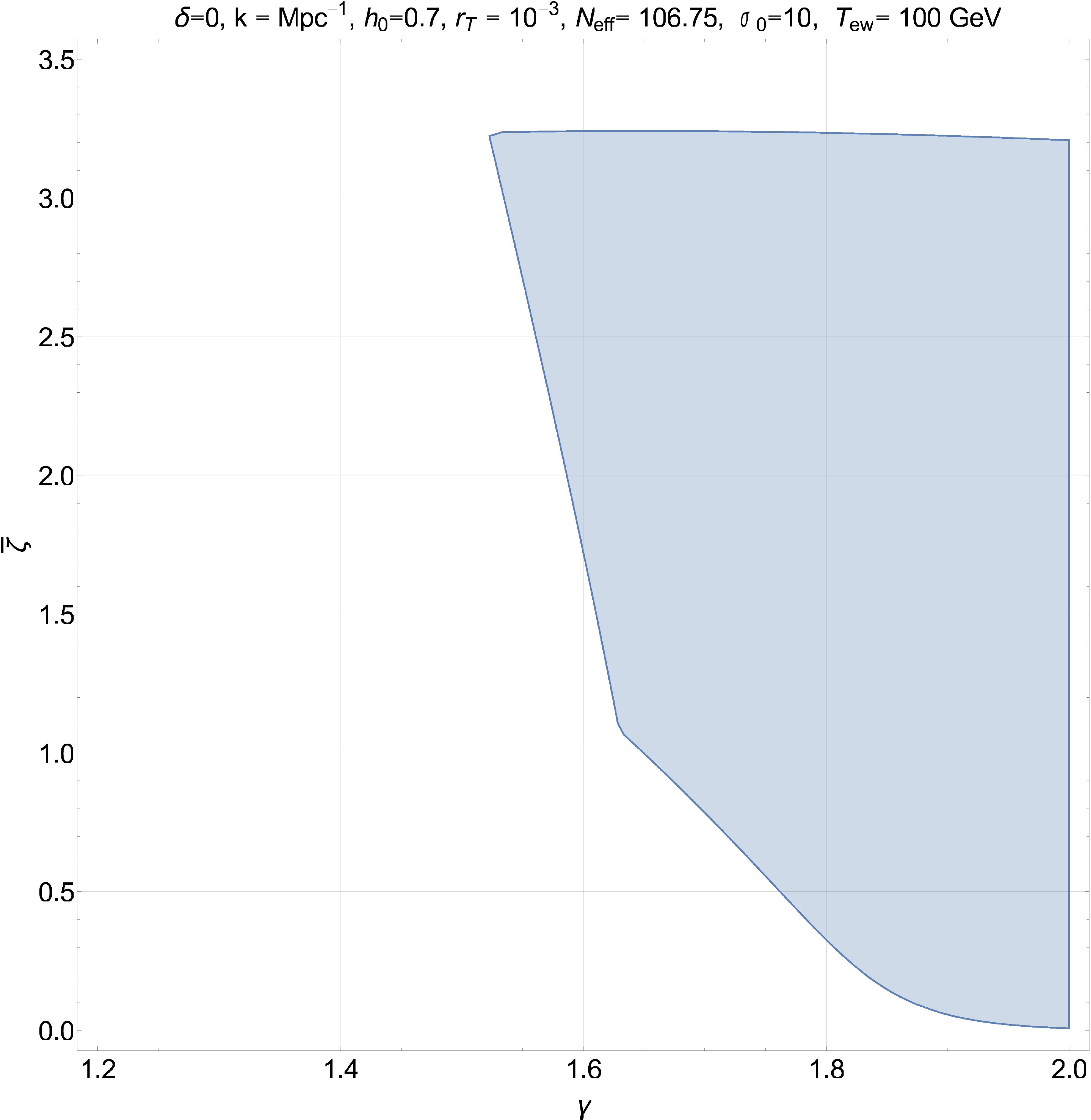}
\caption[a]{The shaded area illustrates the region of the parameter space where the baryogenesis 
and the magnetogenesis constraints are simultaneously satisfied. }
\label{FFF8}      
\end{figure}

In Fig. \ref{FFF6} we instead illustrated the $(\widetilde{\gamma}, \overline{\eta})$ plane: in this case 
the physical features seem roughly the same but there is actually an important  difference.
Both in Figs. \ref{FFF5} and \ref{FFF6} the gauge coupling is ${\mathcal O}(1)$ for $\tau = -\tau_{1}$. However 
 in the case of Fig. \ref{FFF6} the gauge coupling {\em decreases} so that at the initial time of the evolution 
had to be much larger, possibly in the strongly coupled regime of the model. To be compatible with  the strong coupling limit we should therefore 
require that $\sqrt{\lambda_{1}}$ is much larger than $1$. But since $\lambda_{1}$ enters the definition 
of the physical spectrum  the results of Fig. \ref{FFF6} has to be reduced by a factor that
must reduce the initially larger value of the gauge coupling . 
Typically for an inflationary phase characterized by $N = {\mathcal O}(60)$ $e$-folds 
the results of Fig. \ref{FFF6} are suppressed by a factor $e^{ - 2\widetilde{\gamma} N}$. 
This reduction is not present in the case of increasing coupling where, depending on the specific scenario, $e_{1} = \sqrt{4 \pi/\lambda_{1}} ={\mathcal O}(0.01)$. 

Both in Figs. \ref{FFF5} and \ref{FFF6} the typical comoving scale of the spectrum has been fixed 
to the Mpc value. The variation of the comoving scale is considered in Fig. \ref{FFF7} and in the case 
of increasing gauge coupling. 
In Fig. \ref{FFF7} we illustrate two different slices of the parameter space namely the $(k,\,\overline{\zeta})$  and 
the $(k,\,\gamma)$ planes. In both plots we assumed, for the sake of simplicity, $\delta \to 0$.
As expected, for fixed $\overline{\zeta}$ the values of the physical spectra are progressively reduced as 
$\gamma$ decreases. However $\gamma ={\mathcal O}(1.6)$ are still viable. 
Conversely, for fines $\gamma$, the value of the power spectra increase with $\overline{\zeta}$ which is 
however constrained from above by the critical density limits discussed in Fig. \ref{FFF1}.

We finally plot in Fig. \ref{FFF8} the region of the parameter 
space of the scenario where the baryogenesis and the magnetogenesis constraints 
are simultaneously satisfied. As we can see by comparing the two 
classes of requirements, different choices of the parameters lead to similar regions in the $(\gamma, \, \overline{\zeta})$
plane. The shaded areas in Fig. \ref{FFF8} correspond to the situation where 
the values of the BAU are larger than $10^{-10}$ while the magnetogenesis constraints are satisfied. 
We stress that larger values of the BAU are phenomenologically safer since, after the electroweak epoch, different physical 
processes can release entropy and further reduce the generated BAU. 

\newpage 

\renewcommand{\theequation}{7.\arabic{equation}}
\setcounter{equation}{0}
\section{Concluding remarks}
\label{sec7}
The production of hypermagnetic gyrotropy has been investigated when the gauge 
coupling continuously interpolates between a quasi-de Sitter stage of  expansion 
and the radiation phase without entering a strongly coupled regime.
Since during the inflationary epoch the strength of anomalous interactions is constrained 
also the hyermagnetic gyrotropy is indirectly bounded. 

The logic followed in similar situations has been purposely reversed: instead of fixing the couplings and then deducing the 
potential phenomenological implications we treated the  strength of the pseudoscalar 
interactions as a free parameter that must be simultaneously compatible with the critical energy density constraints and with the perturbative evolution of the gauge coupling. Even though the primary focus has been the case where the 
gauge coupling increases and then 
flattens out in the perturbative region, we also found relevant to address 
the dual evolution where at the beginning of inflation the gauge coupling is potentially strong.
From the mutual interplay of these dual scenarios it turns out that 
 the would be  duality symmetry still exchanges hypermagnetic spectra and hyperelectric spectra 
even if the reciprocal transformation is not realized. This collateral observation ultimately determines the late-time 
form of the gauge spectra.

The gauge modes reentering at different times during the radiation-dominated phase 
 impact both the symmetric and the broken phases of the electroweak theory.  
In the plane defined by the strength of the anomalous interaction and by the rate of evolution 
of the gauge coupling the actual weight of the pseudoscalar terms suffices to produce an average 
hypermagnetic gyrotropy whose decay determines the BAU below the electroweak temperature. While the modes contributing to the hypermagnetic gyrotropy are all inside the effective horizon before the electroweak time, the modes reentering just prior to matter-radiation equality also generate large-scale magnetic fields in the nG range. 
According to perspective developed here the strength of the pseudoscalar interactions required 
by the magnetogenesis considerations is compatible with the all the other constraints 
derived both during the quasi-de Sitter stage and at the electroweak epoch. While the present proposal 
is admittedly less conventional than the standard realizations of baryogenesis and leptogenesis
we showed here that it can be implemented in a rather plausible framework which is generally allowed 
by the relevant phenomenological constraints.

\section*{Acknowledgments} 
It is a pleasure to thank T. Basaglia and S. Rohr of the CERN Scientific Information Service for their kind help and assistance.

\newpage

\begin{appendix}

\renewcommand{\theequation}{A.\arabic{equation}}
\setcounter{equation}{0}
\section{Left and Right mode functions and their normalizations}
\label{APPA}
To avoid lengthy digressions, in the two sections of this appendix a general discussion of the circularly polarized mode functions shall be presented. These considerations will be probably useful 
for the interested readers but are technically essential for the derivation of all the major results 
discussed in the bulk of the paper.
In the case of {\em increasing gauge coupling} the evolution of $\sqrt{\lambda}$ follows Eqs. (\ref{TWO1})--(\ref{FIVE2}) and  for $\tau \leq - \tau_{1}$ the explicit form of Eqs. (\ref{MF8})--(\ref{MF9}) is:
\begin{equation}
 f_{k,\, L}^{\,\prime\prime} + \biggl[ k^2 - \frac{\gamma (\gamma -1)}{\tau^2} - \frac{\lambda_{0} \,k \,\gamma}{\tau} \biggr] f_{k,\,L} =0, \qquad f_{k,\, R}^{\,\prime\prime} + \biggl[ k^2 - \frac{\gamma (\gamma -1)}{\tau^2} + \frac{\lambda_{0} \,k \,\gamma}{\tau} \biggr]  f_{k,\, R} =0.
\label{IN2}
\end{equation}
Once the hypermagnetic mode functions have been determined from the normalized solutions of Eq. (\ref{IN2}), the hyperelectric mode functions will be deduced from the following pair 
of relations ultimately coming from the definition of the canonical field operators:
\begin{equation}
g_{k,\, L} = f_{k,\, L}^{\,\prime} - \frac{\gamma}{\tau} f_{k,\, L}, \qquad g_{k,\, R} = f_{k,\, R}^{\,\prime} - \frac{\gamma}{\tau} f_{k,\, R}.
\label{IN3}
\end{equation}
By now introducing the rescaled quantities defined in Eq. (\ref{IN4}) (i.e. $z = 2 \, i \, k \,\tau$ , $\zeta= i\, \lambda_{0} \, \gamma$ and $\mu =| \gamma - 1/2|$) the explicit form of Eq. (\ref{IN6}) is recovered and it coincides with the standardized form of the Whittaker's equation. We follow, in this respect,  the notations of Ref. \cite{abr2} but very similar notations for the Whittaker's equation can be found in other treatments and they all ultimately coincide with the classic discussion of confluent hypergeometric functions of Whittaker 
and Watson \cite{WW}. In the case of {\em decreasing gauge coupling} $\sqrt{\lambda}$ follows instead Eqs. (\ref{ADD1})--(\ref{ADD2}) and for  $\tau \leq - \tau_{1}$ Eqs. (\ref{MF8})--(\ref{MF9}) become:
\begin{equation}
f_{k,\, L}^{\,\prime\prime} + \biggl[ k^2 - \frac{\widetilde{\gamma} (\widetilde{\gamma} +1)}{\tau^2} - \frac{2 \,\lambda_{0} \,k \,\widetilde{\gamma}}{\tau} \biggr] f_{k,\,L} =0, \qquad 
f_{k,\, R}^{\,\prime\prime} + \biggl[ k^2 - \frac{\widetilde{\gamma} (\widetilde{\gamma} +1)}{\tau^2} + \frac{2\,\lambda_{0} \,k \,\widetilde{\gamma}}{\tau} \biggr]  f_{k,\, R} =0.
\label{DEC2}
\end{equation}
We recall here the convention adopted in the bulk of the paper: all the quantities 
with tilde refer to the case of decreasing gauge coupling since this notation turns out to be quite useful 
when discussing the duality properties of the gauge power spectra.
From Eq. (\ref{DEC2}) the hyperelectric mode functions are instead:
\begin{equation}
g_{k,\, L} = f_{k,\, L}^{\,\prime} + \frac{\widetilde{\gamma}}{\tau} f_{k,\, L}, \qquad g_{k,\, R} = f_{k,\, R}^{\,\prime} + \frac{\widetilde{\gamma}}{\tau} f_{k,\, R}.
\label{DEC3}
\end{equation}
From Eqs. (\ref{DEC2})--(\ref{DEC3}) the canonical expressions of the Whittaker's equations (\ref{DEC6}) 
is recovered by using the variables of Eq. (\ref{DEC4}). 
Since the circularly polarized mode functions are related to the 
standardized form of the Whittaker's equation, the most relevant properties will now be listed \cite{abr2}.
If the Whittaker's equation is written in its canonical form as:
\begin{equation}
\frac{d^2 h}{d z^2 } + \biggl[ - \frac{1}{4} + \frac{\alpha}{z} - \frac{\sigma^2 - 1/4}{z^2} \biggr] h=0,
\label{EW1}
\end{equation}
the  two linearly independent solutions of Eq. (\ref{EW1}) are $W_{\alpha,\sigma}(z)$ and $W_{- \alpha,\sigma}(-z)$. For $|z| \gg 1$ the asymptotic limit is $W_{\alpha, \sigma}(z) \to e^{- z/2} z^{\alpha}$. 
According to Eqs. (\ref{IN4}) and (\ref{DEC4}) the variable $z$ is actually defined as $z= 2 i\,k\,\tau$ so that the limit of  $W_{\zeta, \mu}(z)$ for $|z|\gg 1$ is actually a distorted plane wave. In the opposite limit $z\ll1$
\begin{equation}
W_{\alpha, \sigma}(z) = z^{1/2-\sigma } \biggl[\frac{\Gamma(2 \sigma ) }{\Gamma \left(1/2+ \sigma -\alpha\right)}+\frac{\alpha \, z  \Gamma (2 \sigma -1) }{\Gamma \left(1/2 + \sigma -\alpha\right)}+{\mathcal O}\left(z^{5/2}\right)\biggr].
\label{EXP0}
\end{equation}
The Hankel limit  of the Whittaker's functions (i.e. $\alpha \to 0$) allows for a direct connection between 
the, the modified Bessel functions and the Hankel functions \cite{abr2}:
\begin{equation}
\lim_{\alpha\to 0} W_{\alpha, \sigma}(z) = \sqrt{\frac{z}{\pi}} K_{\sigma}(z/2),
\label{EXP1}
\end{equation}
where $K_{\sigma}(z/2)$ are the modified Bessel functions whose analytic continuation is directly 
related to the Hankel functions of first and second kind:
\begin{eqnarray}
K_{\sigma}(z/2) &=& \frac{i \, \pi}{2} e^{i \,\pi\,\sigma/2} \, H_{\sigma}^{(1)}(z e^{i\,\pi/2}/2), \qquad - \pi < \mathrm{arg}z\leq \pi/2,
\label{EXP2}\\
K_{\sigma}(z/2) &=& -\frac{i \, \pi}{2} e^{-i \,\pi\,\sigma/2} \, H_{\sigma}^{(2)}(z e^{-i\,\pi/2}/2), \qquad - \pi/2 < \mathrm{arg}z\leq \pi.
\label{EXP3}
\end{eqnarray}
Equations (\ref{EXP1}) and (\ref{EXP2})--(\ref{EXP3}) are essential to recover 
the correct Hankel limit of various expressions (see e.g. Eqs. (\ref{IN13})).
Furthermore the explicit expressions for the matrix elements reported in Eq. (\ref{AAAin2}) can be swiftly derived by recalling since $H_{\sigma}^{(1)}(w) = J_{\sigma}(w) + i Y_{\sigma}(w)= H_{\sigma}^{(2)\ast}(w)$. When doing the various analytic continuations it is useful to bear in mind that during the quasi-de Sitter stage $\tau$ is {\em negative} while it is {\em positive} in the post-inflationary stage of expansion [see e.g. 
Eq. (\ref{FIVE4}) and discussion thereafter]. When the gauge coupling {\em increases during a 
quasi-de Sitter stage of expansion} the approximate expressions of the mode functions given in Eqs. (\ref{IN8})--(\ref{IN10}) and (\ref{IN11})  can be derived in the limit $ - \, k \tau \ll 1$ 
\begin{eqnarray}
f_{k,\, L}(\tau) &=& \frac{e^{ - \pi \overline{\zeta}/2}\,\, e^{i \pi \mu/2}}{\sqrt{ 2 k}} \, \frac{\Gamma( 2 \mu)\, ( -\, 2 k\tau)^{ 1/2 - \mu}}{\Gamma(1/2 + \mu - i\, \overline{\zeta} )}, 
\nonumber\\
 f_{k,\, R}(\tau) &=& \frac{e^{\pi \overline{\zeta}/2}}\,\,e^{i \pi(\mu -1)/2  }{\sqrt{ 2 k}} \, \frac{\Gamma( 2 \mu)\,( -\, 2 k\tau)^{ 1/2 - \mu}}{\Gamma(1/2 + \mu +  i\, \overline{\zeta})},
\label{IN19}\\
g_{k,\, L}(\tau) &=& \sqrt{ 2 k} \, e^{- \pi \overline{\zeta}/2}\,\,e^{i \pi (\gamma -1/2)/2 } \, \frac{\Gamma( 2 \gamma)\,(- \,2 k \tau)^{- \gamma}}{\Gamma(\gamma - i \overline{\zeta})}, 
\nonumber\\
g_{k,\, R}(\tau) &=& \sqrt{ 2 k} \, e^{\pi \overline{\zeta}/2} \,\, e^{i \pi (\gamma +1/2)/2 } \, \frac{\Gamma( 2 \gamma)\,(- \, 2  k \tau)^{- \gamma}}{\Gamma(\gamma + i \overline{\zeta})};
\label{IN21}
\end{eqnarray}
note that, only for practical reasons, in Eq. (\ref{IN19}) we used directly the quantity $\mu = |\gamma -1/2|$.  The obtained expressions can be further simplified thanks to the standard  
duplication formulas for the Gamma function [i.e. $\Gamma( 2 \sigma) 
= 2^{2\sigma -1/2} \Gamma(\sigma) \Gamma(\sigma+1/2)/\sqrt{2\pi}$]. In the limit $\overline{\zeta}\gg 1$ 
Eqs. (\ref{IN19})--(\ref{IN21}) can also be evaluated with the standard asymptotic expressions valid in the case of the standard Gamma 
functions\footnote{ For instance if  $y$ is a real quantity it is well known that $\bigl| \Gamma( x + i y)\bigr|\to  \sqrt{ 2 \pi} \, |y|^{x -1/2} e^{ - \pi y/2}$
for $y\gg 1$ (see e.g. \cite{abr2}).}. When the gauge coupling {\em decreases during the quasi-de Sitter stage} the limits of the hypermagnetic and hyperelectric mode functions
 for $- k\tau \ll 1$ are instead: 
\begin{eqnarray}
f_{k,\, L}(\tau) &=& \frac{ e^{\pi \overline{\eta}/2} \,\,e ^{i \,\pi (\,\widetilde{\gamma}+1/2)/2}}{\sqrt{ 2 k}}\, ( - \,2  k \tau)^{- \widetilde{\gamma}} \, \frac{\Gamma( 2 \widetilde{\gamma} + 1)}{\Gamma( 1 + \widetilde{\gamma} + i \, \overline{\eta})},
\label{DEC13}\\
f_{k,\, R}(\tau) &=& \frac{ e^{ - \pi \overline{\eta}/2}\,\,e^{i \,\pi (\,\widetilde{\gamma}-1/2)/2} }{\sqrt{ 2 k}}\, ( -\,2  k \tau)^{- \widetilde{\gamma}} \, \frac{\Gamma( 2 \widetilde{\gamma} + 1)}{\Gamma( 1 + \widetilde{\gamma} - i \, \overline{\eta})},
\label{DEC14}\\
g_{k,\, L}(\tau) &=& \frac{\sqrt{ 2 k} \, e^{\pi \overline{\eta}/2}\, e^{i \,\pi (\,\widetilde{\gamma}+1/2)/2}}{\Gamma( 1 + \widetilde{\gamma} + i \overline{\eta})}\,
( \,2  k\tau)^{-\widetilde{\gamma}} \,\biggl[ \overline{\eta} \Gamma( 2 \widetilde{\gamma}) - ( \widetilde{\gamma} -1) ( 2 \overline{\eta}^2 + \widetilde{\gamma})\Gamma( 2 \widetilde{\gamma} -2) (- \,k\tau) + {\mathcal O}(k^2 \tau^2) \biggr]
\nonumber\\
&+& ( - 2 k\tau)^{\widetilde{\gamma}} \biggl[ \frac{\sqrt{2 k}\,e^{\pi\overline{\eta}/2}\, e^{- i \,\pi (\,\widetilde{\gamma}+1/2)/2}\Gamma( -2 \widetilde{\gamma})}{\Gamma(i \overline{\eta} - \widetilde{\gamma}) }+ {\mathcal O}(k\tau) \biggr],
\label{DEC15}\\
g_{k,\, R}(\tau) &=& \frac{\sqrt{ 2 k} \, e^{- \pi \overline{\eta}/2}\,e^{i \,\pi (\,\widetilde{\gamma}-1/2)/2}}{\Gamma( 1 + \widetilde{\gamma} - i \overline{\eta})}\,( - 2  k\tau)^{-\widetilde{\gamma}} \,
\biggl[- \overline{\eta} \Gamma( 2 \widetilde{\gamma}) + ( \widetilde{\gamma} -1) ( 2 \overline{\eta}^2 + \widetilde{\gamma})\Gamma( 2 \widetilde{\gamma} -2) (- k\tau) + {\mathcal O}(k^2 \tau^2) \biggr]
\nonumber\\
&-& ( -\,2 k\tau)^{\widetilde{\gamma}} \biggl[ \frac{\sqrt{2 k}\,e^{ - \pi\overline{\eta}/2} e^{- i \,\pi (\,\widetilde{\gamma}-1/2)/2}\, \Gamma( -2 \widetilde{\gamma})}{\Gamma(- i \overline{\eta} - \widetilde{\gamma})} + {\mathcal O}(k\tau) \biggr].
\label{DEC16}
\end{eqnarray}
Note that in the asymptotic expression of Eqs. (\ref{DEC15})--(\ref{DEC16}) the effective expansion 
parameter is in fact represented by $(- k \tau ) \, \overline{\eta} < 1$. As in the case of Eqs. (\ref{IN19}) and (\ref{IN21}) 
the expansions are valid in the long wavelength limit also when $\overline{\eta} \gg 1$ as long as $ |k\tau| < 1/\overline{\eta}$.

\renewcommand{\theequation}{B.\arabic{equation}}
\setcounter{equation}{0}
\section{Late-time forms of the mode functions}
\label{APPB}
 Both when the gauge coupling increases and decreases the late-time form of the mode functions could be expressed as the superposition of the 
two linearly independent solutions of the homogeneous Whittaker's equation:
\begin{equation}
f_{k,\, X} = A_{-} \,\overline{F}^{(2)}_{k,\,X} + A_{+}\,  \overline{F}^{(1)}_{k,\,X}, \qquad
g_{k,\, L} = A_{-} \,\overline{G}^{(2)}_{k,\,X} + A_{+}\,  \overline{G}^{(1)}_{k,\,X}, 
\label{IN20a}
\end{equation}
where $X = L,\, R$ and  ($F_{k,\,X}^{(1)}$, $F_{k,\,X}^{(2)}$) and ($G_{k,\,X}^{(1)}$, $G_{k,\,X}^{(2)}$) denote  
the two linearly independent solutions for $\tau \geq - \tau_{1}$. 
In the case of {\em increasing gauge coupling} the 
normalized form of the linearly independent solutions follows from Eqs. (\ref{IN21d}) and (\ref{IN21f}). For the $L$-polarization 
the results are
\begin{eqnarray}
&& F_{k\,L}^{(1)}(\tau) = D^{(1)}_{L}(k, \overline{\xi}) \, W_{\xi,\, \nu}(w), \qquad F_{k\,L}^{(2)}(\tau) = D^{(2)}_{L}(k,\overline{\xi}) \, W_{- \xi,\, \nu}(- w),
\label{IN21h}\\
&& G_{k,\,L}^{(1)}(\tau) = 2 \, i\, k \, D^{(1)}_{L}(k, \overline{\xi}) \biggl[ \frac{ w - 2 \xi + 2 \delta}{2 w} W_{\xi,\,\nu}(w) - \frac{W_{1 + \xi,\,\nu}(w)}{w}\biggr],
\label{IN21m}\\
&& G_{k,\,L}^{(2)}(\tau) = 2 \, i\, k \, D^{(2)}_{L}(k, \overline{\xi}) \biggl[ \frac{- w + 2 \xi + 2 \delta}{2 w} W_{-\xi,\,\nu}(-w) - \frac{W_{1 - \xi,\,\nu}(-w)}{w}\biggr],
\label{IN21n}
\end{eqnarray}
In the case of the $R$-polarization the solutions are instead 
\begin{eqnarray}
&& F_{k\,R}^{(1)}(\tau) = D^{(1)}_{R}(k, \overline{\xi}) \, W_{-\xi,\, \nu}(w), \qquad F_{k\,R}^{(2)}(\tau) = D^{(2)}_{R}(k, \overline{\xi}) \, W_{ \xi,\, \nu}(- w),
\label{IN21ha}\\
&& G_{k,\,R}^{(1)}(\tau) = 2 \, i\, k \, D^{(1)}_{R}(k, \overline{\xi}) \biggl[ \frac{(w + 2 \xi + 2 \delta)}{2 w} W_{-\xi,\,\nu}(w) - \frac{W_{1 - \xi,\,\nu}(w)}{w}\biggr],
\label{IN21ma}\\
&& G_{k,\,R}^{(2)}(\tau) = 2 \, i\, k \, D^{(2)}_{R}(k, \overline{\xi}) \biggl[ \frac{(- w - 2 \xi + 2 \delta)}{2 w} W_{\xi,\,\nu}(-w) - \frac{W_{1 + \xi,\,\nu}(-w)}{w}\biggr].
\label{IN21na}
\end{eqnarray}
The normalization appearing in Eqs. (\ref{IN21h})-- (\ref{IN21n}) and (\ref{IN21ha})--(\ref{IN21na}) are:
\begin{eqnarray}
&& D_{L}^{(1)}(k, \overline{\xi}) = \frac{e^{i \pi/4 + \pi \overline{\xi}/2} }{\sqrt{ 2 k}}, \qquad D_{L}^{(2)}(k, \overline{\xi}) = \frac{e^{-i \pi/4 + \pi \overline{\xi}/2} }{\sqrt{ 2 k}},
\nonumber\\
&& D_{R}^{(1)}(k, \overline{\xi}) = \frac{e^{-i \pi/4 - \pi \overline{\xi}/2} }{\sqrt{ 2 k}}, \qquad D_{R}^{(2)}(k, \overline{\xi}) = \frac{e^{i \pi/4 - \pi \overline{\xi}/2} }{\sqrt{ 2 k}}.
\end{eqnarray}
The same procedure can be repeated in the case of {\em decreasing gauge} coupling and the result go as follows:
\begin{eqnarray}
&& \widetilde{F}_{k\,L}^{(1)}(\tau) = \widetilde{D}^{(1)}_{L}(k, \overline{\theta}) \, W_{-\theta,\, \widetilde{\nu}}(w), \qquad \widetilde{F}_{k\,L}^{(2)}(\tau) =\widetilde{D}^{(2)}_{L}(k,\overline{\theta}) \, W_{\theta,\, \widetilde{\nu}}(- w),
\label{IN21hDEC}\\
&& \widetilde{G}_{k,\,L}^{(1)}(\tau) = 2 \, i\, k \, \widetilde{D}^{(1)}_{L}(k, \overline{\theta}) \biggl[ \frac{ w + 2 \theta - 2 \widetilde{\delta}}{2 w} W_{-\theta,\,\widetilde{\nu}}(w) - \frac{W_{1 - \theta,\,\widetilde{\nu}}(w)}{w}\biggr],
\label{IN21mDEC}\\
&& \widetilde{G}_{k,\,L}^{(2)}(\tau) = - 2 \, i\, k \, D^{(2)}_{L}(k, \overline{\theta}) \biggl[ \frac{ w + 2 \theta + 2 \widetilde{\delta}}{2 w} W_{-\xi,\,\widetilde{\nu}}(-w) +\frac{W_{1 + \theta,\,\widetilde{\nu}}(-w)}{w}\biggr],
\label{IN21nDEC}
\end{eqnarray}
In the case of the $R$-polarization the solutions are instead 
\begin{eqnarray}
&& \widetilde{F}_{k\,R}^{(1)}(\tau) = D^{(1)}_{R}(k, \overline{\theta}) \, W_{\theta,\, \widetilde{\nu}}(w), \qquad \widetilde{F}_{k\,R}^{(2)}(\tau) = D^{(2)}_{R}(k, \overline{\theta}) \, W_{ -\theta,\, \widetilde{\nu}}(- w),
\label{IN21haDEC}\\
&& \widetilde{G}_{k,\,R}^{(1)}(\tau) = 2 \, i\, k \, D^{(1)}_{R}(k, \overline{\theta}) \biggl[ \frac{(w - 2 \theta - 2 \widetilde{\delta})}{2 w} W_{\theta,\,\widetilde{\nu}}(w) - \frac{W_{1 + \theta,\,\widetilde{\nu}}(w)}{w}\biggr],
\label{IN21maDEC}\\
&& \widetilde{G}_{k,\,R}^{(2)}(\tau) = - 2 \, i\, k \, D^{(2)}_{R}(k, \overline{\theta}) \biggl[ \frac{( w - 2 \theta + 2 \widetilde{\delta})}{2 w} W_{-\theta,\,\widetilde{\nu}}(-w) - 2 \frac{W_{1 - \theta,\,\widetilde{\nu}}(-w)}{w}\biggr].
\label{IN21naDEC}
\end{eqnarray}
This time the normalizations appearing in Eqs. Eqs. (\ref{IN21hDEC})--(\ref{IN21nDEC})
and (\ref{IN21haDEC})--(\ref{IN21naDEC}) are:
\begin{eqnarray}
&& \overline{D}_{L}^{(1)}(k, \overline{\theta}) = \frac{e^{i \pi/4 - \pi \overline{\theta}/2} }{\sqrt{ 2 k}}, \qquad D_{L}^{(2)}(k, \overline{\theta}) = \frac{e^{-i \pi/4 - \pi \overline{\theta}/2} }{\sqrt{ 2 k}},
\nonumber\\
&& \overline{D}_{R}^{(1)}(k, \overline{\theta}) = \frac{e^{-i \pi/4 + \pi \overline{\theta}/2} }{\sqrt{ 2 k}}, \qquad D_{R}^{(2)}(k, \overline{\theta}) = \frac{e^{i \pi/4 + \pi \overline{\theta}/2} }{\sqrt{ 2 k}}.
\end{eqnarray}
It is finally appropriate, for the present ends, to write explicitly the entries of the matrix $\widetilde{\,{\mathcal M}\,}^{(X)}(w_{1},\, w,\,\widetilde{\delta})$. We remind that the entries of ${\mathcal M}^{(X)}(w_{1},\, w,\,\delta)$
have been already discussed in the bulk of the paper (see Eqs. (\ref{IN22L})--(\ref{IN24L}) and discussion thereafter). For the case  $\widetilde{\,{\mathcal M}\,}^{(X)}(w_{1},\, w,\,\widetilde{\delta})$ the discussion 
is that same so that we shall only give the final results.
Using the mode functions given in Eqs. (\ref{IN21hDEC}) and (\ref{IN21mDEC})--(\ref{IN21nDEC}) we have that the entries of $\widetilde{\,{\mathcal M}\,}^{(L)}(w_{1},\, w,\,\widetilde{\delta})$ are:
\begin{eqnarray}
\widetilde{A}^{(L)}_{f\, f}(w_{1}, w, \widetilde{\delta}) &=& \frac{e^{- \pi \overline{\theta}}}{2 w_1}\biggl\{
(2 \widetilde{\delta} - 2 \theta -w_{1}) W_{\theta ,\widetilde{\nu} }(-w) W_{-\theta ,\widetilde{\nu} }(w_{1})-W_{-\theta ,\widetilde{\nu} }(w) \biggl[(2 \widetilde{\delta}
   + 2 \theta +w_{1}) W_{\theta ,\widetilde{\nu} }(-w_{1})
   \nonumber\\
  &+&2 W_{\theta +1,\widetilde{\nu} }(-w_{1})\biggr]+2 W_{\theta ,\widetilde{\nu} }(-w) W_{1-\theta ,\widetilde{\nu}
   }(w_{1})\biggr\},
 \label{ATFFL}\\
  \widetilde{A}^{(L)}_{f\, g}(w_{1}, w, \widetilde{\delta}) &=& - \frac{i}{2} e^{- \pi \overline{\theta}}
  \biggl\{W_{\theta ,\widetilde{\nu} }(-w) W_{-\theta ,\widetilde{\nu} }(w_{1})-W_{-\theta ,\widetilde{\nu} }(w) W_{\theta ,\widetilde{\nu} }(-w_{1})\biggr\},
  \label{ATFGL}\\
\widetilde{A}^{(L)}_{g\, f}(w_{1}, w, \widetilde{\delta}) &=&  -
\frac{i \, e^{- \pi \overline{\theta}}}{2 w w_{1}} \biggl\{\biggl[-(2 \widetilde{\delta} + 2\theta +w) W_{\theta ,\widetilde{\nu} }(-w)-2 W_{\theta +1,\widetilde{\nu} }(-w)\biggr] \biggl[(-2 \widetilde{\delta} +2 \theta +w_{1}) W_{-\theta ,\widetilde{\nu} }(w_{1})
\nonumber\\
&-& 2 W_{1-\theta ,\widetilde{\nu}
   }(w_{1})\biggr]- \biggl[(2 \theta-2 \widetilde{\delta}  +w) W_{-\theta ,\widetilde{\nu} }(w)-2 W_{1-\theta ,\widetilde{\nu} }(w)\biggr] \biggl[- (2 \widetilde{\delta} +2\theta  +w_{1}) W_{\theta ,\widetilde{\nu} }(-w_{1})
 \nonumber\\  
 &-& 2 W_{\theta +1,\widetilde{\nu} }(-w_{1})\biggr]\biggr\},
   \label{ATGFL}\\
   \widetilde{A}^{(L)}_{g\, g}(w_{1}, w, \widetilde{\delta}) &=& \frac{1}{2 w} e^{- \pi \overline{\theta}}
   \biggl\{W_{\theta ,\widetilde{\nu} }(-w_{1}) \biggl[2 W_{1-\theta ,\widetilde{\nu} }(w)-(2 \theta  -2 \widetilde{\delta} +w) W_{-\theta ,\widetilde{\nu} }(w)\biggr]
  \nonumber\\ 
   &-& W_{-\theta ,\widetilde{\nu} }(w_{1}) \biggl[(2 \widetilde{\delta} +2 \theta +w) W_{\theta
   ,\widetilde{\nu} }(-w)+2 W_{\theta +1,\widetilde{\nu} }(-w)\biggr]\biggr\}.
   \label{ATGGL}
\end{eqnarray}
The results for $\widetilde{\,{\mathcal M}\,}^{(R)}(w_{1},\, w,\,\widetilde{\delta})$ can be obtained, in a similar manner, from Eqs. (\ref{IN21haDEC}) and (\ref{IN21maDEC})--(\ref{IN21naDEC}). This is 
however not necessary since the various entries can be obtained from the above equations 
by flipping the signs of $w$ and $w_{1}$ and by also flipping the sign of $\overline{\theta}$ 
in the normalizations. So for instance we will have that from (\ref{ATFGL}) 
the expression of $\widetilde{A}^{(R)}_{f\, g}(w_{1}, w, \widetilde{\delta})$ will be:
 \begin{equation}
 \widetilde{A}^{(R)}_{f\, g}(w_{1}, w, \widetilde{\delta}) = - \frac{i}{2} e^{ \pi \overline{\theta}}
  \biggl\{W_{\theta ,\widetilde{\nu} }(w) W_{-\theta ,\widetilde{\nu} }(-w_{1})-W_{-\theta ,\widetilde{\nu} }(-w) W_{\theta ,\widetilde{\nu} }(w_{1})\biggr\}.
\end{equation}
Exactly the same strategy can be used to obtain $\widetilde{A}^{(R)}_{f\, f}(w_{1}, w, \widetilde{\delta})$, $\widetilde{A}^{(R)}_{g\, f}(w_{1}, w, \widetilde{\delta})$ and $\widetilde{A}^{(R)}_{g\, g}(w_{1}, w, \widetilde{\delta})$ from the corresponding expressions valid in the case of the $L$-polarization 
[i.e. Eqs. (\ref{ATFFL}), (\ref{ATGFL}) and (\ref{ATGGL})].

\end{appendix}

\end{document}